\documentclass[useAMS,usenatbib,fleqn]{mnras}
\usepackage[utf8]{inputenc}
\usepackage{graphicx,color}
\usepackage{mathtools}	
\usepackage{amsmath}
\usepackage{amssymb}
\usepackage{bm}
\usepackage{pstricks} 
\usepackage{arydshln} 
\usepackage[normalem]{ulem}
   \usepackage[british]{babel}             
   \usepackage{newtxtext}                  
   \usepackage[slantedGreek]{newtxmath}    
   \usepackage[T1]{fontenc}                
\def\apj{ApJ}

\def\mnras{MNRAS}
\def\aap{A\&A}
\def\apjl{ApJ}

\def\apjs{ApJS}

\def\aapr{ARA\&A}

\newcommand{\env}{_\mathrm{env}}					   	

\newcommand{\gas}{_\mathrm{gas}}

\newcommand{\soft}{_\mathrm{soft}}

\newcommand{\init}{_\mathrm{i}}
\newcommand{\final}{_\mathrm{f}}

\newcommand{\mass}{_\mathrm{m}}

\newcommand{\Rsun}{\,\mathrm{R_\odot}}
\newcommand{\Msun}{\,\mathrm{M_\odot}}

\newcommand{\amb}{_\mathrm{amb}}

\newcommand{\CE}{_\mathrm{CE}}

\newcommand{\bind}{_\mathrm{bind}}

\newcommand{\unb}{_\mathrm{unb}}

\newcommand{\kB}{k_\mathrm{B}}
\newcommand{\mH}{m_\mathrm{H}}
\newcommand{\mHe}{m_\mathrm{He}}
\newcommand{\rec}{_\mathrm{rec}}
\newcommand{\thm}{_\mathrm{thm}}
\newcommand{\HeI}{\mathrm{He\,I}}
\newcommand{\HeII}{\mathrm{He\,II}}
\newcommand{\HeIII}{\mathrm{He\,III}}
\newcommand{\HI}{\mathrm{H\,I}}
\newcommand{\HII}{\mathrm{H\,II}}
\newcommand{\Hmol}{\mathrm{H_2}}

\newcommand{\rad}{_\mathrm{rad}}

  \newcommand{\cm}{\,{\rm cm}}

  \newcommand{\erg}{\,{\rm erg}}
  \newcommand{\g}{\,{\rm g}}
  \newcommand{\gcmcmcm}{\,{\rm g\,cm^{-3}}}

  \newcommand{\K}{\,{\rm K}}

  \newcommand{\s}{\,{\rm s}}
  \newcommand{\yr}{\,{\rm yr}}     
  \newcommand{\da}{\,{\rm d}}     
  \newcommand{\dyncmcm}{\,{\rm dyn\,cm^{-2}}}

  \newcommand{\amu}{\,\mathrm{amu}}

\title[Recombination in common envelope evolution] 
{How negative feedback and the ambient environment limit the influence of recombination in common envelope evolution}
\author[L.~Chamandy et al.]{Luke Chamandy,$^{1,2}$\thanks{lchamandy@niser.ac.in}
Jonathan Carroll-Nellenback,$^{3,2}$\thanks{jcarrol5@ur.rochester.edu}
Eric G.~Blackman,$^2$\thanks{blackman@pas.rochester.edu}
Adam Frank,$^2$\thanks{afrank@pas.rochester.edu}
\newauthor 
Yisheng Tu,$^{4,2}$\thanks{yt2cr@virginia.edu}
Baowei Liu,$^{3,2}$\thanks{baowei.liu@rochester.edu}
Yangyuxin Zou$^2$\thanks{yzou5@ur.rochester.edu}
and Jason Nordhaus$^{5,6}$\thanks{nordhaus@astro.rit.edu
}
\\
$^1$National Institute of Science Education and Research, An OCC of Homi Bhabha National Institute, Bhubaneswar 752050, Odisha, India\\
$^2$Department of Physics and Astronomy, University of Rochester, Rochester NY 14627, USA\\
$^3$Center for Integrated Research Computing, University of Rochester, Rochester NY 14627, USA\\
$^4$Department of Astronomy, University of Virginia, Charlottesville, VA 22904, USA\\
$^5$Center for Computational Relativity and Gravitation, Rochester Institute of Technology, Rochester, NY 14623, USA\\
$^6$National Technical Institute for the Deaf, Rochester Institute of Technology, Rochester, NY 14623, USA
}

\begin{document}


\maketitle

\begin{abstract}
We perform 3D hydrodynamical simulations to study recombination and ionization 
during the common envelope (CE) phase of binary evolution, 
and develop techniques to track the ionic transitions in time and space.  
We simulate the interaction of a $2\Msun$ red giant branch primary and a $1\Msun$ companion modeled as a particle.  
We compare a run employing a tabulated equation of state (EOS) that accounts for ionization and recombination,  
with a run employing an ideal gas EOS.
During the first half of the simulations, 
$\sim15$ per cent more mass is unbound in the tabulated EOS run due to the release of recombination energy,
but by simulation end the difference has become negligible. We explain this as being a consequence of  
(i)~the tabulated EOS run experiences a shallower inspiral and hence smaller orbital energy release at late times
because recombination energy release expands the envelope and reduces drag,
and (ii)~collision and mixing between expanding envelope gas, ejecta and circumstellar ambient gas assists in unbinding the envelope, 
but does so less efficiently in the tabulated EOS run where some 
of the energy transferred to bound envelope gas is used for ionization.
The rate of mass unbinding is approximately constant in the last half of the simulations 
and the orbital separation  steadily decreases  at late times.
A simple linear extrapolation predicts a CE phase duration of $\sim2\yr$, after which the envelope would be unbound.
\end{abstract}

\defcitealias{Chamandy+18}{Paper~I}
\defcitealias{Chamandy+19a}{Paper~II}
\defcitealias{Chamandy+19b}{Paper~III}
\section{Introduction}
\label{sec:intro}
In the common envelope (CE) scenario, 
a star engulfs a typically much smaller companion and the companion and core of the primary star
spiral in together until either the envelope is ejected or the core and companion merge \citep{Paczynski76}.
The CE phase is short-lived and hence difficult to observe.
However, 
understanding CE evolution (CEE) is crucial for understanding phenomena 
as wide-ranging as neutron star-neutron star mergers, 
type Ia supernovae and planetary nebulae (for recent reviews see \citealt{Ivanova+20} and \citealt{Roepke+Demarco22}).
While theoretical work on CEE has come a long way, simulating CEE up to a realistic end state 
(e.g. complete envelope ejection and stabilization of the orbit) has still not been  achieved.
The overarching reason  is that simulating such a large dynamic range of spatial, temporal and density scales remains  challenging.

Nevertheless, such calculations have been useful in constraining the effects of various physical processes during CEE.
One such process is the recombination of ions and electrons as the envelope expands and cools,
which releases recombination energy.
As has become somewhat standard in the CE literature, 
we define recombination energy as the \textit{latent} energy contained by a plasma that would be released upon recombination.

While the recombination energy content of the initial envelope is generally substantial and thus could potentially
play an important role in envelope unbinding, its efficacy remains  unclear.
This is partly because some released recombination energy radiates away
\citep{Soker+Harpaz03,Ivanova+13b,Sabach+17,Ivanova18,Grichener+18,Soker+18,Reichardt+20,Lau+22a}.
How much is radiated is not addressed in the present work.
Rather, we ask:
what difference does recombination energy make to CEE, 
assuming (optimistically) that released recombination energy is thermalized locally?
In particular:  how much extra envelope mass does recombination unbind at any given time, 
where in the envelope does this extra unbinding happen,
and what are the relative contributions of the various ionic species?
Even in the optimistic case that assumes local absorption and thermalization,
some of the recombination energy  would be released into already unbound gas, and play  no further role in envelope unbinding.  Therefore, there is an efficiency associated with the transfer of released recombination energy 
to binding energy of the remaining envelope.

Sometimes authors count gas as unbound if its total energy density,  
including that of the recombination energy, is positive.
This is misleading because it greatly overestimates the role of recombination energy in envelope unbinding.
First, the recombination  energy is still latent.
Second, upon release, it might  contribute little to unbinding due to the inefficiencies mentioned above.
A more justifiable  approach is to include only kinetic (bulk and perhaps thermal) 
and potential energy terms in assessing whether material is unbound,
and compare the unbound mass in a simulation with a realistic tabulated EOS 
that includes recombination energy with a simulation that does not.
If cooling is neglected, it is appropriate to use a $\gamma=5/3$ ideal gas EOS for the latter simulation.
When such a comparison has been made \citep{Ohlmann16,Reichardt+20,Sand+20,Lau+22a,Gonzalez-bolivar+22}, 
it has been found that the unbound envelope mass is significantly higher in the tabulated EOS run.
Moreover, the fractional difference in the unbound mass curves between the runs usually increases with time,
indicating that the effect becomes more important at late times.%

The details of how and how much recombination energy can really assist envelope unbinding are still emerging.
\citet{Reichardt+20} included a spatial analysis 
where they mapped out the ionization fractions of various species using the Saha equation.
This allowed them to identify recombination fronts and analyze their evolution.
They then compared this data with data of where unbinding is happening 
in both $\gamma=5/3$ and tabulated EOS runs, allowing them to infer how recombination affects unbinding.
In this work we  in addition employ gas tracers to track envelope gas of a given ionization species at $t=0$. 
Comparing the present ionization state calculated using the Saha equation with 
the original ionization state recorded by the tracers
enables us to map out the net change in the ionization state of gas in the simulation,
as a function of both space and time.
This method also makes it possible to directly compute the net energy 
absorbed by ionization or released by recombination for some subset of the gas -- for example, 
gas which is presently bound.

We also find that we need to assess the role played by the ambient medium in our simulations,
i.e.~extra gas that fills the simulation box.
We thus compare two $\gamma=5/3$ simulations, 
one with a dense ambient medium and one with a low-density ambient medium.
This helps to place bounds on effects that would stem from the presence of a circumbinary torus,
formed in nature by mass loss that precedes the CE interaction.

The structure of the paper is as follows.
In Section~\ref{sec:methods}, we explain the methods used in the setup, 
running and analysis of the simulations,
and describe the physical model, 
including the initial conditions and the parameters of the runs performed.
Results are then presented in Section~\ref{sec:results}.
In Section~\ref{sec:unbinding} we present the evolution of the unbound mass with time,
in Section~\ref{sec:energization} we analyze the transfer of energy between the various components,
in Section~\ref{sec:recombination} we explore the ionic evolution and the release of recombination energy,
and in Section~\ref{sec:ambient_role} we discuss the role of the ambient medium.
We summarize our results and conclude in Section~\ref{sec:conclusions}.

\section{Methods and Models}\label{sec:methods}
We first clarify the terminology we use, which might otherwise cause confusion.
``Envelope'' or ``envelope gas'' refers to gas that was originally part of the stellar envelope,
and ``unbound envelope gas'' refers to such gas which has become unbound 
(and hence is no longer technically part of the envelope).
Envelope gas is separate from ambient gas, 
which is extra gas in the simulation not included in the original spherically symmetric stellar model used in our initial condition.
The gravitating point particles are referred to as core particles, separate from the gas
(even though they can represent gaseous astrophysical bodies).

Our study involves a binary system consisting of a $1.96\Msun$, $48.1\Rsun$ red giant branch (RGB) primary star 
with a $0.366\Msun$ core and a $0.978\Msun$ secondary representing a main sequence star, white dwarf, 
or even a very low-mass neutron star.
The primary core and secondary are modeled as gravitation-only point particles.
A spline function is used to model the potential around the point particles,
which becomes Newtonian outside the spline softening sphere, with radius $r\soft$.
The initial density and pressure profiles of the primary are mapped to our 3D grid
from a 1D Modules for Experiments in Stellar Astrophysics (MESA) snapshot 
\citep{Paxton+11,Paxton+13,Paxton+15,Paxton+19}.
The snapshot is taken from a MESA release 12778 simulation.
It matches almost exactly the snapshot used for our previous RGB simulations using release 8845
\citep{Chamandy+18}, 
except that here we increased the radial resolution by a factor $\sim20$ to smooth the profile.
We utilize the adaptive mesh refinement (AMR) code AstroBEAR \citep{Cunningham+09,Carroll-Nellenback+13}, 
and employ an HLLC Riemann solver.
Simulations are carried out in a centre of momentum frame with the grid 
centred on the initial location of the centre of the primary star.
The simulation box of side length $L_\mathrm{box}=1150\Rsun$ is discretized into $512^3$ AMR level zero cells, 
corresponding to a base resolution of $\delta_0\approx2.25\Rsun$.
For further numerical method details, see \citet{Chamandy+18,Chamandy+19a,Chamandy+19b,Chamandy+20}.

Initially, the envelope and some of the ambient medium surrounding the RGB star are resolved at AMR level $4$, 
or $\delta_4\approx0.140\Rsun$,
and this refinement zone gradually reduces as the particle separation $a$ decreases.
However, unlike in our previous RGB simulations, 
AMR level $5$, with $\delta_5\approx0.070\Rsun$, 
was added around the point particles out to slightly farther than the softening sphere.
This extra level of refinement helps to conserve energy and to avoid artificial reduction 
of the central density and pressure during the simulation.
Buffer zones with $16$ cells were included to smoothly transition between AMR levels.
At $t=25.23\da$, the softening radius around the particles was halved to $\approx1.2\Rsun$ 
and a sixth level of refinement with $\delta_6\approx0.035\Rsun$ was added around each particle.
At $t=50.46\da$ the softening radius was again halved to $\approx0.6\Rsun$, 
and AMR level $7$ was added, with $\delta_7\approx0.018\Rsun$.
The enhanced refinement zones are approximately spherical in shape and contain uniform resolution.

\subsection{Initial orbit}\label{sec:io}
The binary is initialized in a circular orbit at a separation $a\init=49\Rsun$, 
slightly outside the stellar surface ($R_1=48.1\Rsun$).
The primary has no initial rotation.
The initial conditions are almost identical to those in our previous works,
and similar to those of \citet{Ohlmann+16a} (see also \citealt{Ohlmann16}) and \citet{Prust+Chang19},
which allows for quantitative comparison (see \citealt{Chamandy+19a} and App.~\ref{sec:comparison}).

\subsection{Ambient medium}\label{sec:ambient}
The pressure scale height near the core and at the stellar surface of the MESA profile are too small for our 3D simulation to resolve.
Therefore, we cut out the RGB core and replaced it with a spline-softened gravitating point particle,
with softening length $r\soft=2.41\Rsun$, equal to the cut radius,
along with a core density and pressure profile obtained by solving a modified Lane-Emden equation \citep{Ohlmann+17}.
This solution also incorporates an iteration over the particle mass to ensure that the mass below the cut radius is preserved \citep{Chamandy+18}.%
\footnote{We performed a short simulation starting from an initial condition 
that does not include this iterative step and found that the results are similar, as expected.}
The softening radius of the secondary is the same as that of the RGB core particle.

The ambient medium has uniform density $\rho\amb=1.0\times10^{-9}\gcmcmcm$ 
and uniform pressure $P\amb=1.0\times10^5\dyncmcm$;
the ambient pressure is \textit{added} everywhere on the grid to ensure a smooth transition in pressure at the stellar surface,
while the density undergoes a sharp transition at the stellar surface.
The value of $P\amb$ is chosen such that the RGB pressure profile effectively 
truncates (but smoothly) just inside the outer radius of the star 
to avoid the very small pressure scale height at that location in the MESA profile.
The value of $\rho\amb$ is about $7$ times smaller than the density at the outer radius of the star $\rho(R_1)$;
smaller values are possible but would result in a higher ambient temperature and prohibitively small timesteps.
We find that we require at least eight resolution elements per scale height to adequately resolve the initial stellar profile.
This number was determined by studying the smoothness and stability of both the core and surface, 
where the scale heights are smallest, during the first $\sim1\da$ of the simulations.%
\footnote{We did not perform a preliminary run to prepare the initial condition for the simulation.
Including such a relaxation run makes little difference \citep{Chamandy+18},
and our results agree 
with similar simulations by other authors who employed a relaxation run, as discussed in \citet{Chamandy+19a}.
Moreover,  artifacts due to the Cartesian grid can be magnified when a relaxation run is used 
because the star is motionless.}

To ascertain the effect of the ambient medium on the results,
we perform one run for which we drastically reduce the density and pressure of the ambient gas.
We reduce the ambient medium near the beginning of an ideal gas run,
at a time chosen as a compromise between reducing the ambient as early as possible 
and retaining the stability of the stellar surface. Also, the later the restart from the fiducial, the less the computational cost.%
\footnote{It is not possible to start with these values at $t=0$ 
because numerical artifacts occur just outside the stellar surface,
propagate into the ambient medium, and crash the simulation.
We also  tried a hydrostatic atmosphere that transitions to a uniform ambient medium at larger radius 
where the density and pressure are low.
However, this  was numerically  unstable.}
This ambient reduction procedure is made possible by the tracer (Sec.~\ref{sec:tracers}) 
assigned to the ambient gas during the  initial setup at $t=0$.
Thus, between $t_1=19.9\da$ and $t_2=24.5\da$, 
the ambient density and pressure are reduced by three orders of magnitude according to the expressions
\begin{equation}
  \rho\amb = \rho_\mathrm{amb,0} \exp\left\{\ln(\epsilon)\left[1-\exp\left(-x^n\right)\right]\right\}
\end{equation}
and 
\begin{equation}
  P\amb = P_\mathrm{amb,0} \exp\left\{\ln(\epsilon)\left[1-\exp\left(-x^n\right)\right]\right\},
\end{equation}
where $x\equiv (t-t_1)/\tau$, $\epsilon=10^{-3}$, $n=4$ and $\tau=2.3\da$.
The parameter $\tau$ determines how gradual the transition is.
The code was  stable for test runs with $\tau$ an order of magnitude shorter,
but we used the above value to be conservative.

\subsection{Tracers}\label{sec:tracers}
In the initial setup, tracers are added to track gas that is initially in the envelope ($0< r\init < R_1$) or ambient ($r\init>R_1$).
Tracking the ambient gas allows us to exclude it in post-processing \citep{Zou+22} 
or to remove it during the simulation (Sec.~\ref{sec:ambient}).
By employing tracers, 
the relative mass of each component (e.g.~envelope or ambient gas) in \textit{any given AMR cell} can be accurately determined.
Thus, the method remains effective if there is mixing between various components.

Separate tracers are also added to track the initial hydrogen and helium ionization states of the gas. 
The number density profiles of the key ionic species, 
as computed from the Saha equation, are shown in Fig.~\ref{fig:ne_r12778_O}.
Then, by comparing the ionization state from the Saha equation at time $t$
with the original ionization state from MESA, we  deduce how the ionization state  evolves in the simulation.

\begin{table*}
  \begin{center}
  \caption{List of runs
          \label{tab:runs}
          }
  \begin{tabular}{@{}cccc@{}}
    \hline
    Run		&EOS						&Ambient reduction	&Fate of released recombination energy	\\
    \hline
    A		&Ideal gas, $\gamma=5/3$			&No			&Immediate radiative loss (effectively) \\
    B		&MESA, excluding radiation energy		&No			&Immediate local thermalization 	\\
    C		&MESA, including only gas thermal energy	&No			&Immediate radiative loss (effectively) \\
    D		&Ideal gas, $\gamma=5/3$  			&Yes			&Immediate radiative loss (effectively) \\
    \hline                  	 
  \end{tabular}
  \end{center}
\end{table*}

\subsection{Equation of State}\label{sec:eos}
Our tabulated EOS is essentially the ``DT2'' part of the MESA tabulated EOS \citep[see][]{Paxton+19}, 
which is based on the OPAL EOS \citep{Rogers+Nayfonov02} and the SCVH EOS \citep{Saumon+95}.
For the parameter space of the simulations, OPAL is much more relevant than SCVH.
These tables have been adapted for use in AstroBEAR;
in some cases, this requires inverting tables to change the independent variables.
The tables cover the full parameter space encountered by the simulations presented.%
\footnote{To make the tables rectangular in $\log_{10}$ of the independent variables 
$\rho$ and $T$, $\rho$ and specific energy $e$, or $\rho$ and $P$,
we extrapolate them using an ideal gas law with mean atomic weight of metals set equal to the value $17.4\amu$. 
This is the value obtained from the MESA model. 
However, the extrapolated regions correspond to regions of the parameter space that do not arise in the simulations.}

For our 3D simulation, we modified the tabulated EOS by subtracting the radiation specific energy $a\rad T^4/\rho$,
where $T$ is temperature, $\rho$ is gas density and $a\rad$ is the radiation constant.
In the original MESA RGB profile, the ratio of this component to the local gas thermal specific energy $3\kB T/2\mu \mH$,
where $\kB$ is the Boltzmann constant and $\mu$ is the mean molecular mass units of the hydrogen mass $\mH$,
reaches a maximum of $20\%$ at $r\approx0.6\Rsun$ and is $<7\%$ outside $r\soft=2.41\Rsun$.
The ratio of the total radiation energy to that of the total thermal energy over the entire star is $<0.1\%$. 
If included, the radiation energy would lead to a high internal energy density in the high-temperature ambient medium,
which could  artificially help unbind the envelope through mixing. 
The choice to remove radiation energy is also consistent with the fact that radiation \textit{pressure} 
is not included in the tabulated EOS, although it is included separately in the stellar models computed by MESA. 
\footnote{When preparing the simulation initial condition,
we chose to make the gas pressure equal to the total pressure (gas plus radiation) in the MESA 1D RGB profile.}

That said, we noticed during post-processing that the subtraction of the radiation energy was imperfect,
due to a slightly different value of the radiation constant $a\rad$ used for the subtraction 
compared to that implied by the tables.
This resulted in $1.5\times10^{-4}$ of the radiation energy remaining in the simulation.
This leftover radiation energy is rather negligible in the envelope, 
but makes up about two thirds of the energy of the ambient medium at $t=0$; 
the gas thermal energy makes up about 90 per cent of the remaining one third, and the rest is recombination energy.
Thus, the energy density of the ambient medium is a few times higher in the tabulated EOS run 
as compared to the ideal gas run.
We account for this leftover radiation energy in the analysis below.

\subsection{Runs}\label{sec:runs}
Our runs are summarized in Table~\ref{tab:runs}.
Model~A employs an ideal gas EOS, with $\gamma=5/3$.
This is an updated version of the fiducial RGB CE model 
from our previous works \citep{Chamandy+18,Chamandy+19a,Chamandy+19b,Chamandy+20},
with higher resolution and with initial ambient density reduced by a factor of about seven.
Model~B makes use of the MESA EOS with radiation energy removed, 
as explained in Sec.~\ref{sec:eos}, but is otherwise the same as Model~A.
It assumes that recombination energy is thermalized locally when it gets released.
Model~C is like Model~B except that the internal energy is replaced by the gas thermal energy alone;
since both recombination and radiation energy are removed, 
the results are expected to be closer to those of Model~A than Model~B.
Model~D is like Model~A except that the ambient density and pressure are reduced by a factor of $1000$ early in the simulation
(Sec.~\ref{sec:ambient}).
Models~A and B were run for $115.7\da$, Model~C for $50.5\da$, and Model~D for $142.4\da$.

\begin{figure*}
\includegraphics[scale=0.145,clip=true,trim=50  170 200 200]{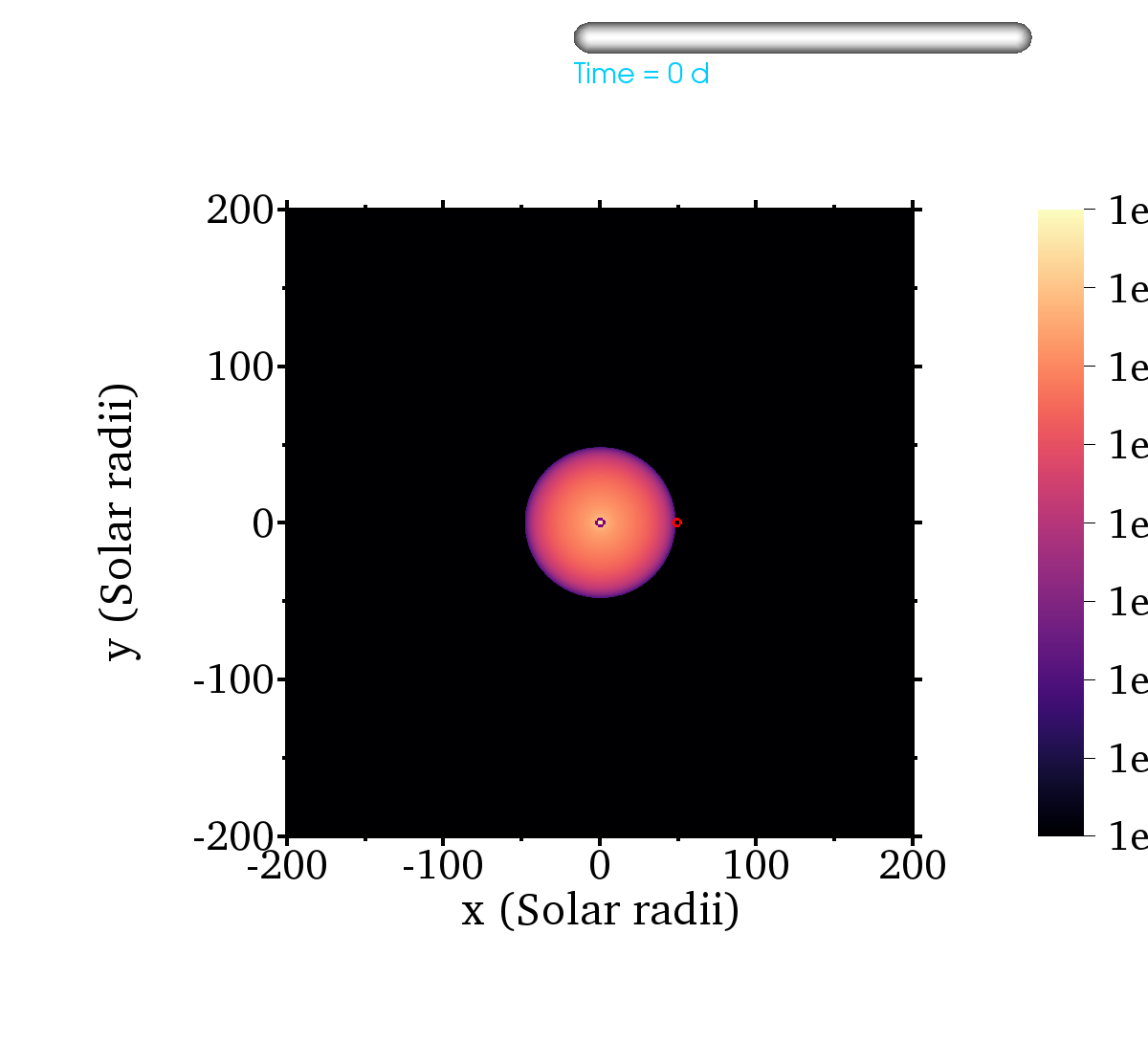}
\includegraphics[scale=0.145,clip=true,trim=38  170 200 200]{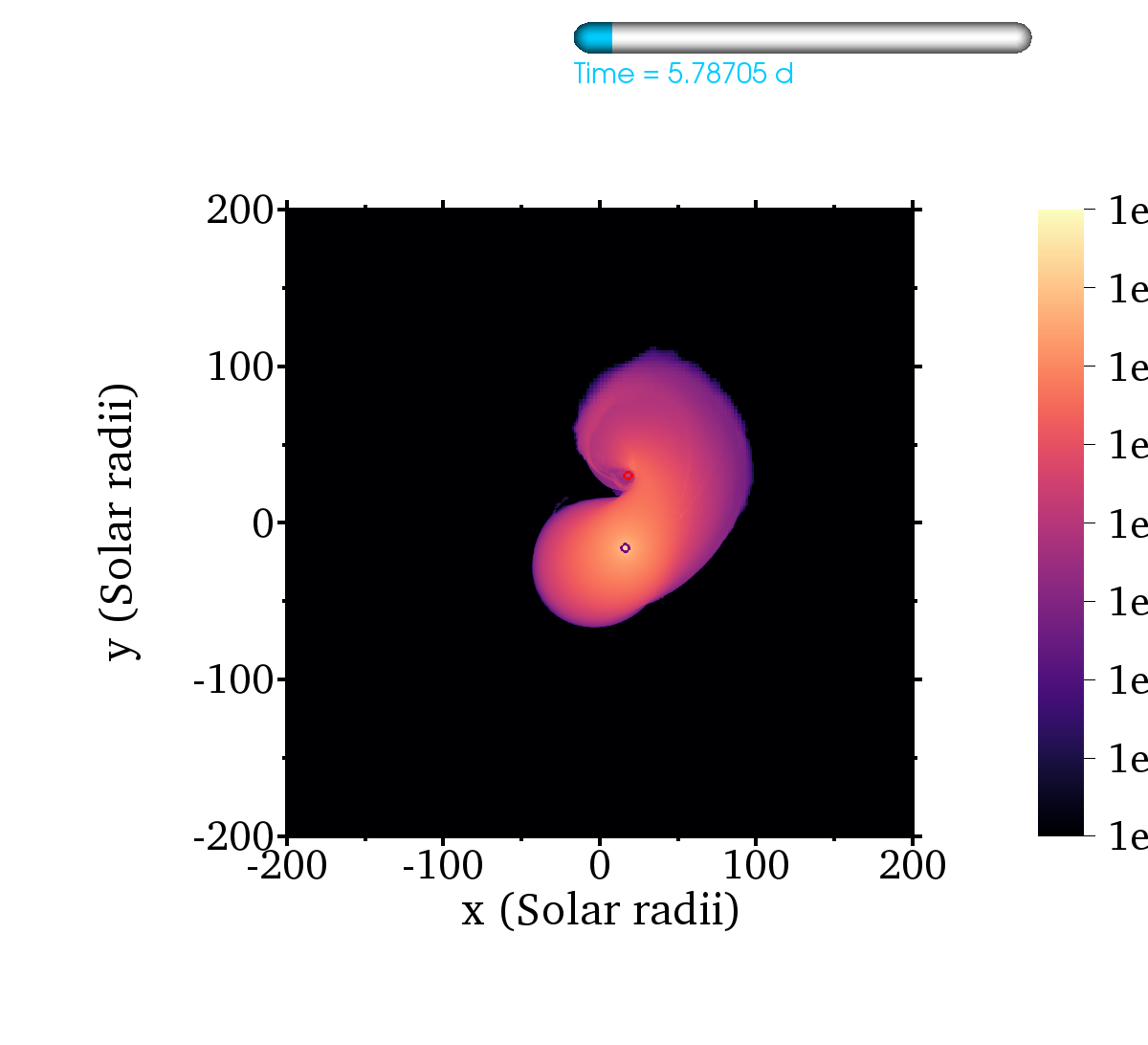}
\includegraphics[scale=0.145,clip=true,trim=38  170   0 200]{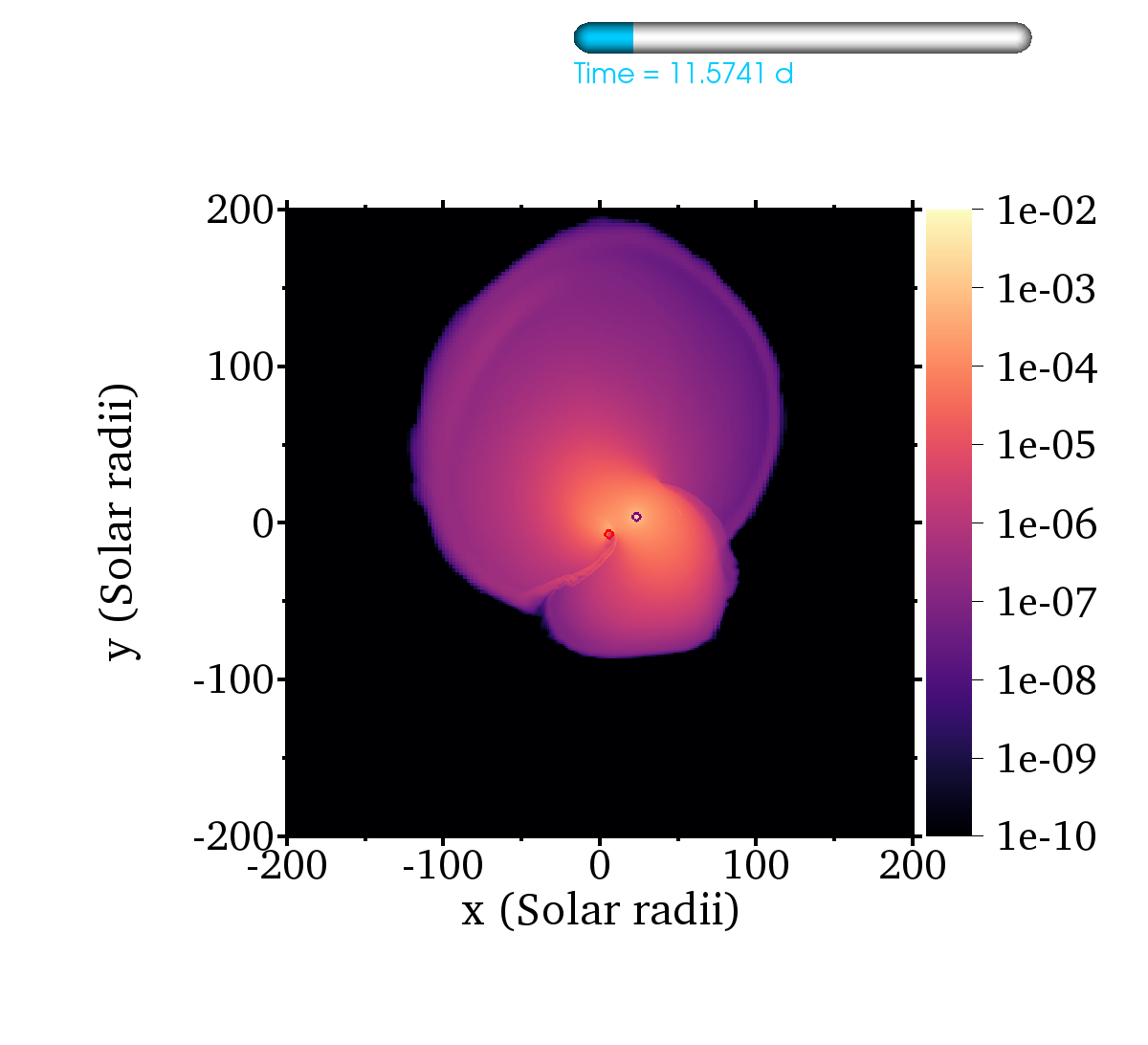}
\\
\includegraphics[scale=0.145,clip=true,trim=50  170 200 200]{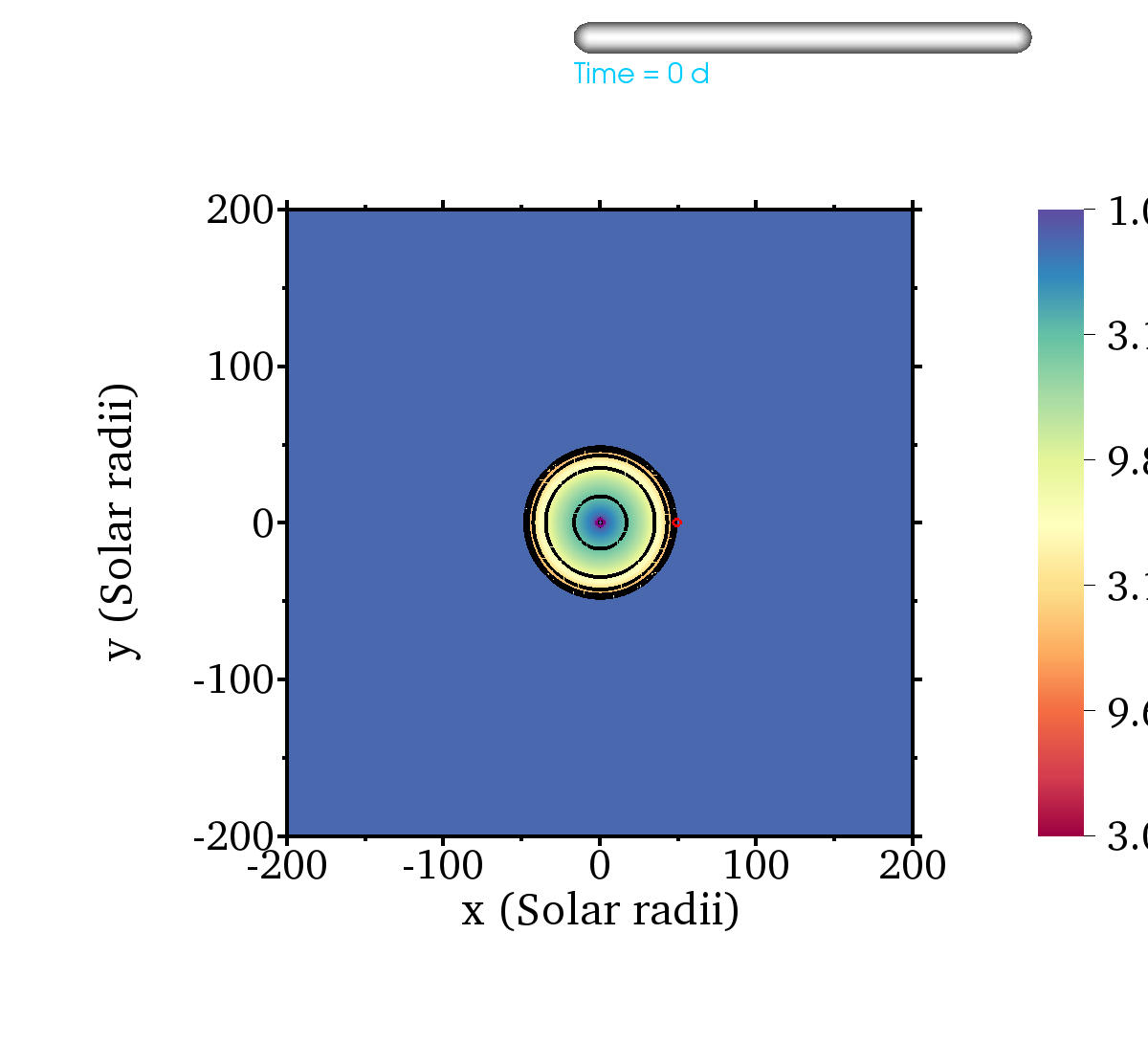}
\includegraphics[scale=0.145,clip=true,trim=38  170 200 200]{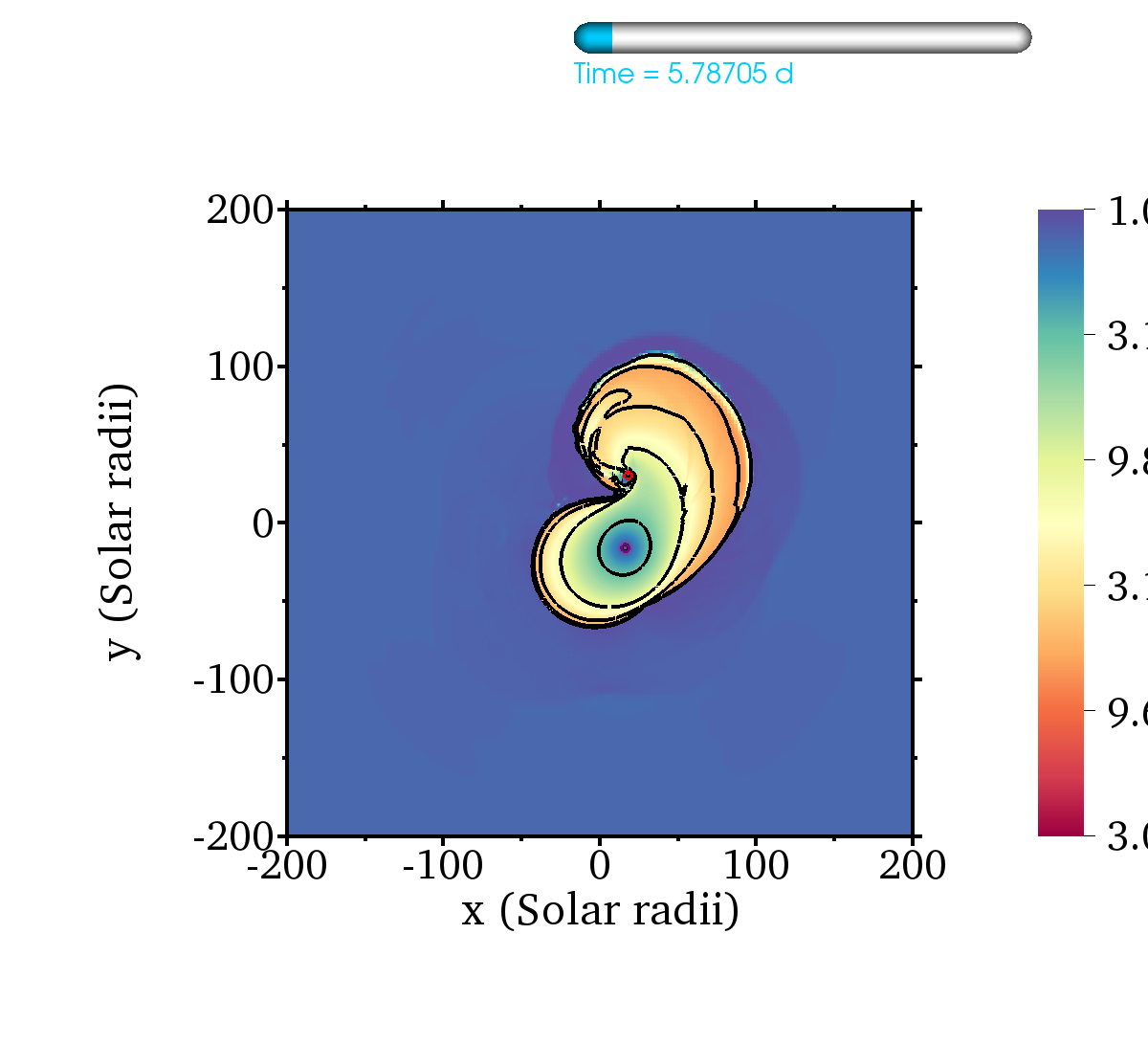}
\includegraphics[scale=0.145,clip=true,trim=38  170   0 200]{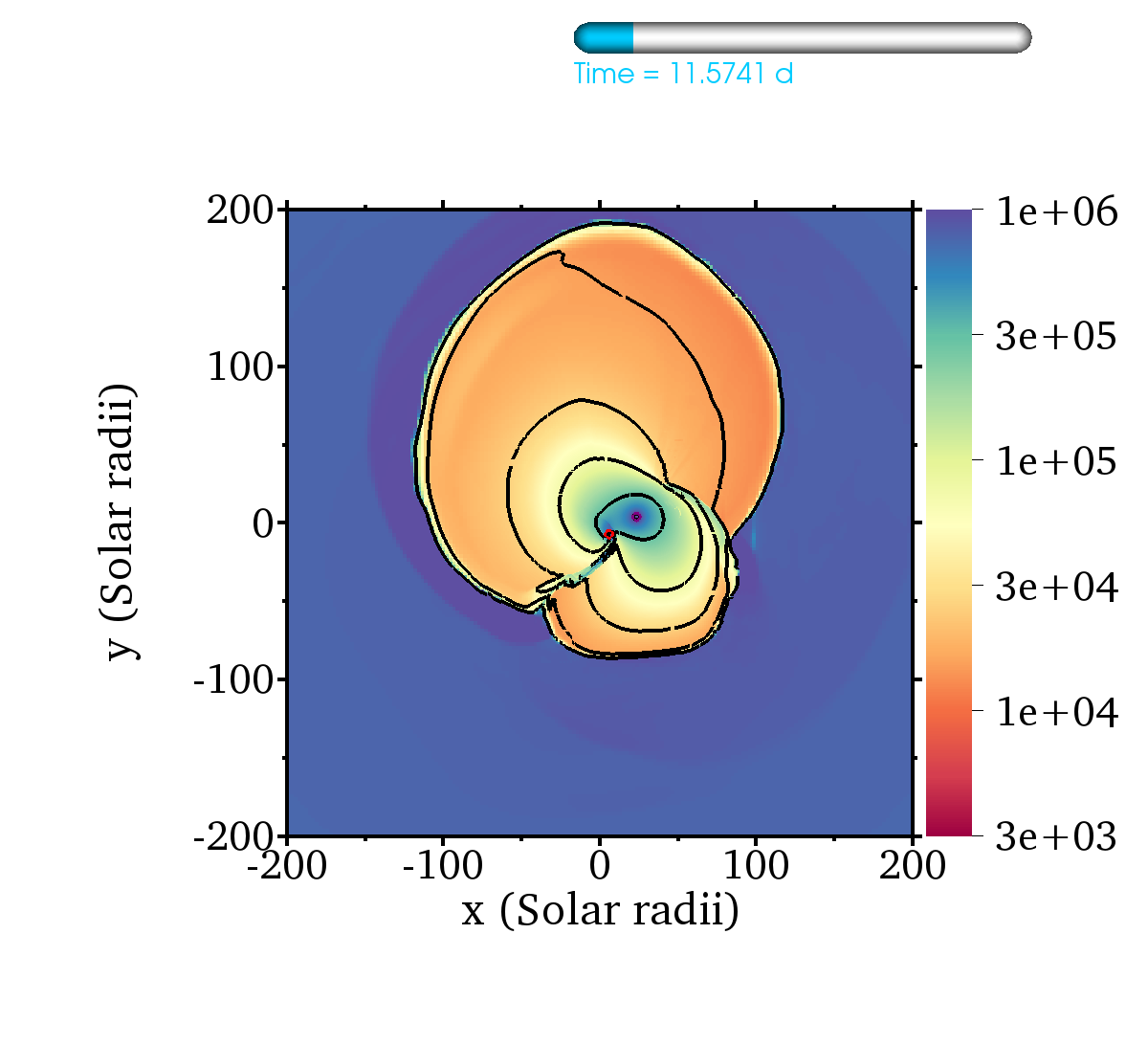}
\\
\includegraphics[scale=0.145,clip=true,trim=50   85 200 100]{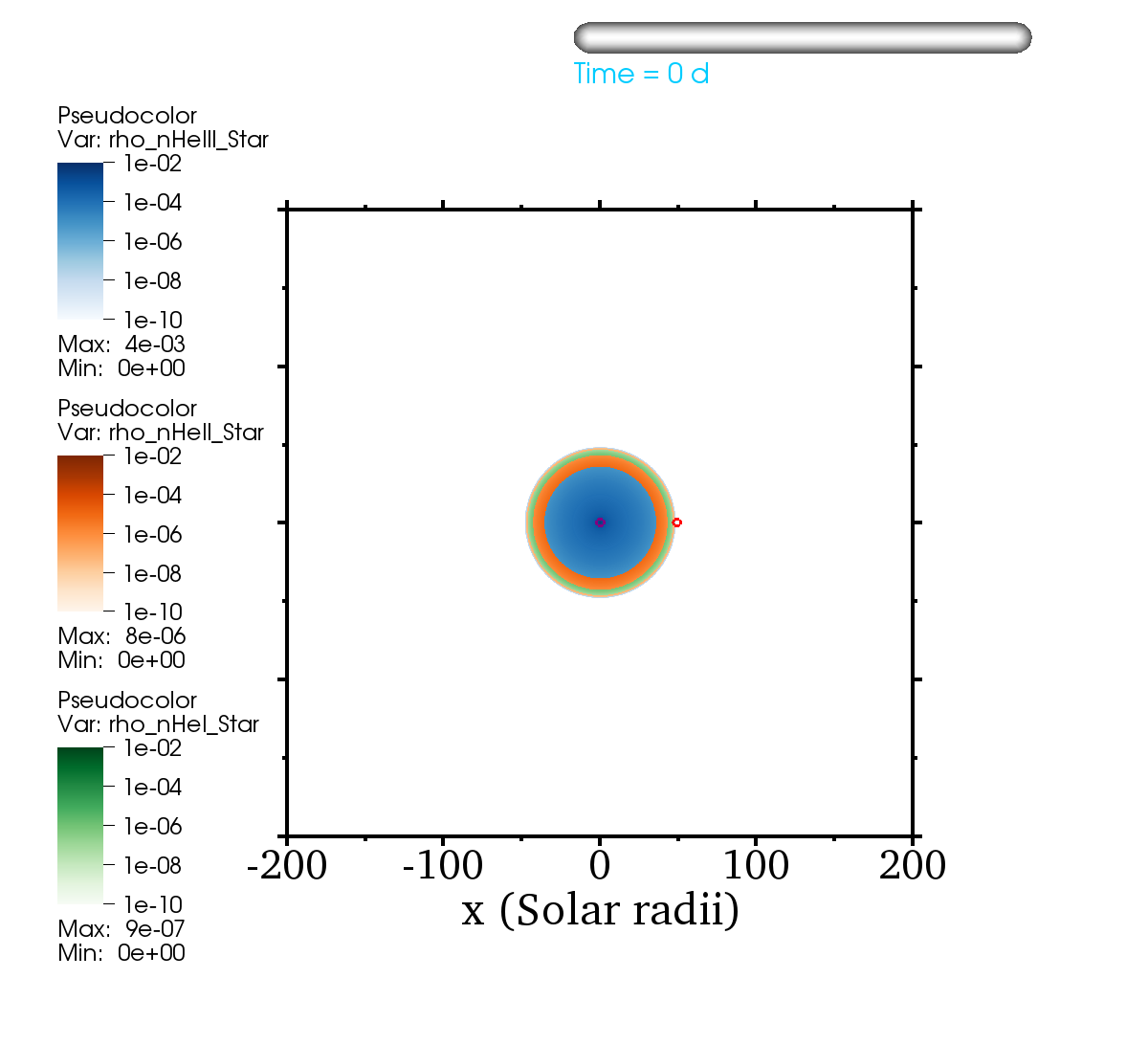}
\includegraphics[scale=0.145,clip=true,trim=38   85 200 100]{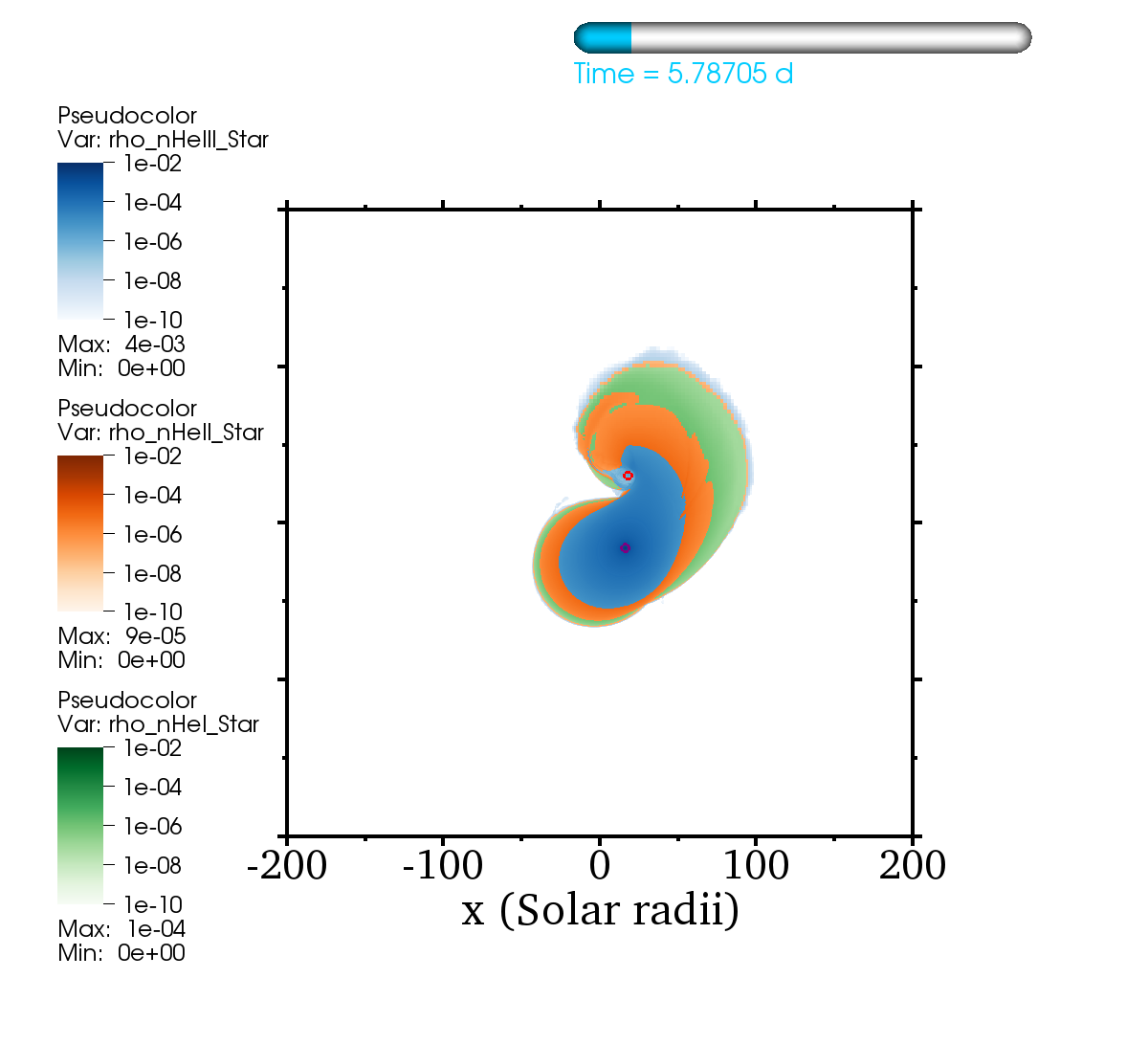}
\includegraphics[scale=0.145,clip=true,trim=38   85   0 100]{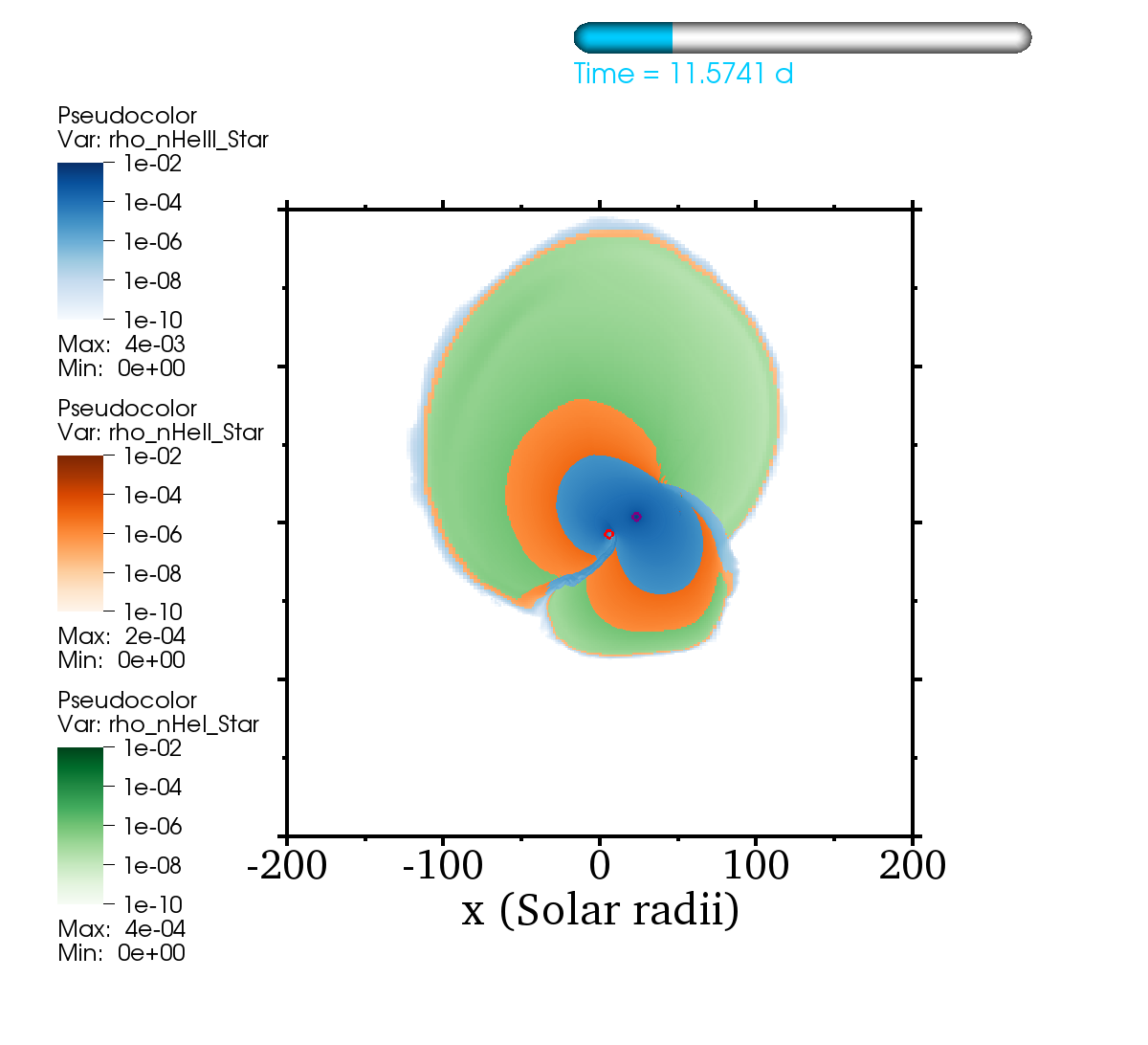}
\\
\includegraphics[scale=0.145,clip=true,trim=50   85 200 100]{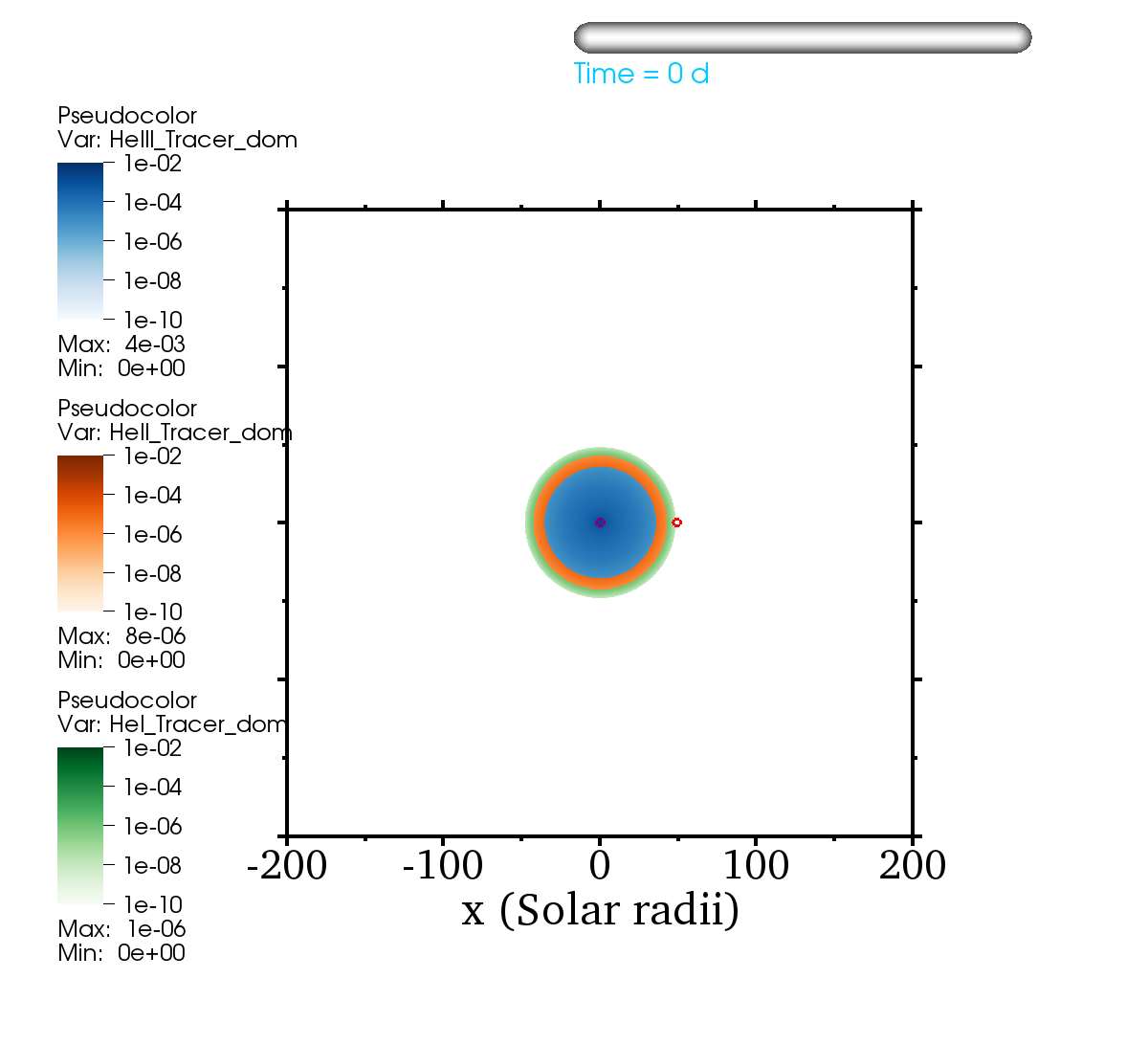}
\includegraphics[scale=0.145,clip=true,trim=38   85 200 100]{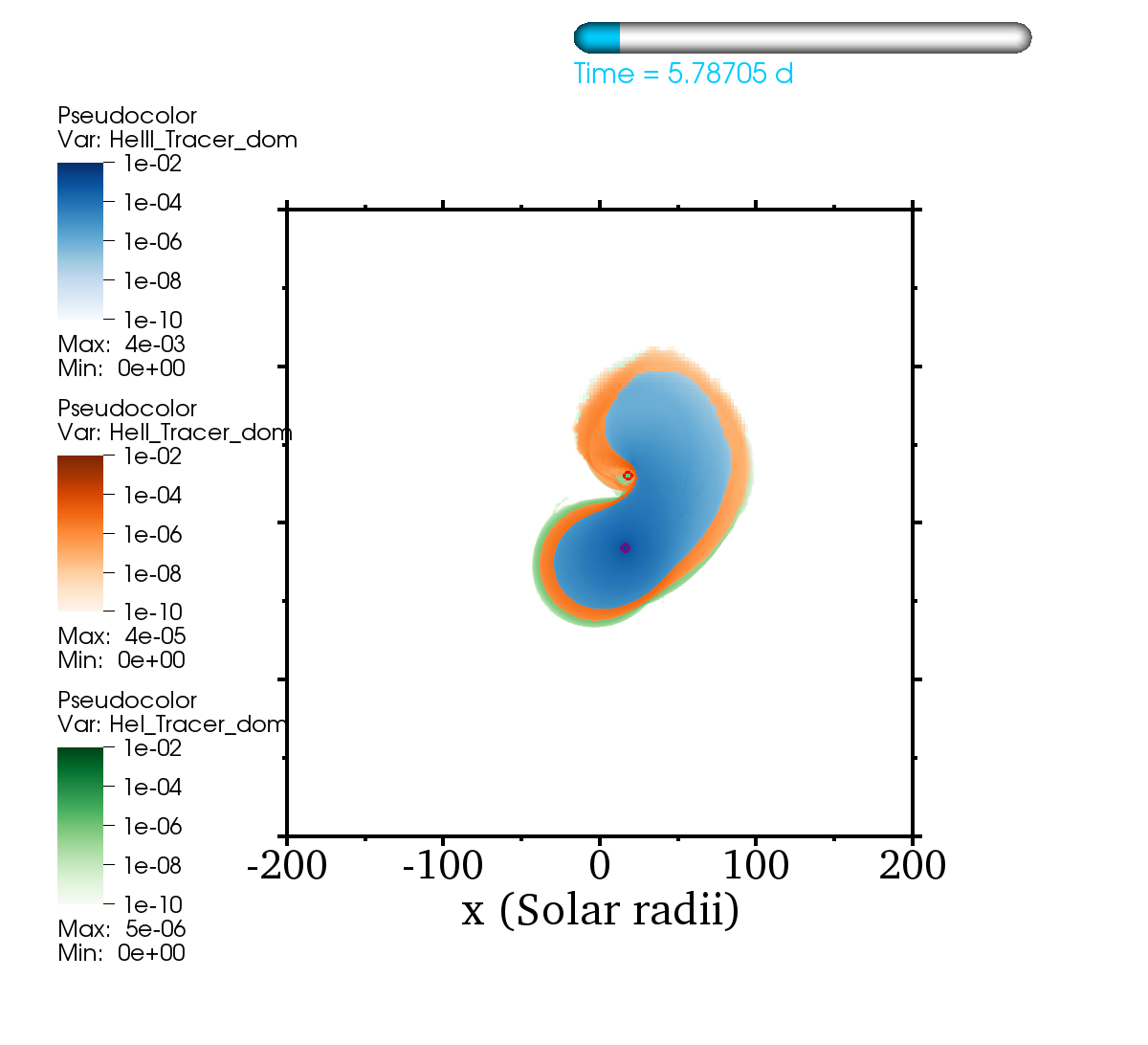}
\includegraphics[scale=0.145,clip=true,trim=38   85   0 100]{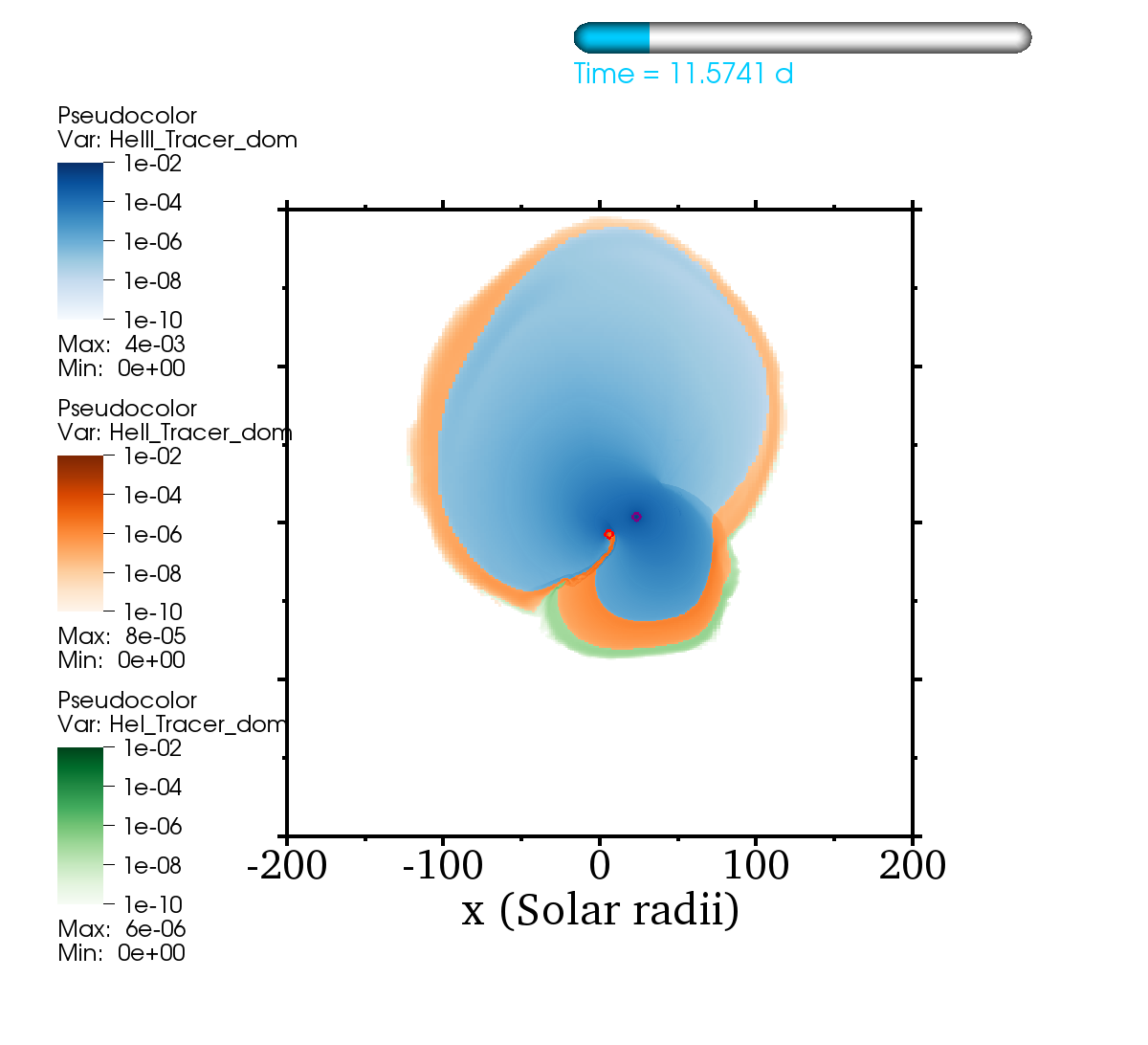}
\\
\includegraphics[scale=0.145,clip=true,trim=50  204 238 180]{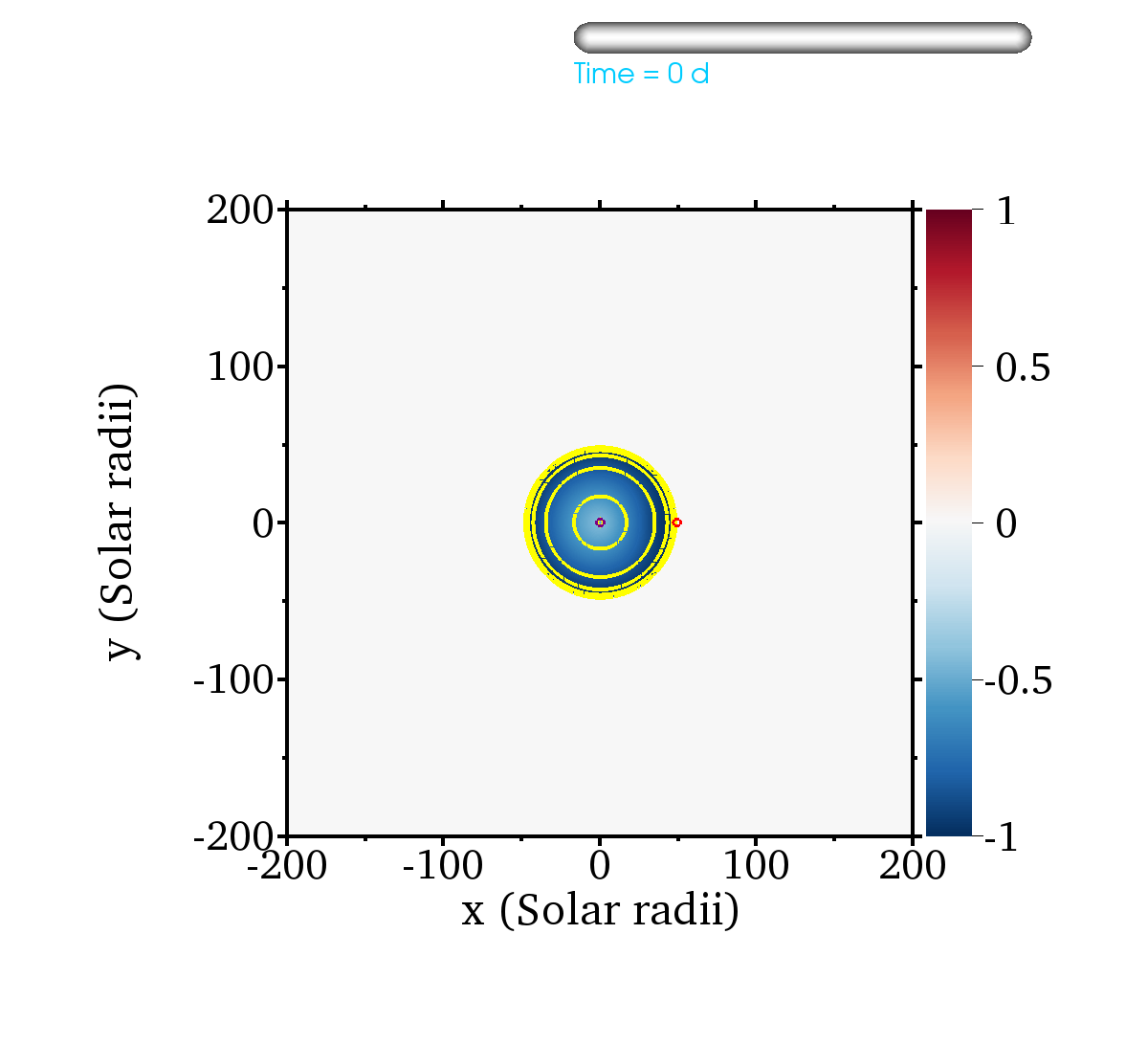}
\includegraphics[scale=0.145,clip=true,trim= 0  204 238 180]{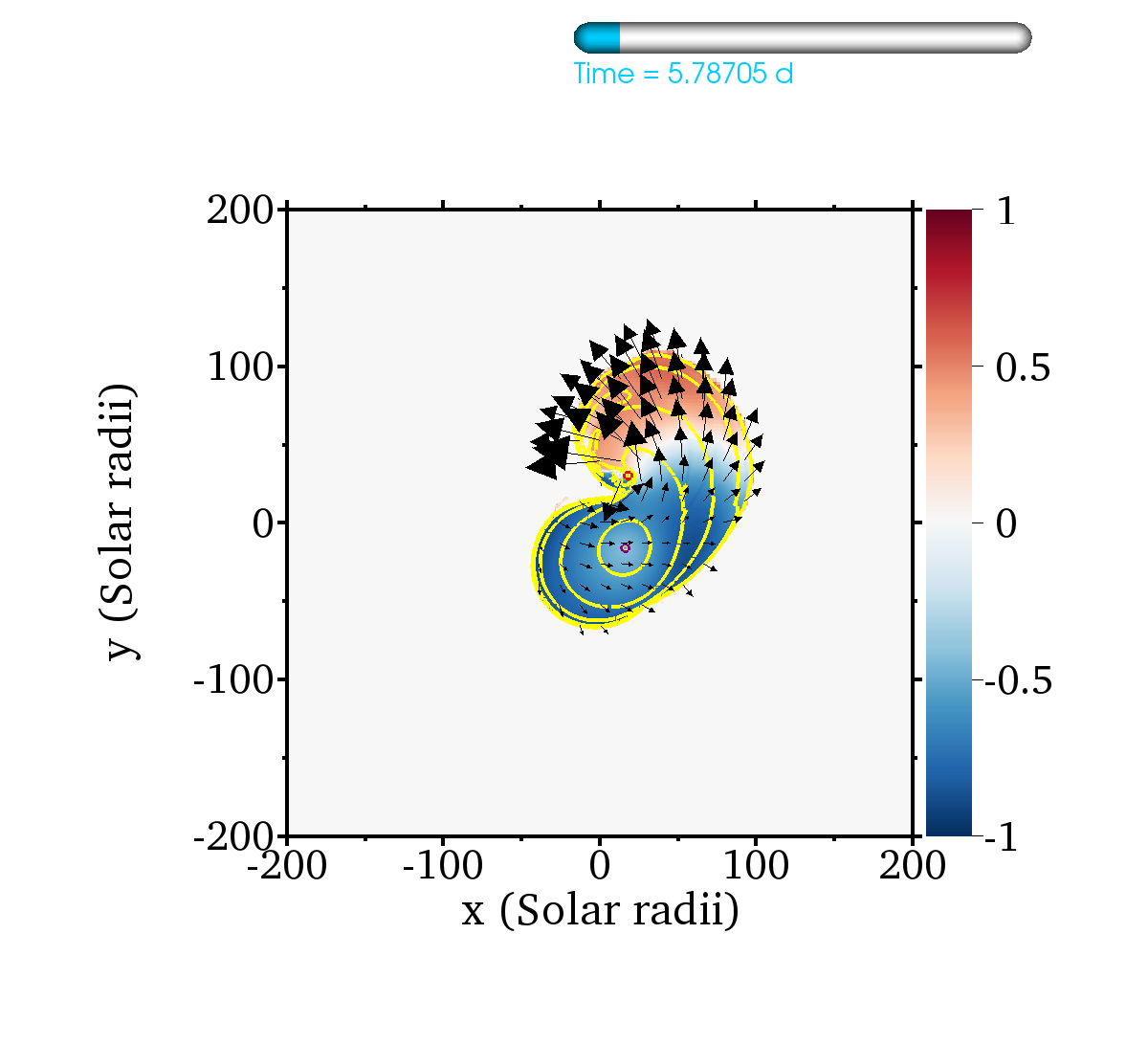}
\includegraphics[scale=0.145,clip=true,trim= 0  204   0 180]{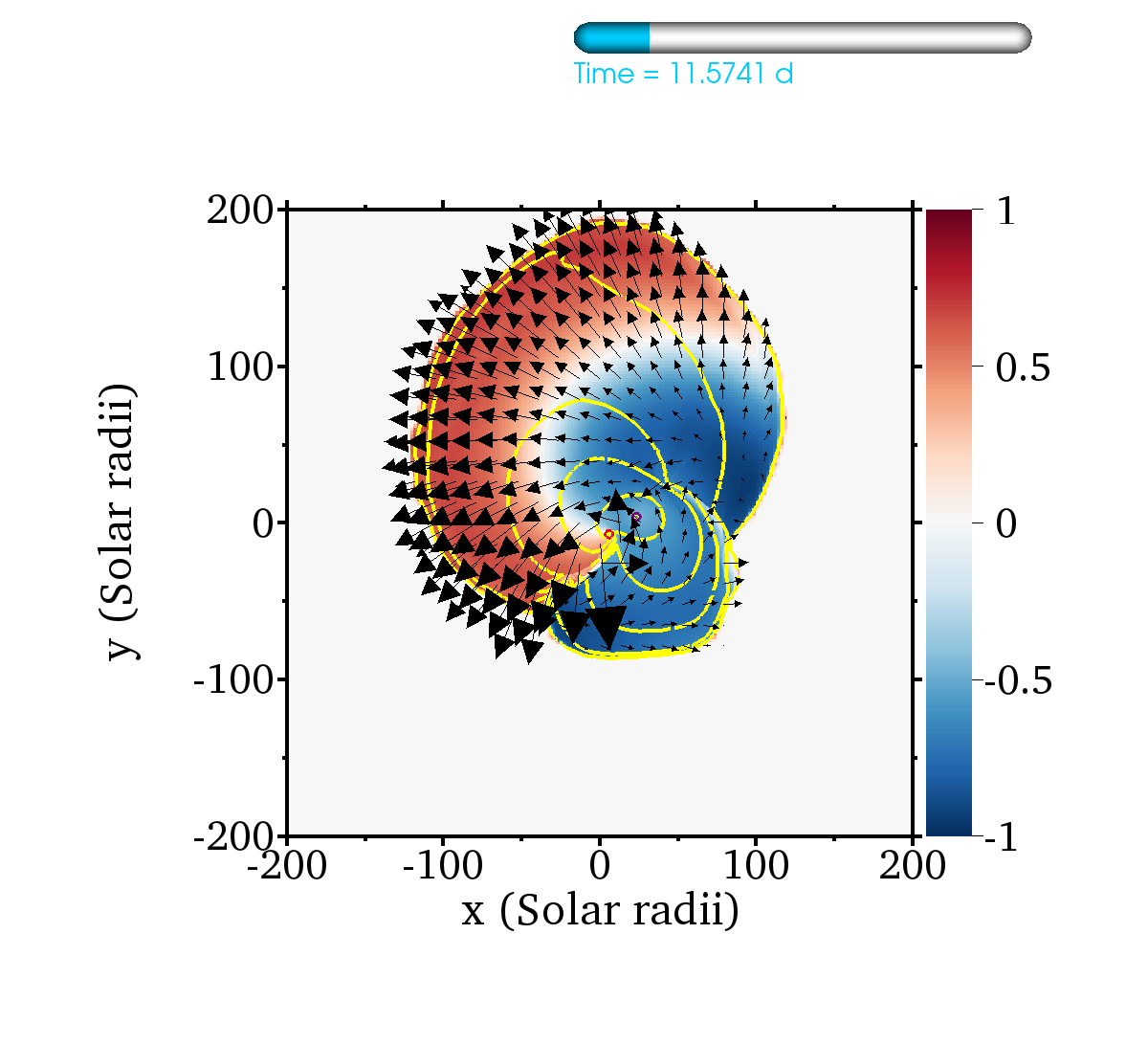}
\\
\includegraphics[scale=0.145,clip=true,trim=50  203 238 180]{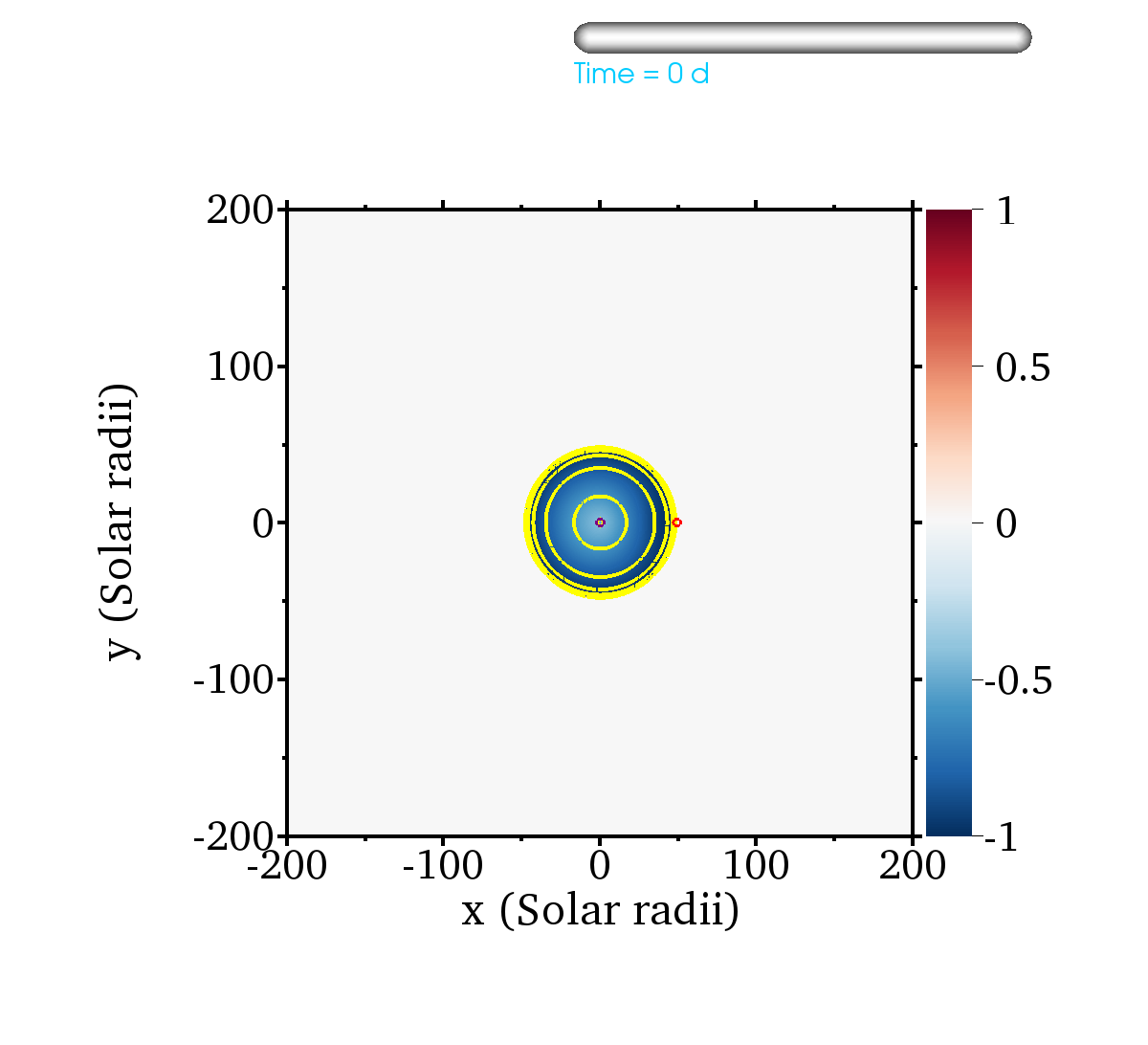}
\includegraphics[scale=0.145,clip=true,trim= 0  203 238 180]{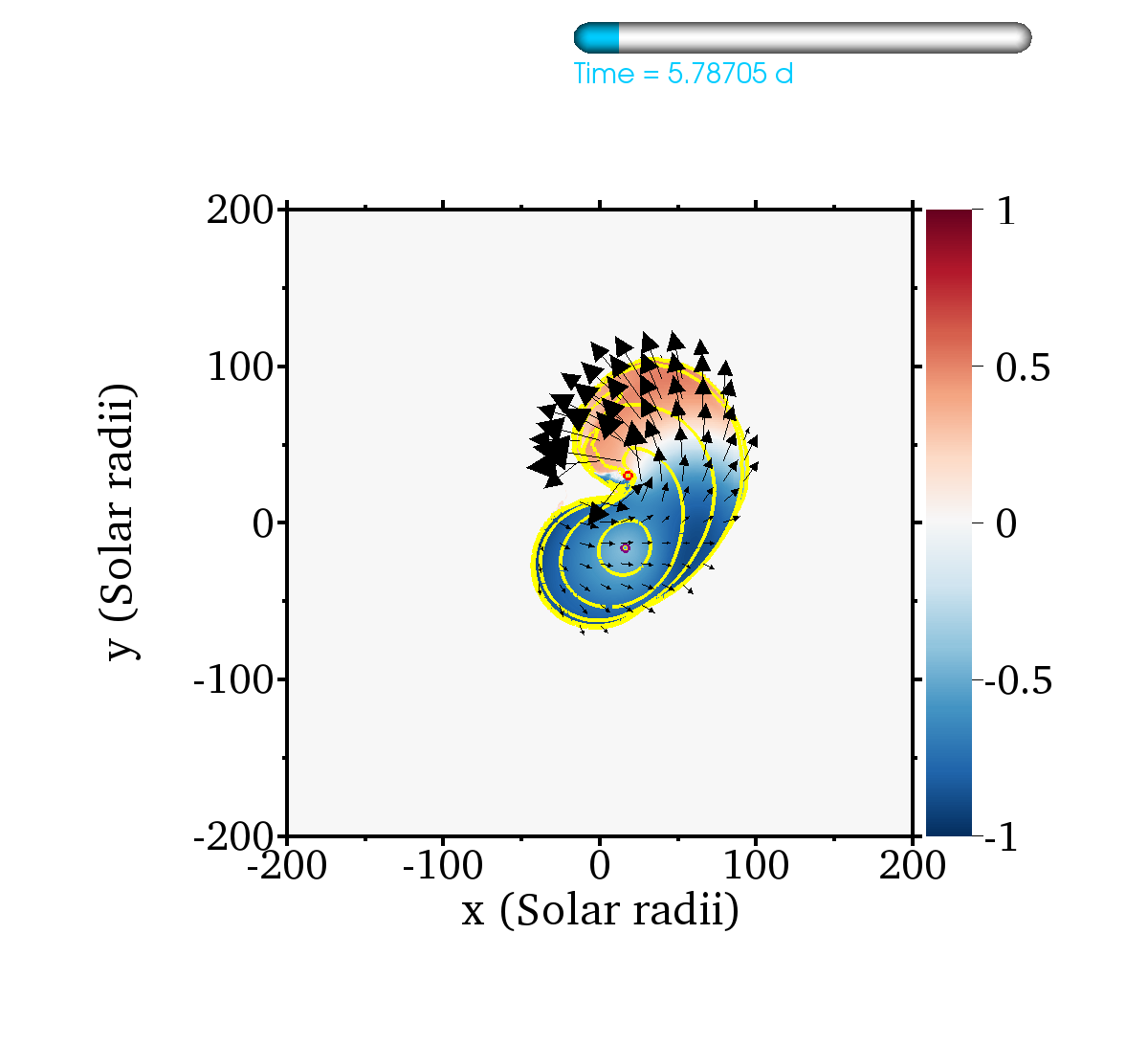}
\includegraphics[scale=0.145,clip=true,trim= 0  203   0 180]{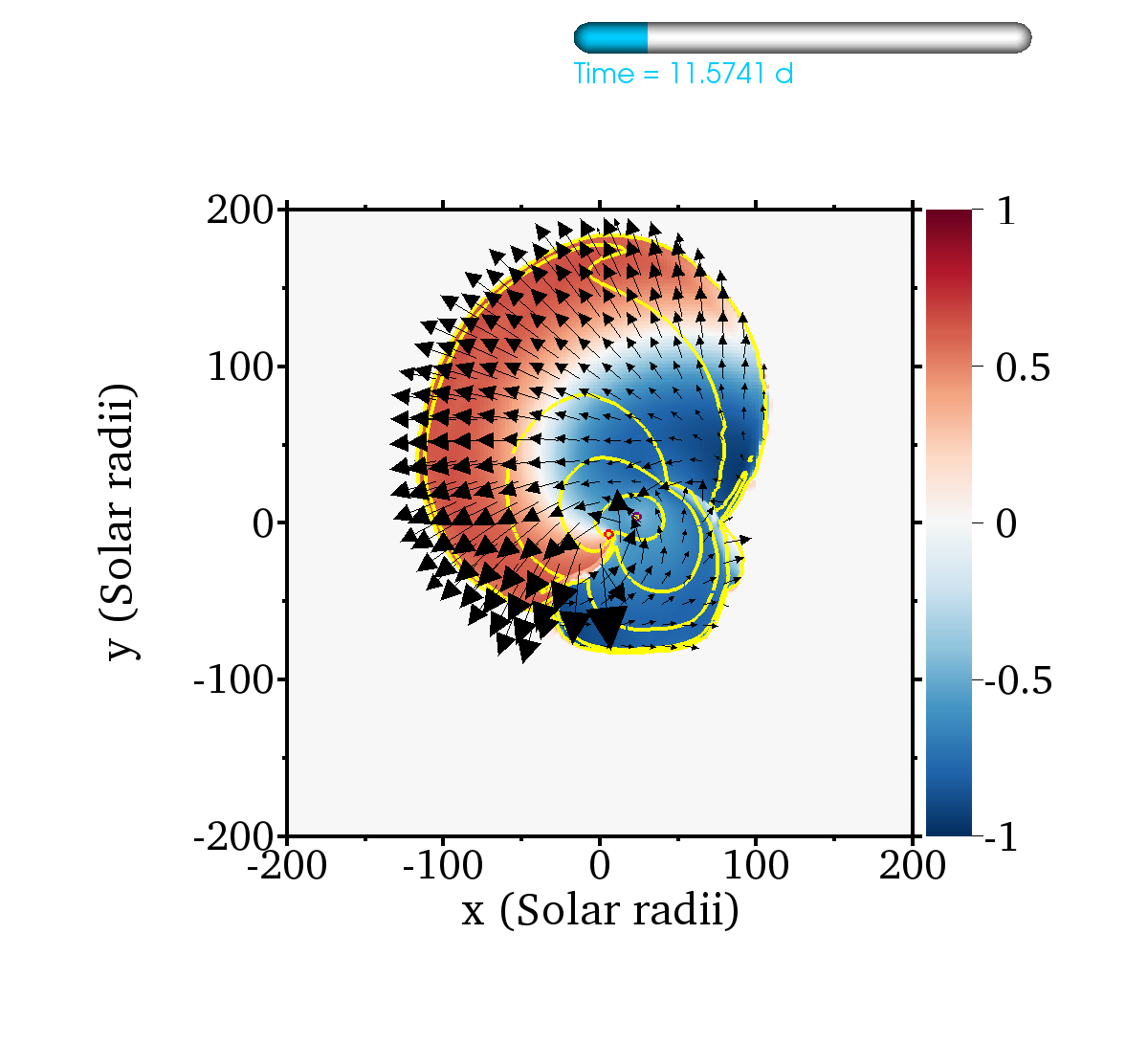}
\\
\vspace{-0.2cm}
\caption{Columns show $t=0$, $5.8$ and $11.6\da$.
         Rows $1$--$5$ show Model~B (MESA EOS) and row $6$ shows Model~A (ideal gas):
         (1)~Mass density of envelope gas;
         (2)~Temperature; 
         (3)~Mass density of locally dominant He ion
         -- HeI (green), HeII (orange) or HeIII (blue);
         (4)~Mass density of He ion that dominated this gas at $t=0$;
         (5)~Normalized binding energy density of envelope gas 
showing bound (blue) and unbound (red) gas;
         (6)~Same as (5) but for Model~A.
        }            
\label{fig:He_face-on_0-50}
\end{figure*}

\begin{figure*}
\includegraphics[scale=0.145,clip=true,trim=50  170 238 200]{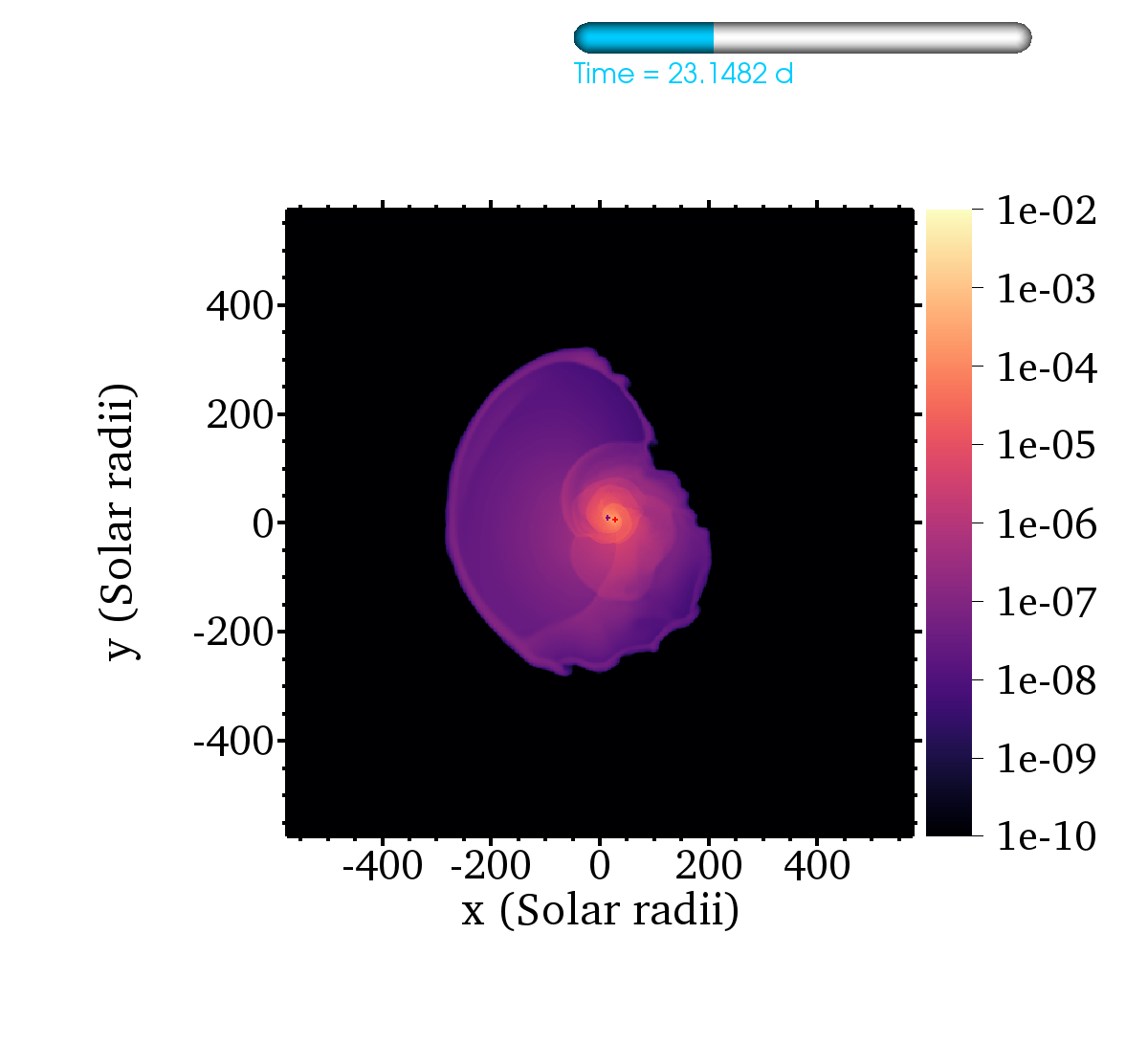}
\includegraphics[scale=0.145,clip=true,trim= 0  170 238 200]{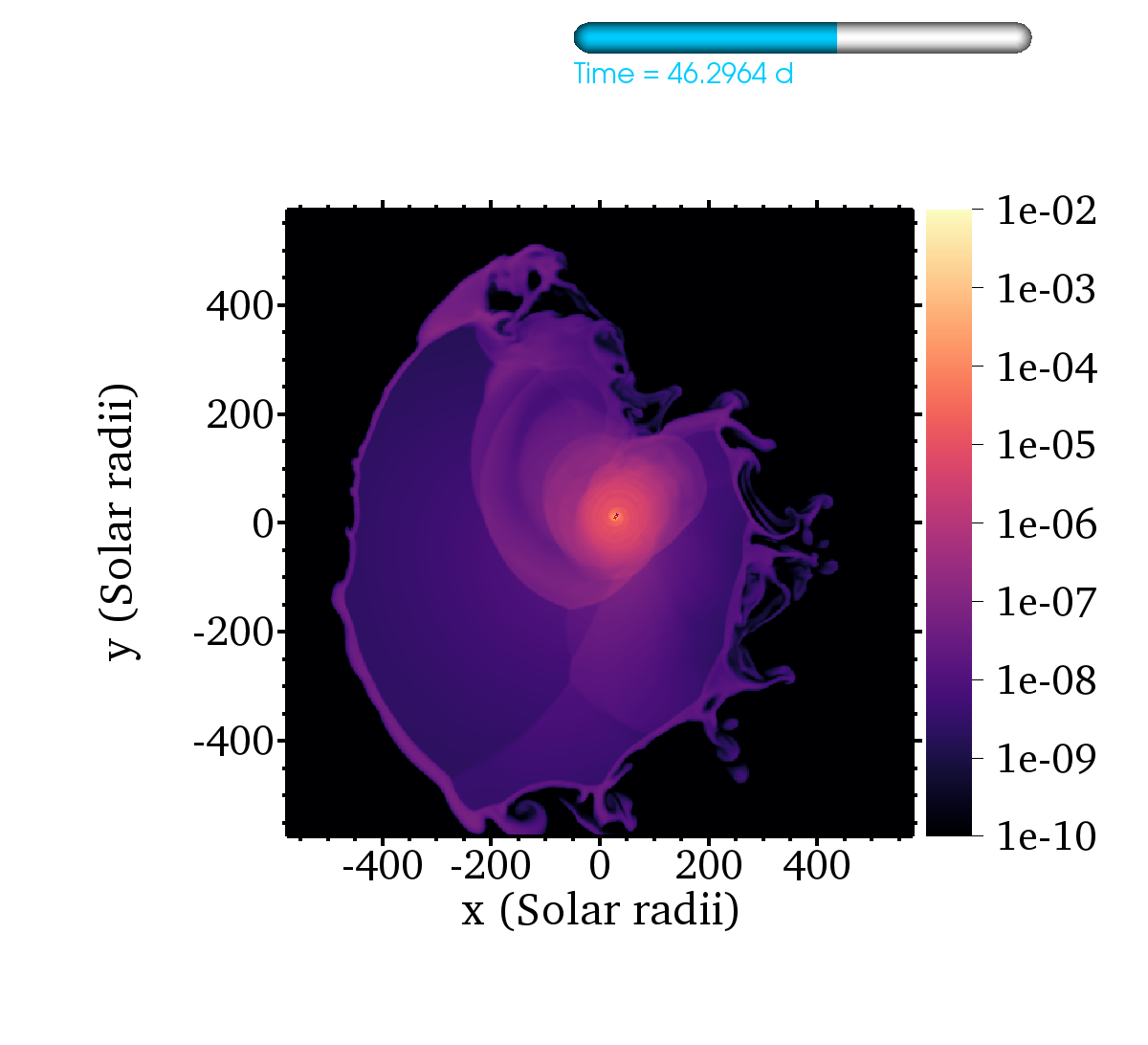}
\includegraphics[scale=0.145,clip=true,trim= 0  170   0 200]{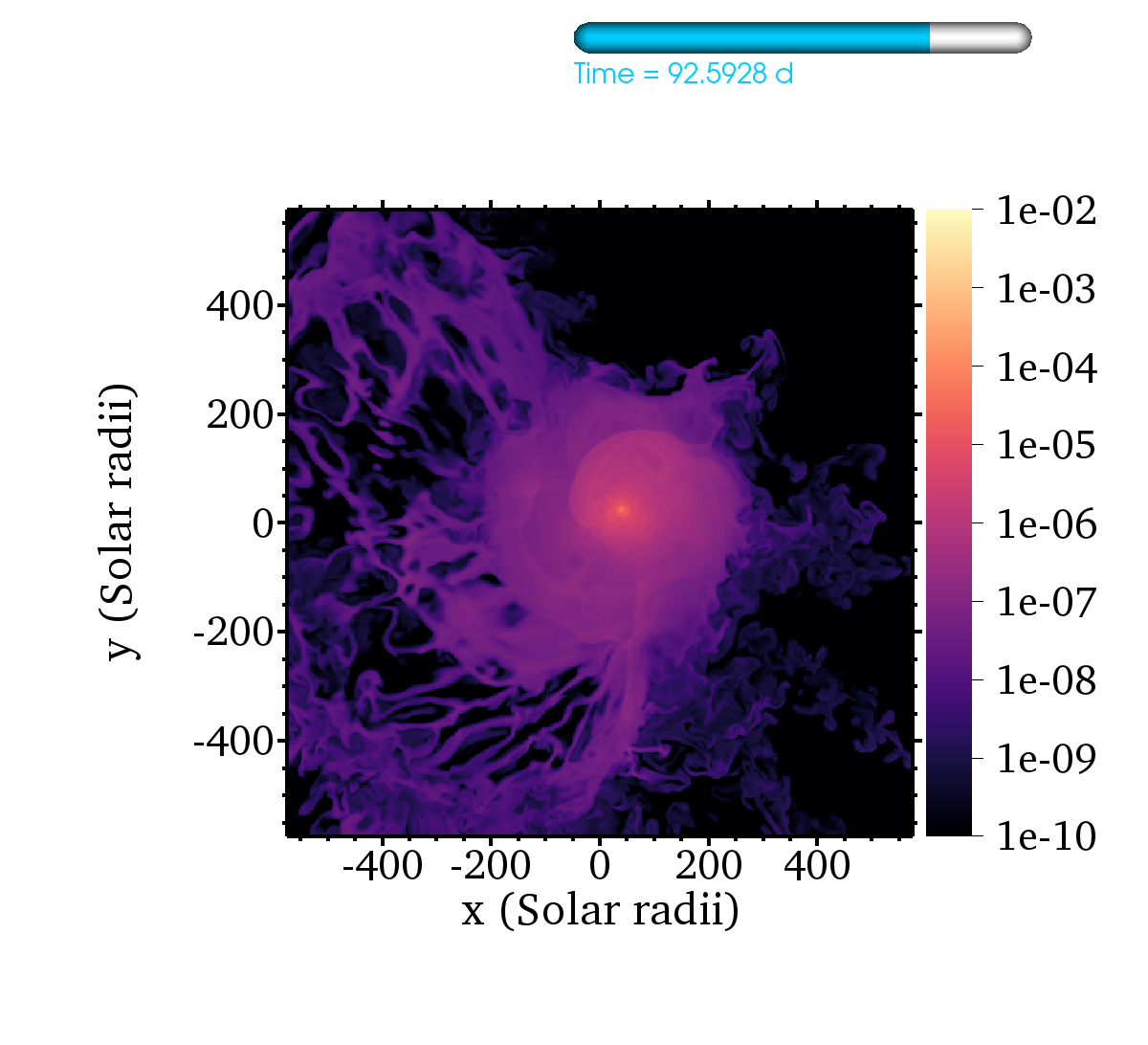}
\\
\includegraphics[scale=0.145,clip=true,trim=50  170 238 200]{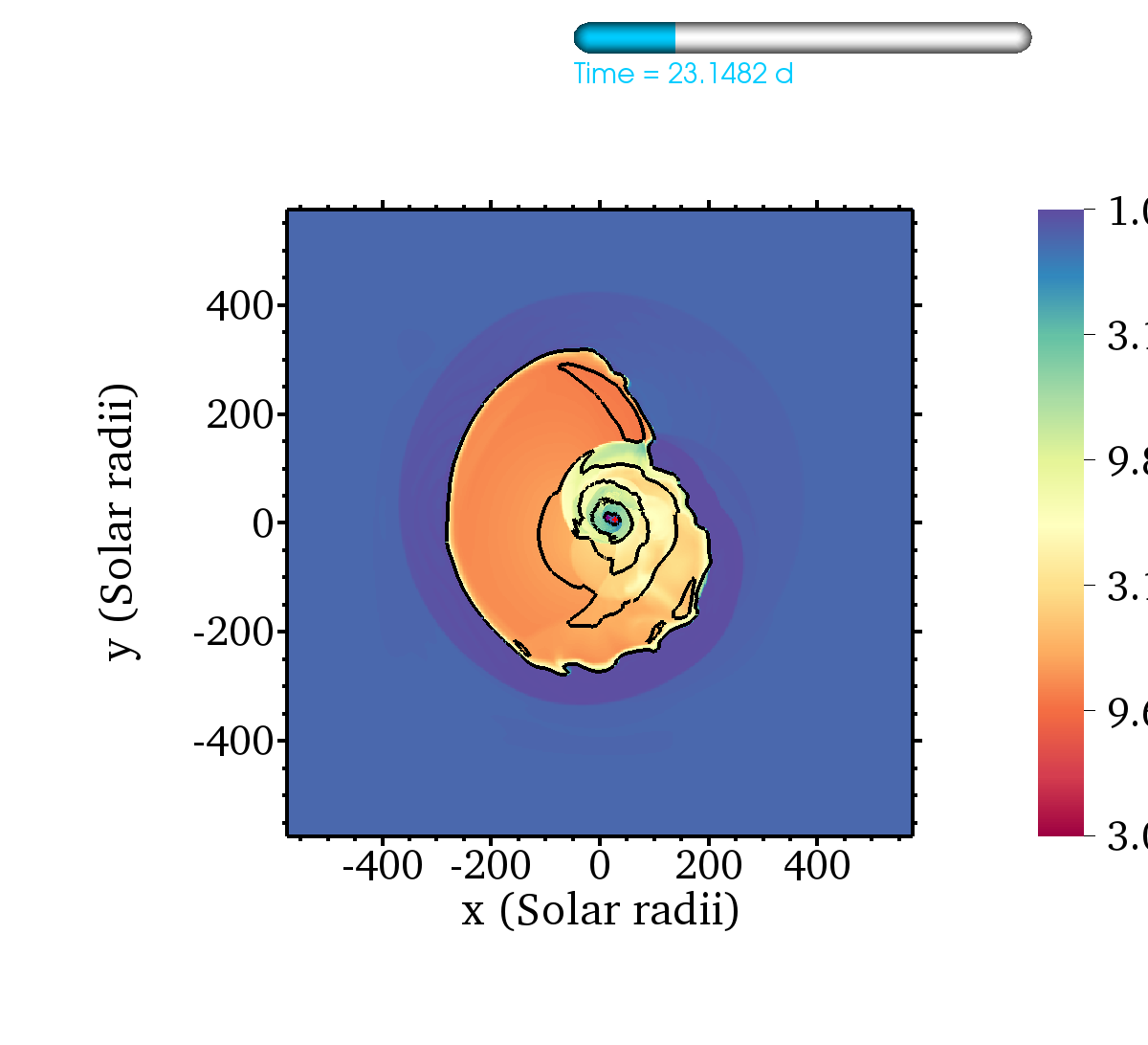}
\includegraphics[scale=0.145,clip=true,trim= 0  170 238 200]{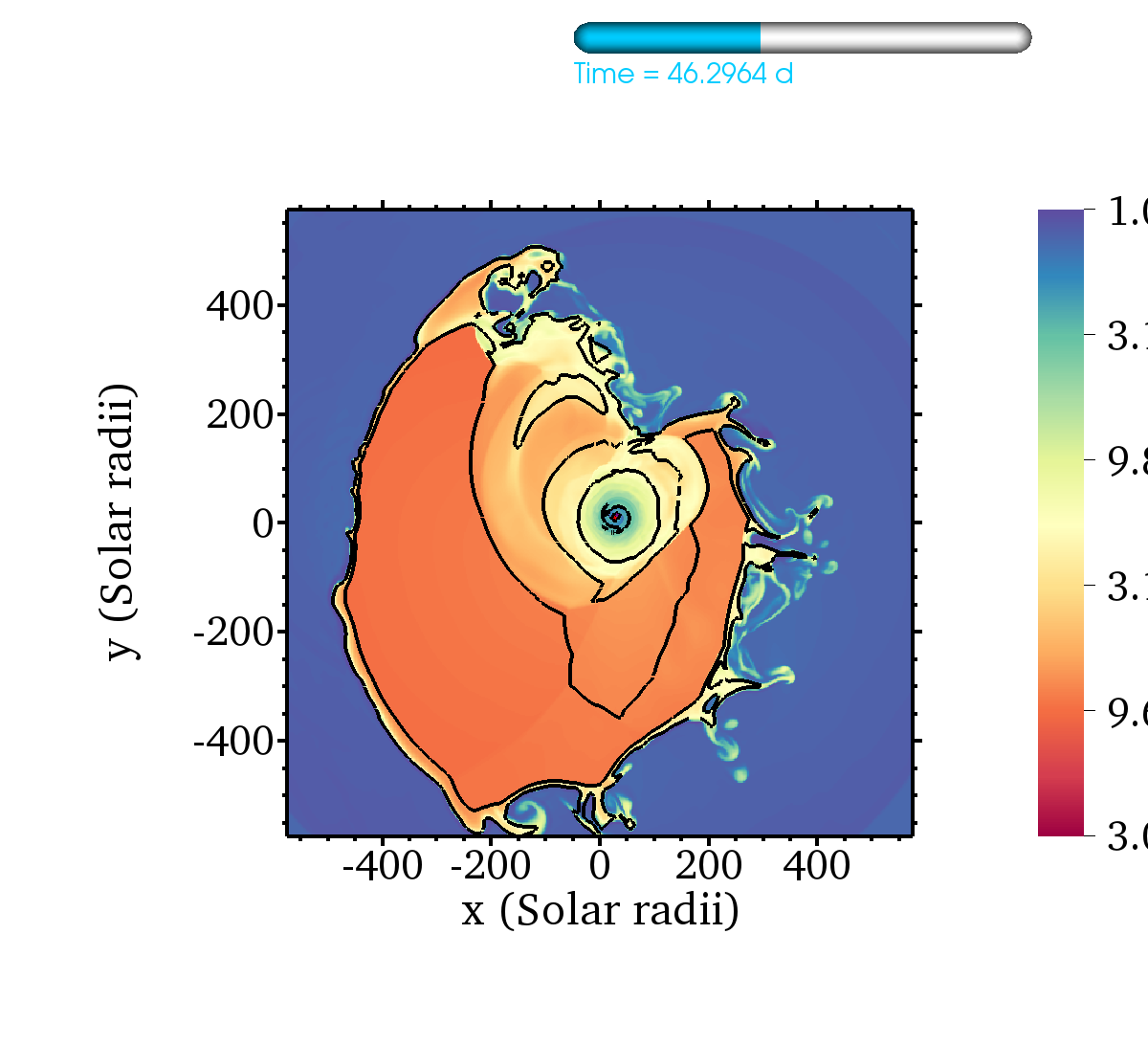}
\includegraphics[scale=0.145,clip=true,trim= 0  170   0 200]{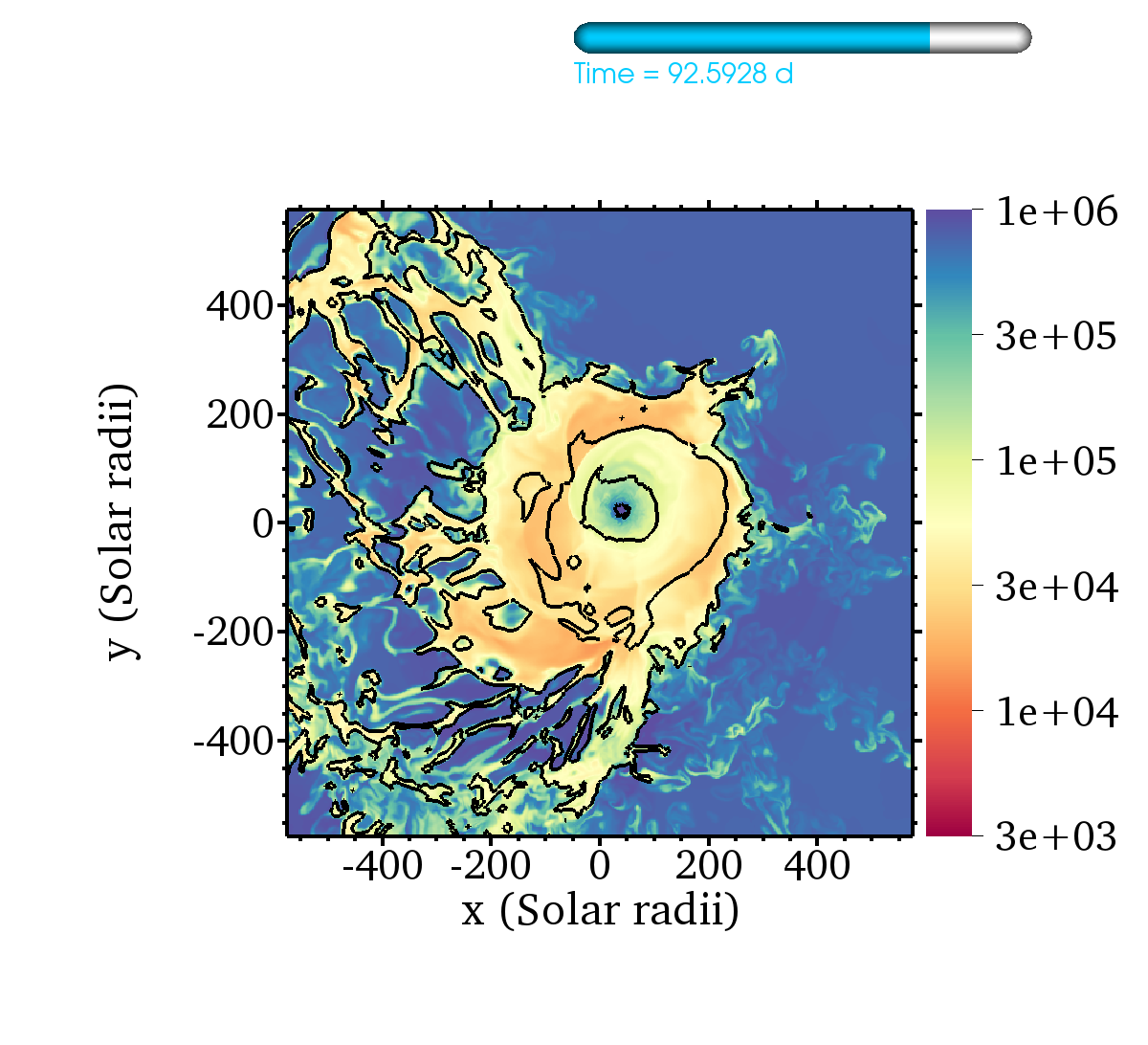}
\\
\includegraphics[scale=0.145,clip=true,trim=50   85 238 100]{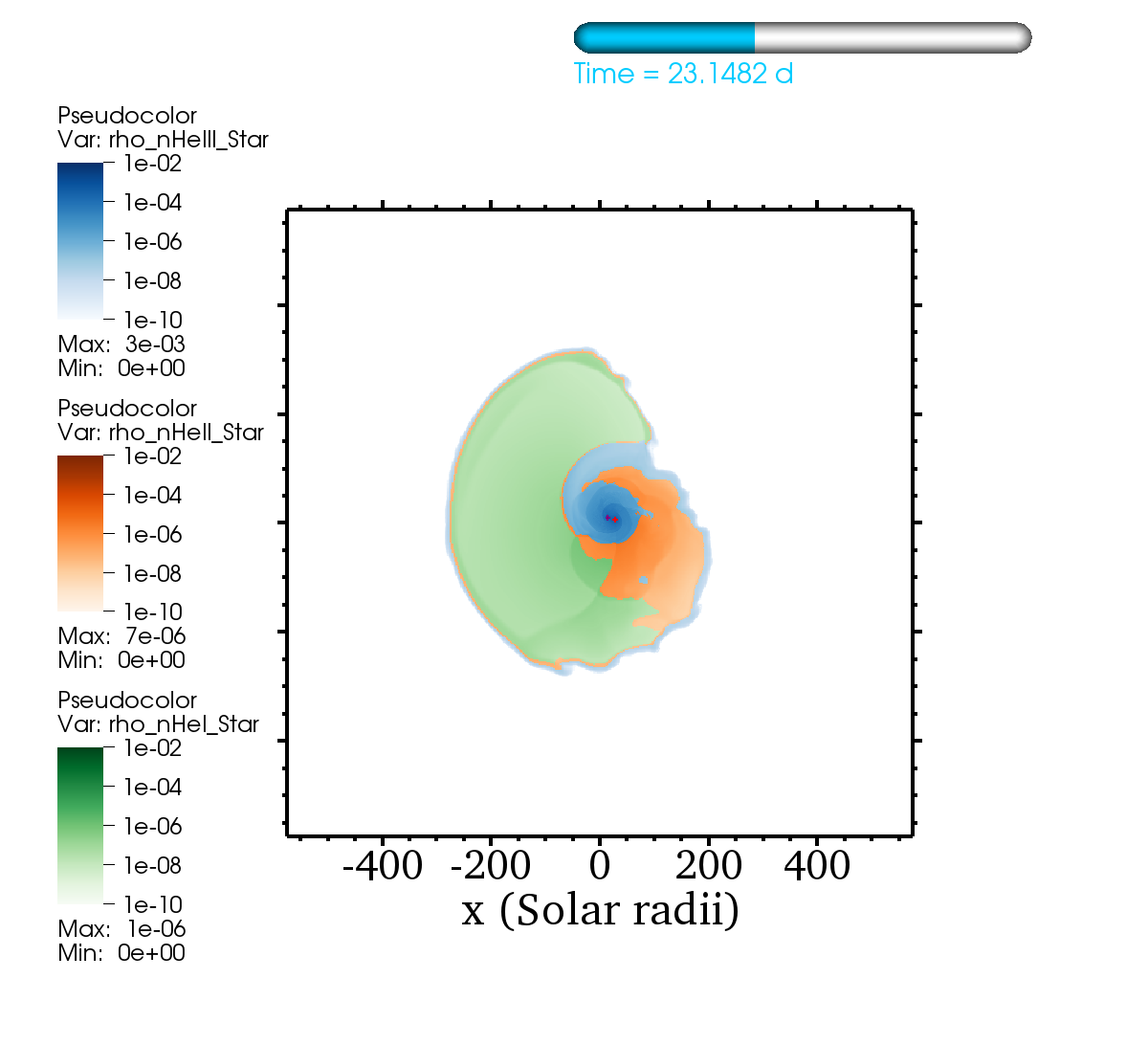}
\includegraphics[scale=0.145,clip=true,trim= 0   85 238 100]{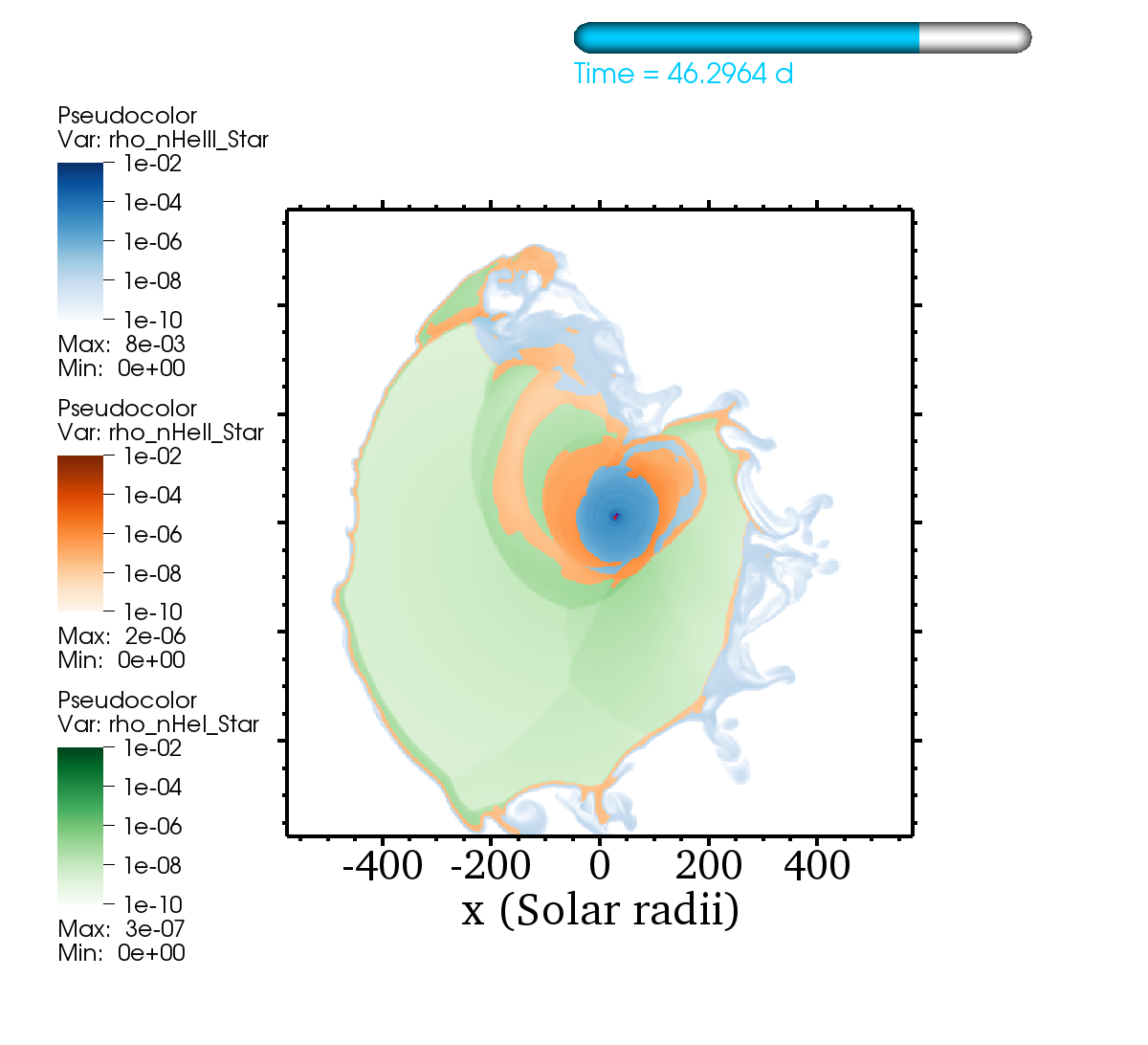}
\includegraphics[scale=0.145,clip=true,trim= 0   85   0 100]{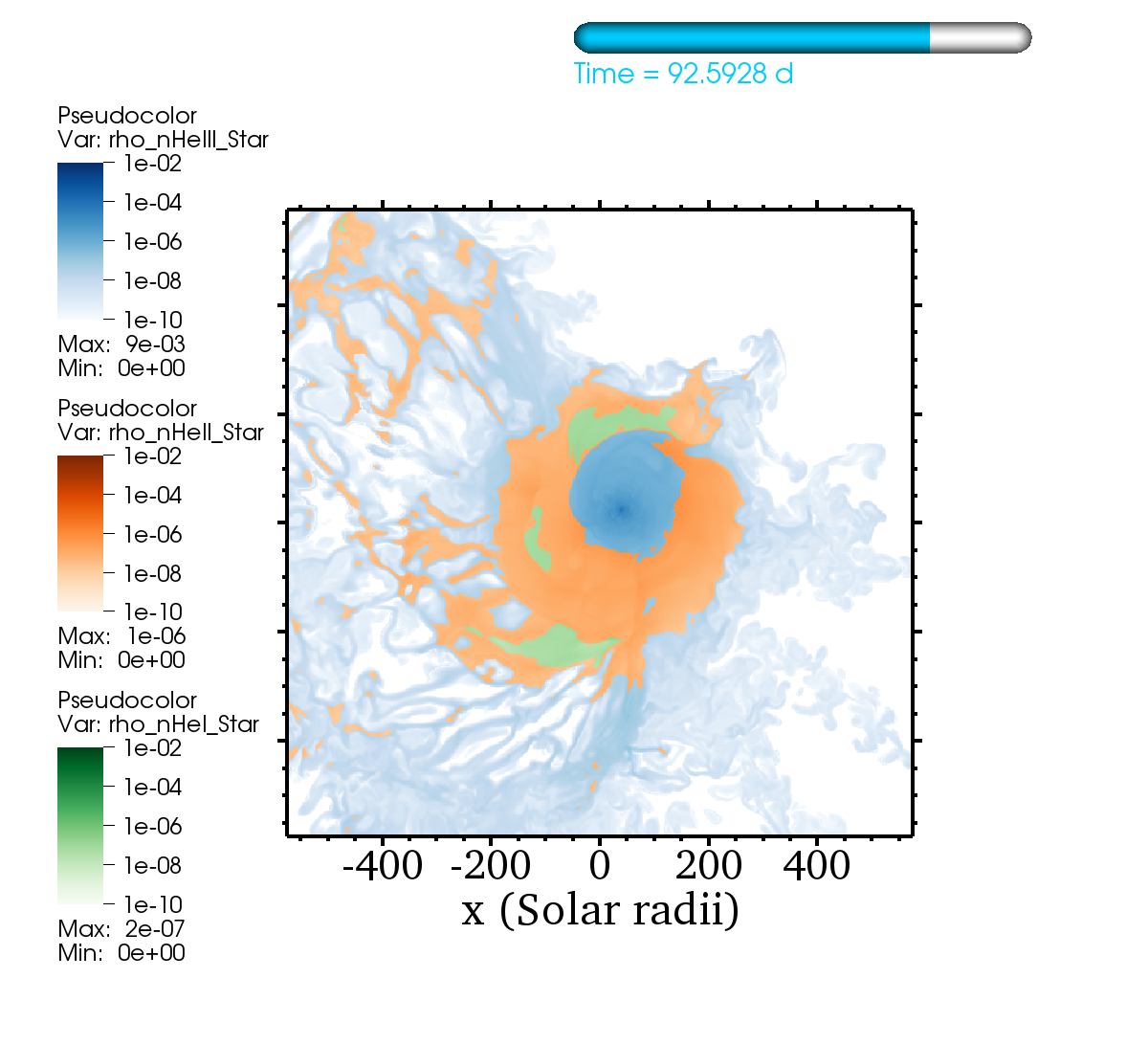}
\\
\includegraphics[scale=0.145,clip=true,trim=50   85 238 100]{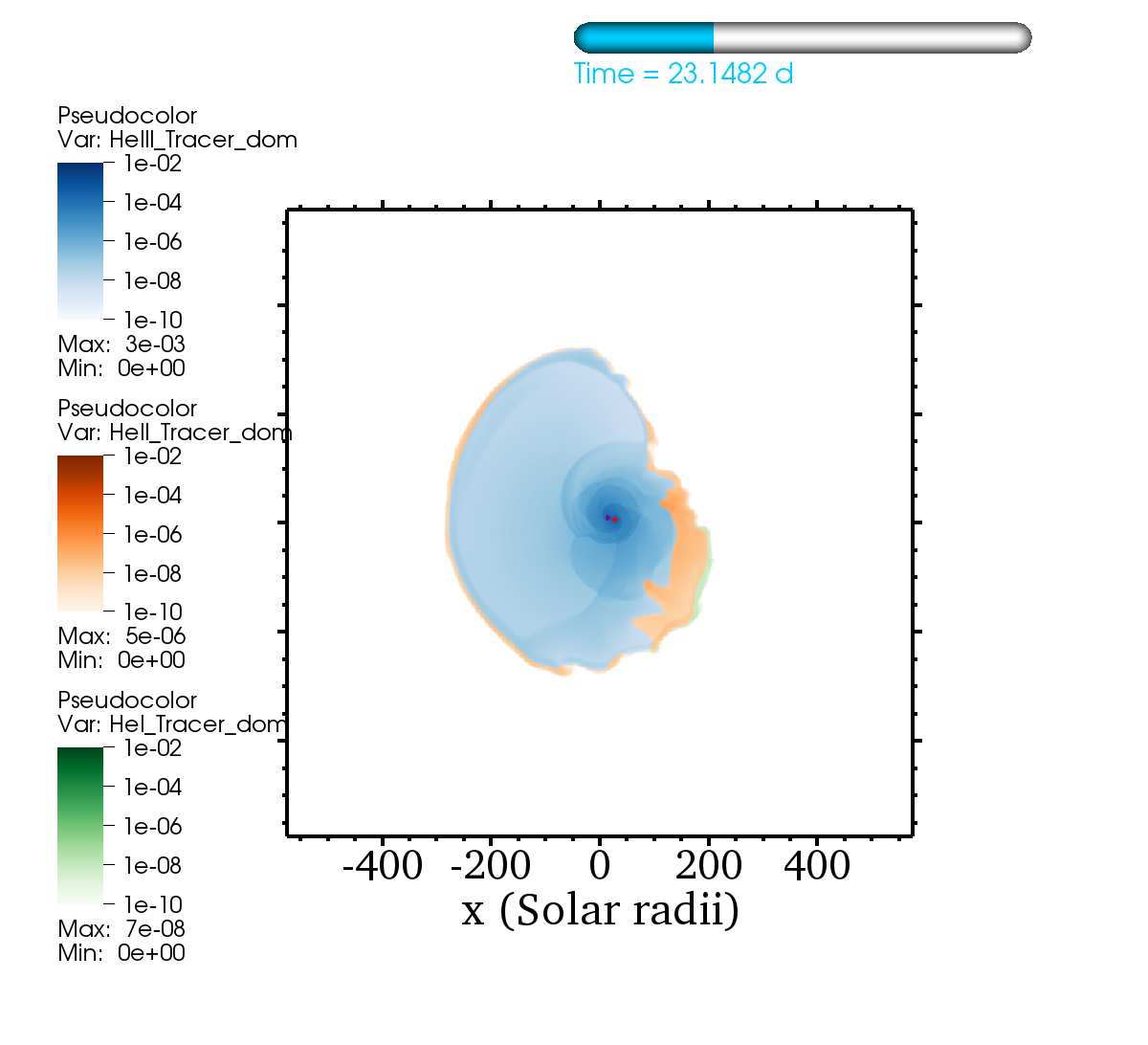}
\includegraphics[scale=0.145,clip=true,trim= 0   85 238 100]{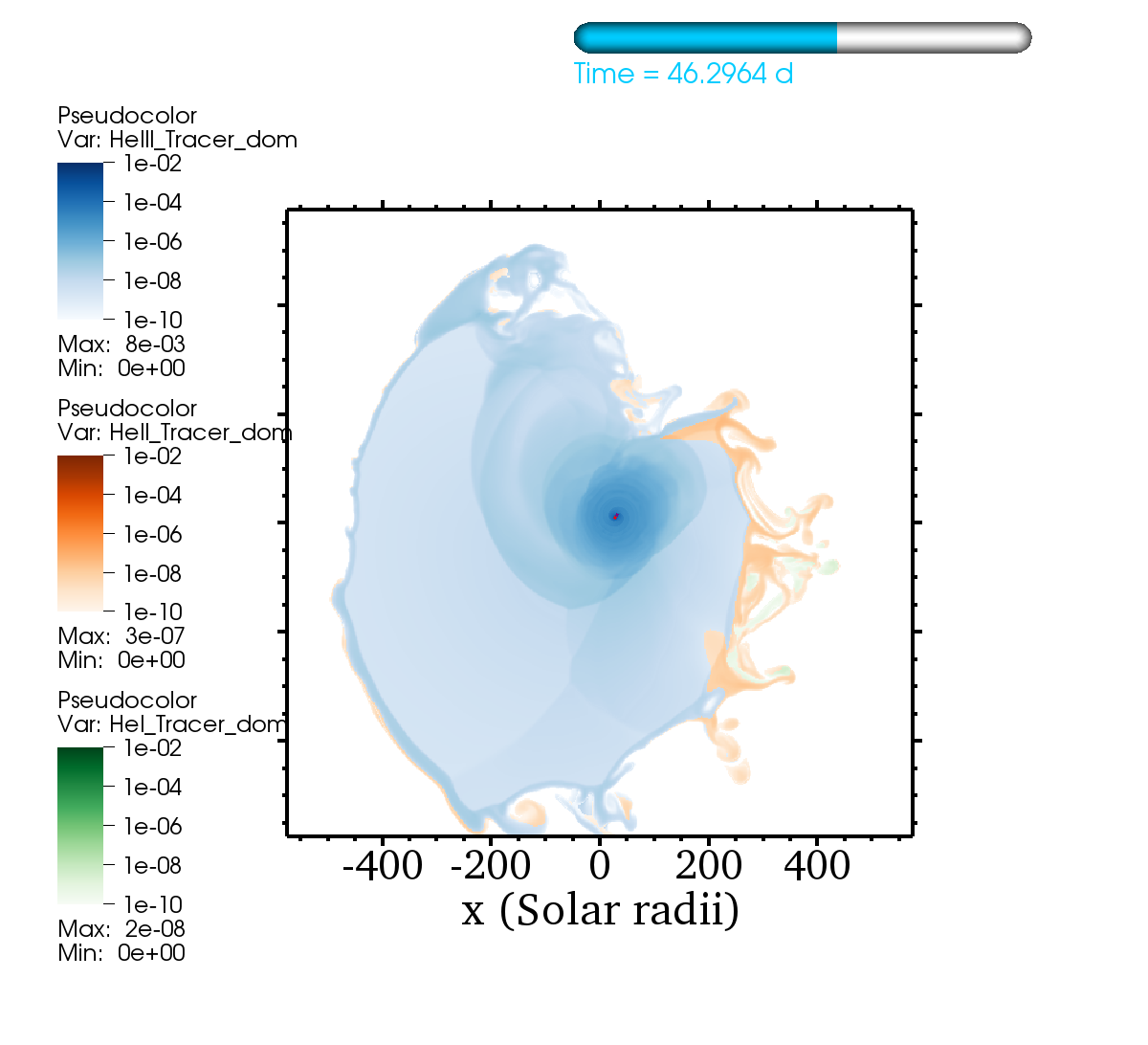}
\includegraphics[scale=0.145,clip=true,trim= 0   85   0 100]{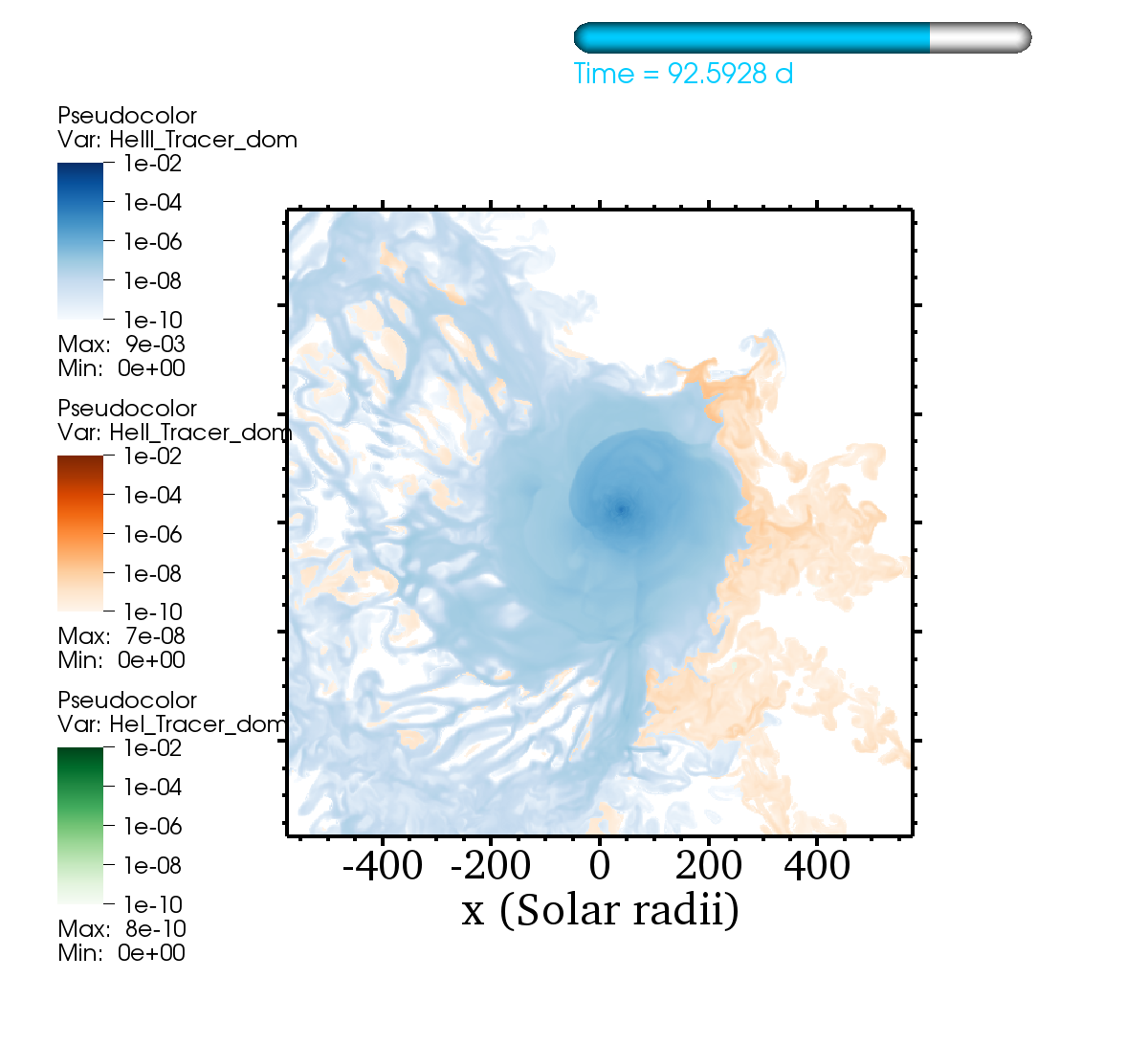}
\\
\includegraphics[scale=0.145,clip=true,trim=50  204 238 180]{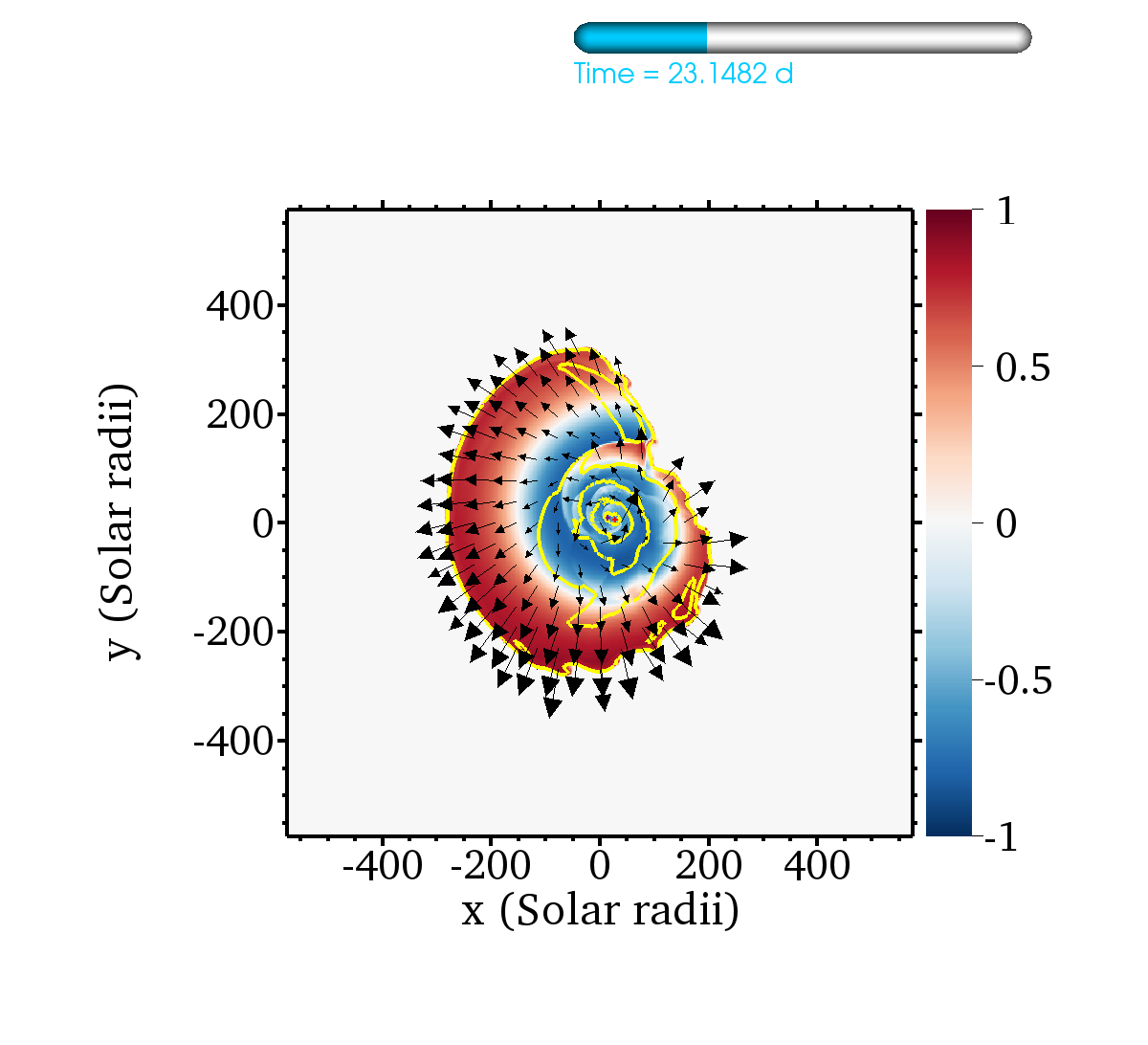}
\includegraphics[scale=0.145,clip=true,trim= 0  204 238 180]{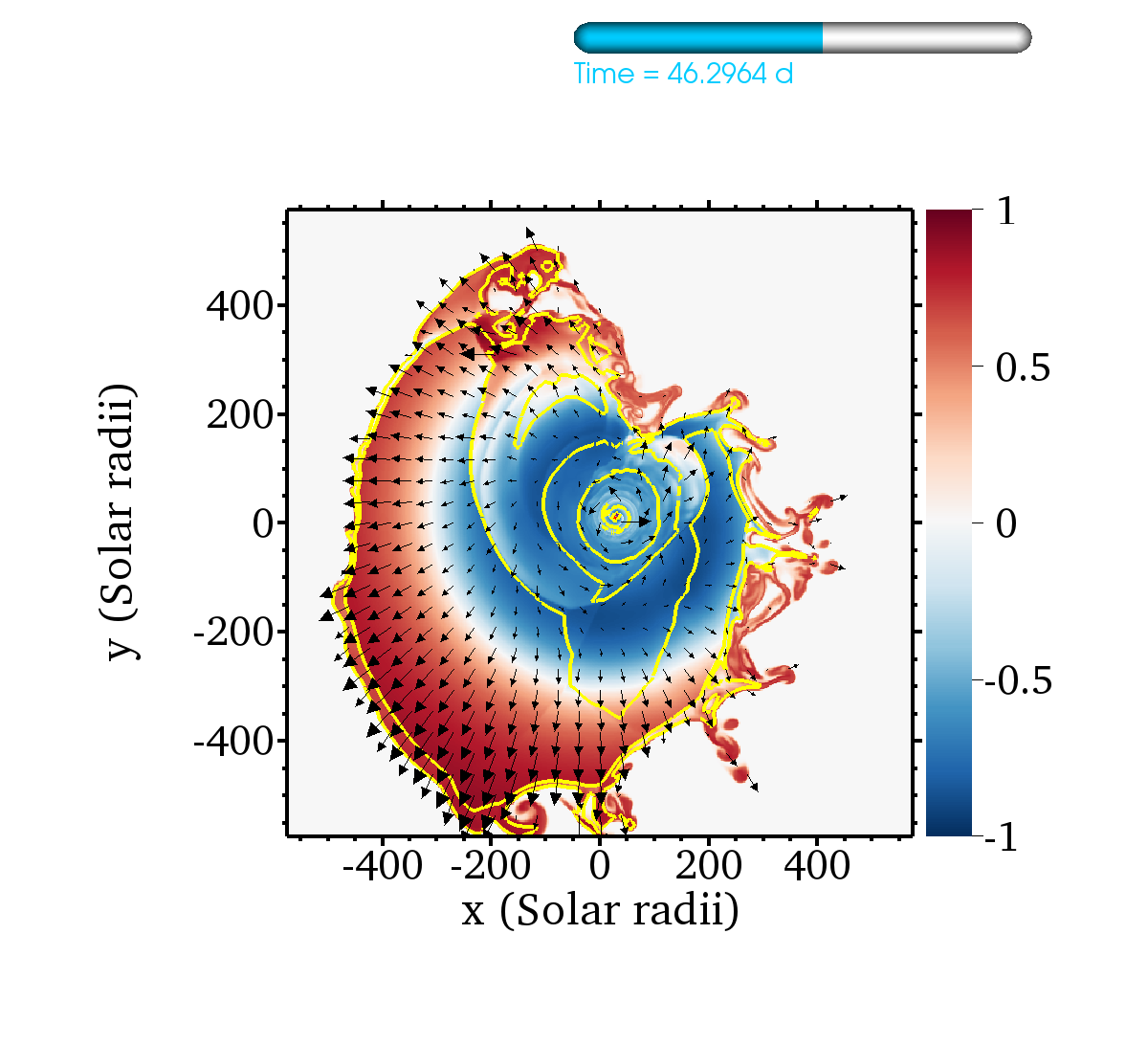}
\includegraphics[scale=0.145,clip=true,trim= 0  204   0 180]{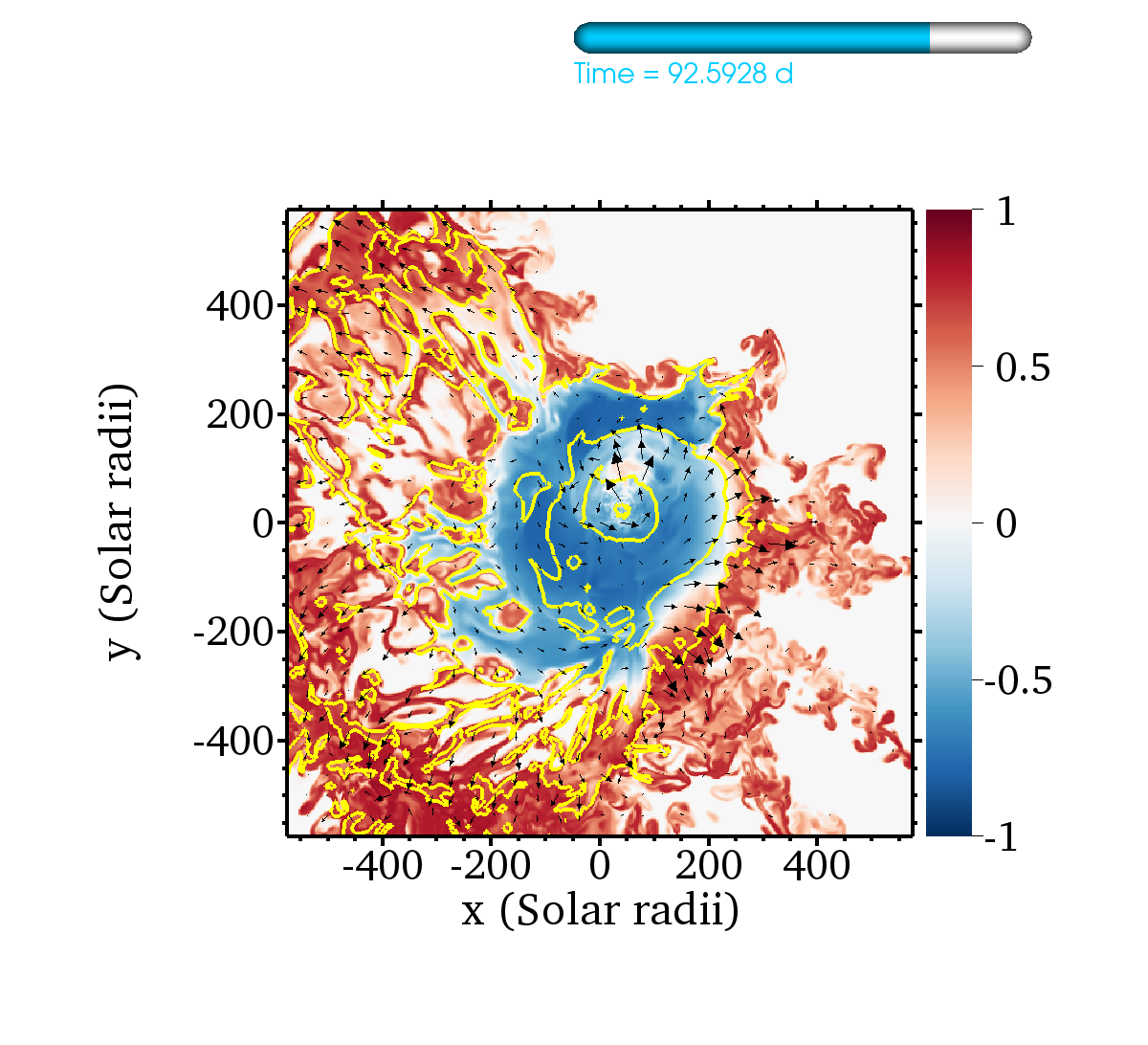}
\\
\includegraphics[scale=0.145,clip=true,trim=50  203 238 180]{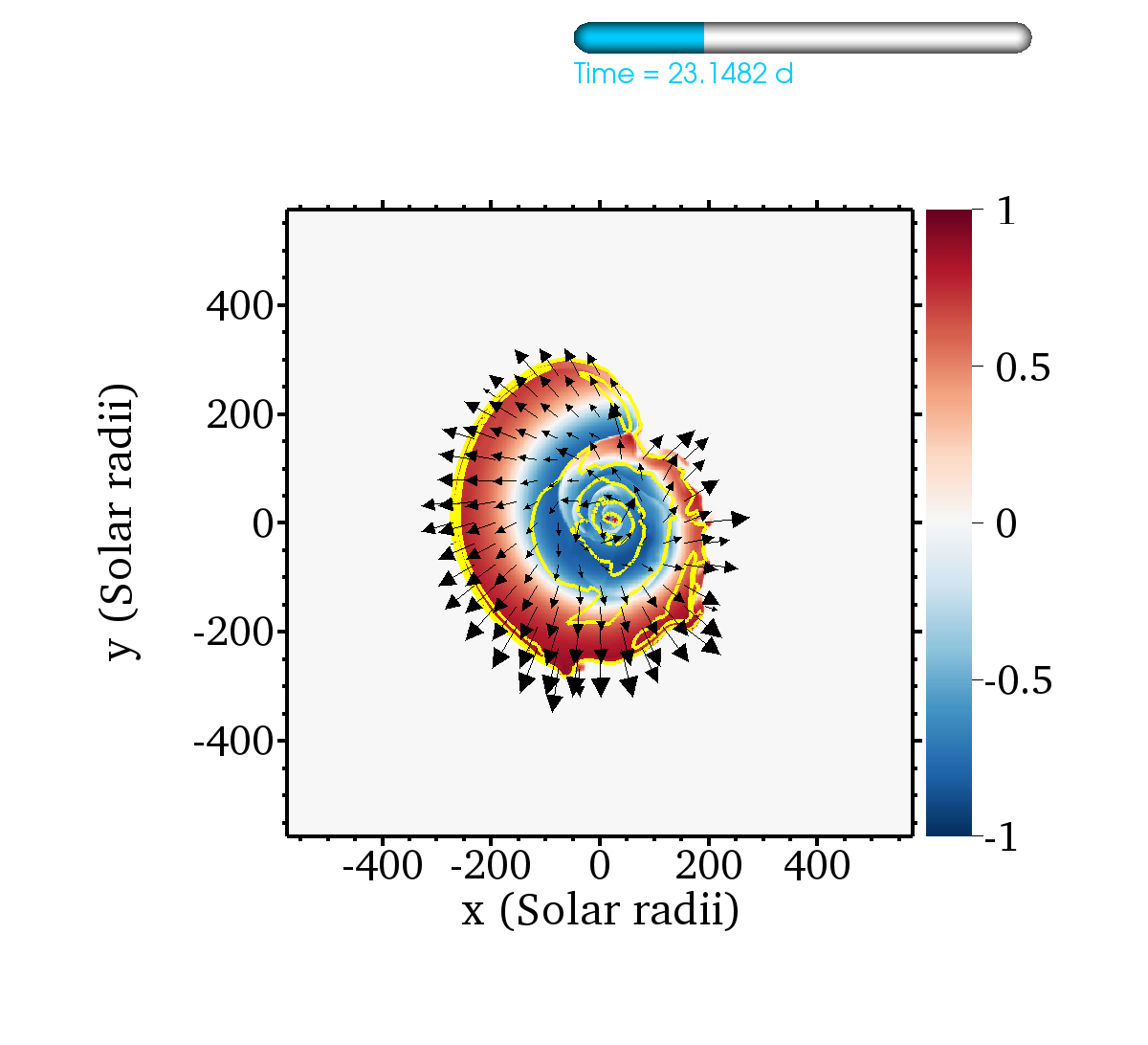}
\includegraphics[scale=0.145,clip=true,trim= 0  203 238 180]{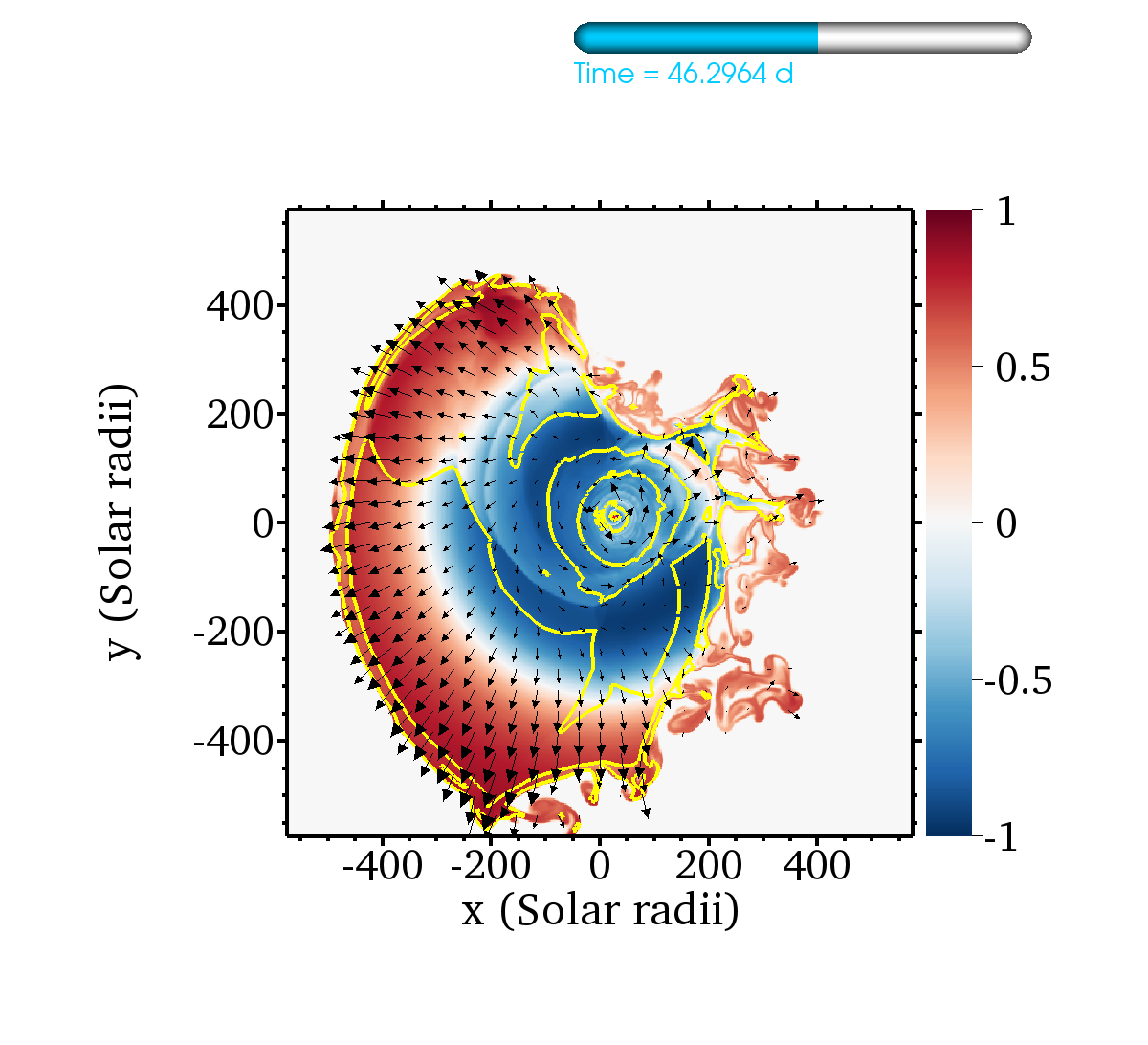}
\includegraphics[scale=0.145,clip=true,trim= 0  203   0 180]{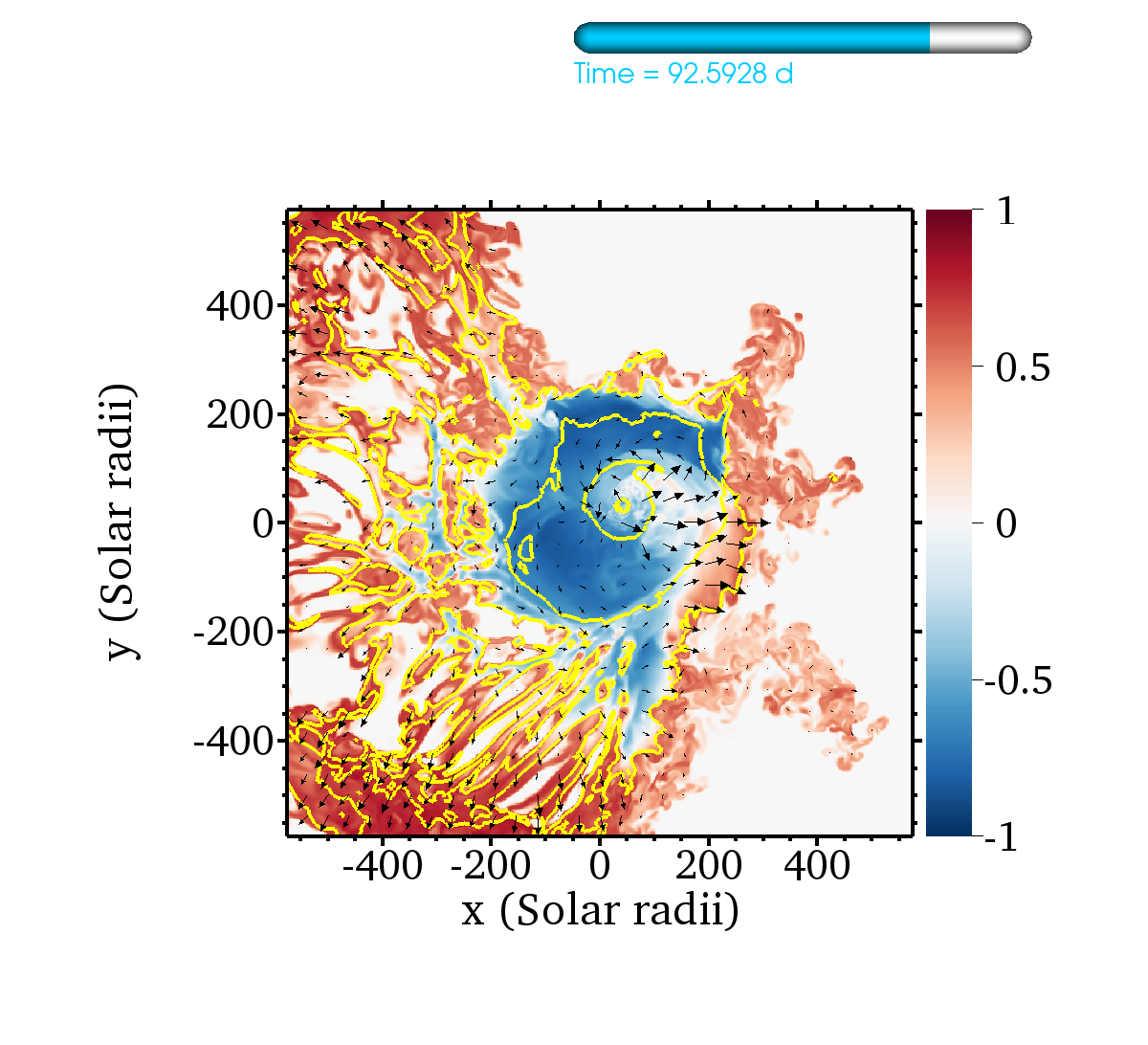}
\\
\vspace{-0.2cm}
\caption{Continuation of Fig.~\ref{fig:He_face-on_0-50} 
(but zoomed out to show the slice for the entire simulation domain) 
for times $t=23.1$, $46.3$ and $92.6\da$.
        }            
\label{fig:He_face-on_100-400}
\end{figure*}

\section{Results}\label{sec:results}
This presentation of results is organized as follows.
We first present 2D snapshots that illustrate the evolution of certain key quantities in Sec.~\ref{sec:spatial}.
We then discuss the orbital evolution of the core particles in Sec.~\ref{sec:orbit}
and analyze the unbinding of the envelope with time in Sec.~\ref{sec:unbinding}.
In Sec.~\ref{sec:energization} we attempt to shed light on the envelope unbinding process and role of recombination energy
by focussing on the energy transfer in the bound envelope gas.
We then study the ionic transitions and their energetics in detail in Sec.~\ref{sec:recombination}.
In Sec.~\ref{sec:ambient_role} we try to determine what role the ambient medium is playing in the simulations.

\subsection{Overall evolution and spatial dependence}\label{sec:spatial} 
In this section we focus on the tabulated EOS run, Model~B.
Snapshots are shown in Figs.~\ref{fig:He_face-on_0-50} and \ref{fig:He_face-on_100-400}.
Slices are through the orbital plane and axes are those of the lab frame.
From top to bottom, the first row shows the gas density $\rho$, 
the second row the temperature $T$, 
the third row the density of each helium ionization species 
($\HeI$ in green, $\HeII$ in orange and $\HeIII$ in blue),
the fourth row the density of helium ion tracer gas, indicating initial ionization state
(using the same colour scheme), 
the fifth row a measure of the unbound mass, discussed below, 
and the sixth row is the same as the fifth but for Model~A instead of Model~B, for comparison.
The left column of the first figure shows snapshots at $t=0$ and the middle column shows $t=5.79\da$.
Each subsequent column, 
starting with the right column of Fig.~\ref{fig:He_face-on_0-50} 
and moving left to right in Fig.~\ref{fig:He_face-on_100-400},
has a time equal to double that of the previous column: $11.57$, $23.15$, $46.30$ and $92.59\da$.
Note that the panels of Fig.~\ref{fig:He_face-on_0-50} 
are more zoomed in than those of Fig.~\ref{fig:He_face-on_100-400}.

\subsubsection{Density}\label{sec:density_spatial}
The top row of Figs.~\ref{fig:He_face-on_0-50} and \ref{fig:He_face-on_100-400} 
shows the mass density in $\gcmcmcm$ for Model~B.
The primary core particle (secondary) softening sphere is visible in the first four columns as a purple (red) circle.
Since the softening radius decreases by a factor of four during the simulation, 
the softening spheres are too small to be plotted in the final two snapshots.
We do not include colour-plots of density for the other run, Model~A, because the results are very similar to Model~B,
but density contours for Model~A are shown in the bottom row.

\subsubsection{Temperature}\label{sec:temperature_spatial}
The second row shows the gas temperature for Model~B, with density contours overplotted. 
At first (Fig.~\ref{fig:He_face-on_0-50}) the local temperature 
does not undergo much change as the star deforms.
Then the outer envelope shows a modest amount of cooling (Fig.~\ref{fig:He_face-on_100-400} 
left and middle columns),
and eventually it heats up again owing to mixing with ambient gas (Fig.~\ref{fig:He_face-on_100-400} 
right column).

\subsubsection{Recombination and ionization}\label{sec:recombination_spatial}
The third row of Figs.~\ref{fig:He_face-on_0-50} and \ref{fig:He_face-on_100-400} 
shows the helium ionization state of the gas,
as computed from the Saha equation. 
This is done by plotting the mass density of each ionic species:
$\HeIII$ (blue), $\HeII$ (orange) and $\HeI$ (green).
We focus on helium rather than hydrogen because, as we will see,
helium releases more recombination energy during the simulation.
For clarity, we plot only the ionic species whose density is highest at a given position.
In other words, for ease of presentation, we do not attempt to plot overlapping colours.
However, 
the calculations below include contributions from every hydrogen and helium ionic species at every location.%
\footnote{Note that at $t=0$ there is a very thin (barely visible) 
layer at the stellar surface where $\HeII$ dominates 
(orange, outside of the green).
This is caused by the higher temperature, which is in turn caused by our prescription for the ambient.
This extra ionization is accounted for in the calculations below and does not greatly affect the results.}

To get a sense of where and when ionization and recombination are occurring, 
we need to understand how the ionization state of each fluid element changes over time.
The fourth row shows the density of tracer gas for each of the helium ion species (Sec.~\ref{sec:tracers}).
These plots tell us where envelope gas that was originally dominated 
by one or the other helium ionic species at $t=0$ is located at the present time $t$.
Visually comparing the third and fourth rows allows one 
to broadly understand where ionization and recombination have taken place.%
\footnote{Note that the thin outer layer seen in the third row at $t=0$ 
and mentioned in the previous footnote is not present in the fourth row
because the ion tracers were defined based on the MESA input profile, not the simulation initial condition;
this is accounted for in the analysis below.}

As an example, let us first consider the snapshot at $t=11.6\da$ (Fig.~\ref{fig:He_face-on_0-50}, right column), 
which is just before the first periastron passage time of $t=13.0\da$.
Clearly, much of the $\HeIII$ tracer gas (4th row, blue) has recombined 
to become $\HeII$ (3rd row, orange) or $\HeI$ (3rd row, green).
Additionally, much of the $\HeII$ tracer (4th row, orange) has recombined to become $\HeI$ (3rd row, green).
However, the bulk of the gas remains $\HeIII$ (3rd row, blue).
Interestingly, ionization has also occurred in some places.
In particular, it appears that some of the $\HeII$ tracer gas gets dragged down by the secondary to hotter regions,
where it gets ionized.
This is visible in the thin wake of dense (first row) and hot (second row) material 
attached to the secondary (red circle) 
that is orange in the fourth row (originally dominated by $\HeII$) but blue in the third row 
(now dominated by $\HeIII$).
To the right of the primary core particle, 
there is another stream of material within which ionization of $\HeII$ has occurred.

Ionization also happens in envelope gas that is in direct contact with the ambient medium.
At the early times shown in Fig.~\ref{fig:He_face-on_0-50}, 
this seems to be caused by shock heating at the envelope-ambient interface.
This can be seen along the outer edge of the structure, 
where blue $\HeIII$ has replaced orange $\HeII$ and green $\HeI$.
However, this gas is very 
low-density,
so this ionization does not have a large influence on the energy budget at this stage.
At a later stage (Fig.~\ref{fig:He_face-on_100-400}), 
mixing with the ambient medium heats and reionizes gas in the outer envelope,
and we revisit this in Sec.~\ref{sec:ambient_role}.

\begin{figure*}
\includegraphics[width=0.95\textwidth,clip=true,trim= 0 0 0 0]{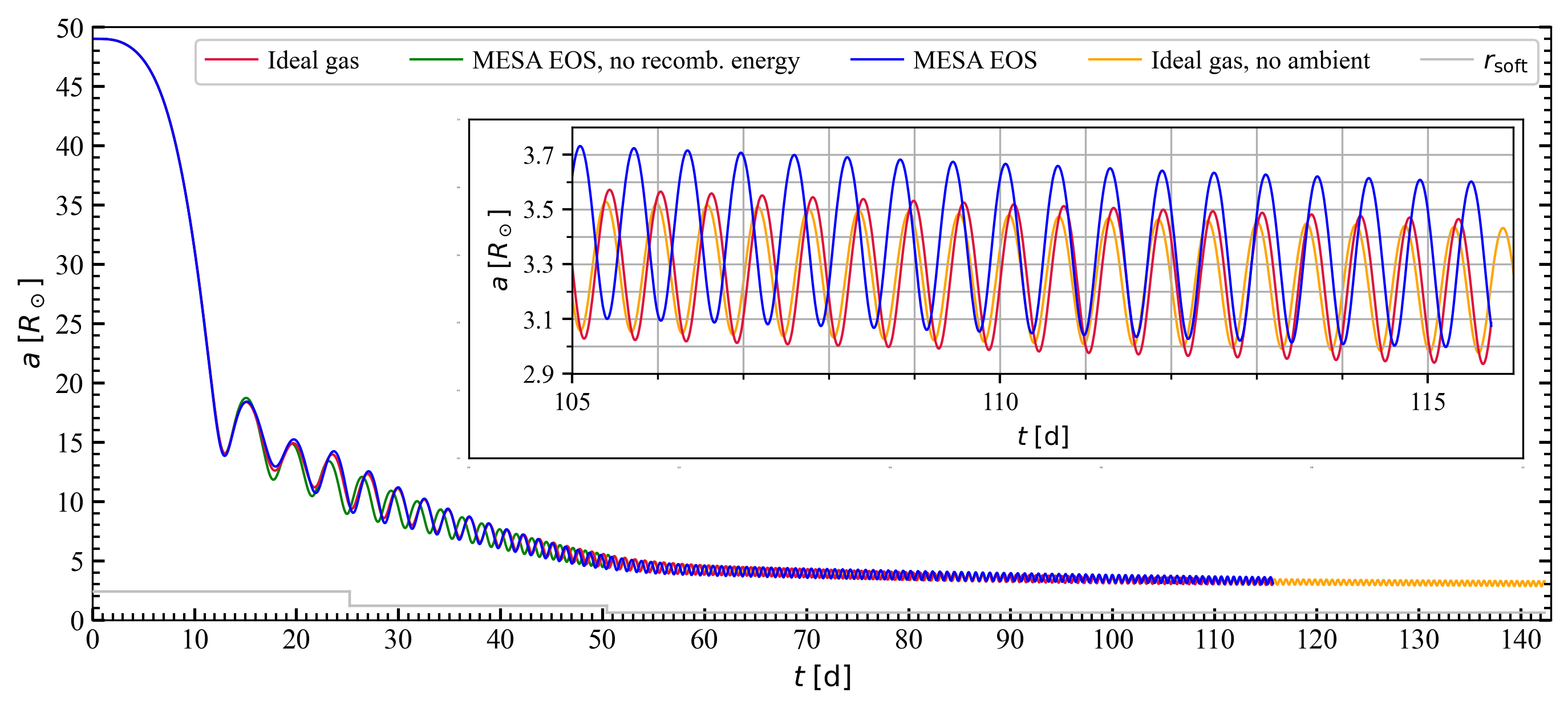}
\caption{Evolution of the orbital separation of the core particles for the three runs.
The softening radius is shown for reference.
The inset shows the evolution of the separation at late times.
}            
\label{fig:separation}
\end{figure*}

\subsubsection{Envelope unbinding}\label{sec:unbinding_spatial}
The fifth row of Figs.~\ref{fig:He_face-on_0-50} and \ref{fig:He_face-on_100-400} 
shows the differentiation of the envelope gas into bound (blue) and unbound (red) components,
along with velocity vectors in the orbital plane and density contours, for Model~B. 
The sixth row shows the same snapshots, but now for the comparable ideal gas run, Model~A.
Gas is defined to be unbound if 
\begin{equation}\label{unbound}
  \mathcal{E}\env\equiv
  \mathcal{E}_\mathrm{blk,env}
  +\mathcal{E}_\mathrm{thm,env}
  +2\mathcal{E}_\mathrm{pot,env-1}
  +2\mathcal{E}_\mathrm{pot,env-2}
  +\mathcal{E}_\mathrm{env-gas}>0.
\end{equation}
Here, $\mathcal{E}_\mathrm{blk,env}$ is the bulk kinetic energy density of envelope gas, 
$\mathcal{E}_\mathrm{thm,env}$ is the thermal energy density of envelope gas, 
$2\mathcal{E}_\mathrm{pot,env-1}$ is twice the potential energy density of envelope gas associated with the RGB core particle,
$2\mathcal{E}_\mathrm{pot,env-2}$ is twice the potential energy density of envelope gas associated with the secondary,
and $\mathcal{E}_\mathrm{env-gas}$ is the potential energy density due to self-gravity.%
\footnote{The factors of $2$ are included to account for the half of the potential energy nominally located inside the core particles.
The self-gravity term is labeled ``env-gas'' because it uses the gravitational potential from all the gas (envelope+ambient)
to determine whether envelope gas is bound or unbound. 
Omitting the ambient from the potential would reduce its magnitude by $\lesssim1\%$,
and this small difference would be comparable for Models~A, B and C.
No definition of unbound predicts with accuracy whether fluid elements will ultimately become unbound, 
because the outcome will depend on other factors \citep[e.g.][]{Ivanova+13a}. 
The definition used here is fairly conservative. 
We assess the sensitivity to the definition of ``unbound'' in Sec.~\ref{sec:unbinding}; see also \citet{Chamandy+19a}.}
The quantity plotted in the fifth row of Figs.~\ref{fig:He_face-on_0-50} and \ref{fig:He_face-on_100-400} 
is 
\begin{equation}\label{widetildeE}
  \widetilde{\mathcal{E}}\env\equiv\mathcal{E}\env/\max(\mathcal{E}_\mathrm{kin,env},-\mathcal{E}_\mathrm{pot,env}),
\end{equation}
where $\mathcal{E}_\mathrm{kin,env}\equiv \mathcal{E}_\mathrm{blk,env} +\mathcal{E}_\mathrm{thm,env}$ and 
$\mathcal{E}_\mathrm{pot,env}\equiv 2\mathcal{E}_\mathrm{pot,env-1} +2\mathcal{E}_\mathrm{pot,env-2} +\mathcal{E}_\mathrm{env-gas}$;
this definition ensures that $-1\le\widetilde{\mathcal{E}}\env\le1$. 

To begin with, comparing the panels in the fifth (Model~B, tabulated EOS) and sixth (Model~A, ideal gas EOS) rows 
of Figs.~\ref{fig:He_face-on_0-50} and \ref{fig:He_face-on_100-400} 
shows that the envelope unbinding process is extremely similar in the two runs,
but that Model~B has a slightly faster expansion.
In addition, the unbound mass is slightly higher in Model~B;
this is most evident when comparing the panels in the left column of Fig.~\ref{fig:He_face-on_100-400} ($t=23.1\da$).
The faster expansion and greater unbinding are consistent with the expectation that the energy released into the gas
owing to recombination assists envelope unbinding.
 However, the similarity between the runs shows that the release of orbital energy, 
rather than recombination energy, is the main driver of mass unbinding at this stage.

\textit{How much} of the released recombination energy helps to unbind the envelope?
Recombination in gas which is \textit{already unbound} will likely not assist unbinding. 
By comparing the third, fourth and fifth rows of Figs.~\ref{fig:He_face-on_0-50} and \ref{fig:He_face-on_100-400}, 
we can infer how much recombination happens in bound versus unbound envelope gas.
Recombination is significant in both, 
which confirms that only \textit{part} of the released recombination energy is utilized to unbind the envelope.
We address this  further in Sec.~\ref{sec:recombination}.

\subsection{Orbital Evolution of the core particles}\label{sec:orbit}
Fig.~\ref{fig:separation} shows the  evolution of the core particles' separation $a$.
The inset shows a zoom-in of the late stage of the simulation.
For the first $\sim10$ orbits ($\sim40\da$), 
the particle separations are very similar in Models~A ($\gamma=5/3$ EOS; red) and B (tabulated EOS; blue).
Then Model~B acquires a \textit{slightly}
smaller one period-averaged mean separation than Model~A.
This can be seen at $t=50\da$.
After that, the difference in the mean separations gradually reduces and  flips sign near $t=65\da$. 
By the end of the simulation, Model~B has a slightly larger mean separation than Model~A, as seen in the inset.
The larger separation in the tabulated EOS run as compared to the ideal gas run 
is qualitatively consistent with \citet{Sand+20}, \citet{Lau+22a}, \citet{Gonzalez-bolivar+22}, 
which report on systems with different binary parameter values.
The results of \citealt{Reichardt+20} were less conclusive on this point.

The separation curve of Model~C is quite similar to those of Models~A and B,
but shows less correspondence with Model~A than does Model~B.
Recall that in Model~C we replaced the internal energy of the EOS tables
with the thermal energy alone, akin to Model~A, 
so the  distinct evolutions between between Model~C and Model~A must be due to other model differences.

The separation curve for the reduced ambient run Model~D agrees very closely with that of Model~A.
This verifies that the orbital evolution is insensitive to the density and total mass of ambient gas.

Finally, note that the mean separation continues to decrease monotonically at the end of the simulations,
which shows that the orbital inspiral has not ended. 
This is consistent with the fact that envelope gas is still present around the core particles. 
We expand on this point in Sec.~\ref{sec:extrapolation}.

\begin{figure*}
\includegraphics[width=0.95\textwidth,clip=true,trim= 0 0 0 0]{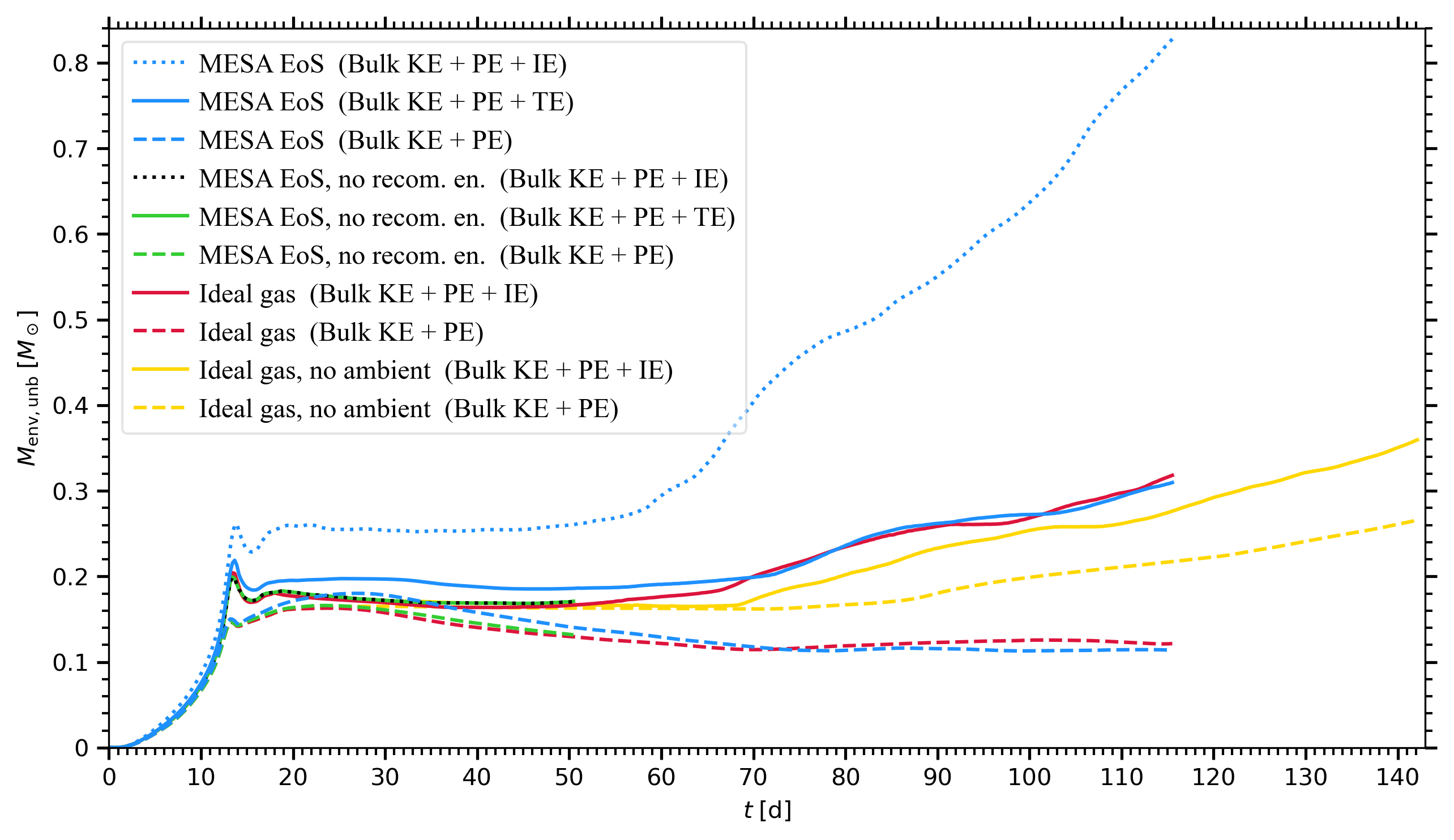}\\
\caption{Evolution of the unbound envelope mass for different definitions of ``unbound.''
Model~A (ideal gas EOS) is plotted in red, Model~B (MESA EOS) is plotted in blue, 
and Model~C (MESA EOS with recombination energy removed) is plotted in green or dotted black. 
Unbound mass according to fiducial definition of equation~\eqref{unbound} (solid),
as equation~\eqref{unbound} but excluding the thermal energy density (dashed), 
or as equation~\eqref{unbound} but also including the recombination energy density (dotted).
Note the excess unbound mass in Model~B as compared to Model~A between $t\approx13\da$ and $t\approx70\da$,
caused by thermalization of released recombination energy. 
}            
\label{fig:EnvMass}
\end{figure*}

\subsection{Evolution of the unbound mass}\label{sec:unbinding}
Fig.~\ref{fig:EnvMass}, shows the unbound mass as a function of time (see also Sec.~\ref{sec:unbinding_spatial}).
To aid interpretation and facilitate comparison with the literature, 
we present the unbound mass using different definitions of ``unbound.''
Our fiducial definition is given by inequality~\eqref{unbound},
and the corresponding unbound mass is represented by solid lines in Fig.~\ref{fig:EnvMass}.
Model~A is shown in red, Model~B in blue, Model~C in green (or dotted black), and Model~D in yellow.
Spatial information to go with the blue and red solid lines are provided in the fifth and sixth rows, respectively,
of Figs.~\ref{fig:He_face-on_0-50} and \ref{fig:He_face-on_100-400}.

The dashed lines in 
Fig.~\ref{fig:EnvMass} do not include the term $\mathcal{E}_\mathrm{thm}$.
This is thus a more  conservative definition of ``unbound.''
The dotted lines show a more liberal definition, where $\mathcal{E}_\mathrm{thm,env}$ 
is replaced by $\mathcal{E}_\mathrm{int,env}$.
As discussed in Sec.~\ref{sec:intro} however, the latter definition is unrealistic because it effectively assumes that \textit{all}
of the recombination energy will be released into gas that is still bound,
or into marginally unbound gas that would otherwise  rebind.
In reality, the process will not be perfectly efficient because some of this energy will be released in gas that is already unbound,  
or radiate away.
This is also why  orbital energy cannot be utilized with $100\%$ efficiency to unbind the envelope \citep{Chamandy+19a}. 
An efficiency $\alpha\rec<1$ can be assigned analogous to $\alpha\CE<1$ for orbital energy \citep[e.g.][]{Zorotovic+10}.
In addition, recombination energy is \textit{latent} not  kinetic, 
so it is misleading to include it unless released as kinetic energy. 

\begin{figure*}
\includegraphics[width=0.95\textwidth,clip=true,trim= 0 0 0 0]{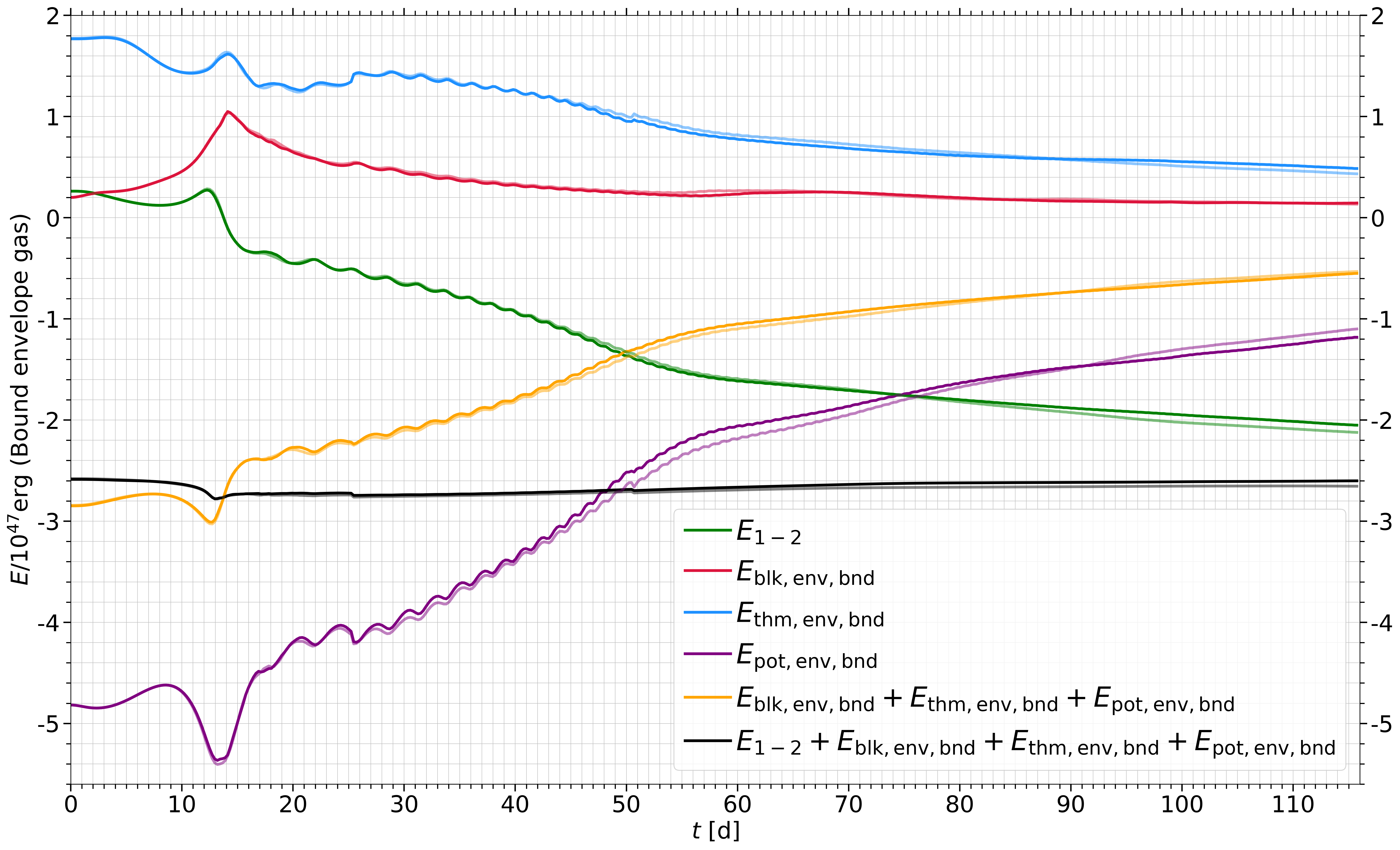}
\caption{Contributions to the energy of bound envelope gas where equation~\eqref{unbound} defines unbound;
Model~B (MESA EOS, opaque lines) and Model~A (ideal gas EOS, semi-transparent lines).
}            
\label{fig:Env_E_terms_bnd}
\end{figure*}

The solid and dashed curves of Fig.~\ref{fig:EnvMass} tell a similar story.
Model~B (tabulated EOS) unbinds $\sim(10$--$20)\%$ more mass than Model~A (ideal gas EOS) starting at the first periastron passage
until $t\approx40\da$.
This is expected based on previous  work (Sec.~\ref{sec:intro}),
and because the envelope acquires  latent energy released by recombination, 
some marginally bound envelope material unbinds and some marginally unbound material is prevented from rebinding.
However, at $t\approx40\da$ the difference in unbound mass between the runs  declines,
nearly vanishing by $t\approx70\da$ and later. 
We will later address why this difference in unbound mass vanishes at late times 
even though recombination occurs in Model~B but not Model~A.

Next consider the solid and dashed green curves, and dotted black curves, showing Model~C,
where the internal energy was replaced with thermal energy in the tabulated EOS.
As expected, the curve that includes the internal energy (dotted) precisely traces the curve that includes thermal energy only, giving confidence that the EOS was  modified correctly.
Also, the solid and dashed curves closely trace   their Model~A counterparts.
The general agreement between Models~A and C is expected, 
and suggests that the differences in the unbound mass between Models~A and B are due to release of recombination energy.
That said, the agreement is not perfect, and Model~C  unbinds slightly more mass than Model~A.

\begin{figure*}
\includegraphics[height=0.25\textwidth,clip=true,trim= 0 0 0 0]{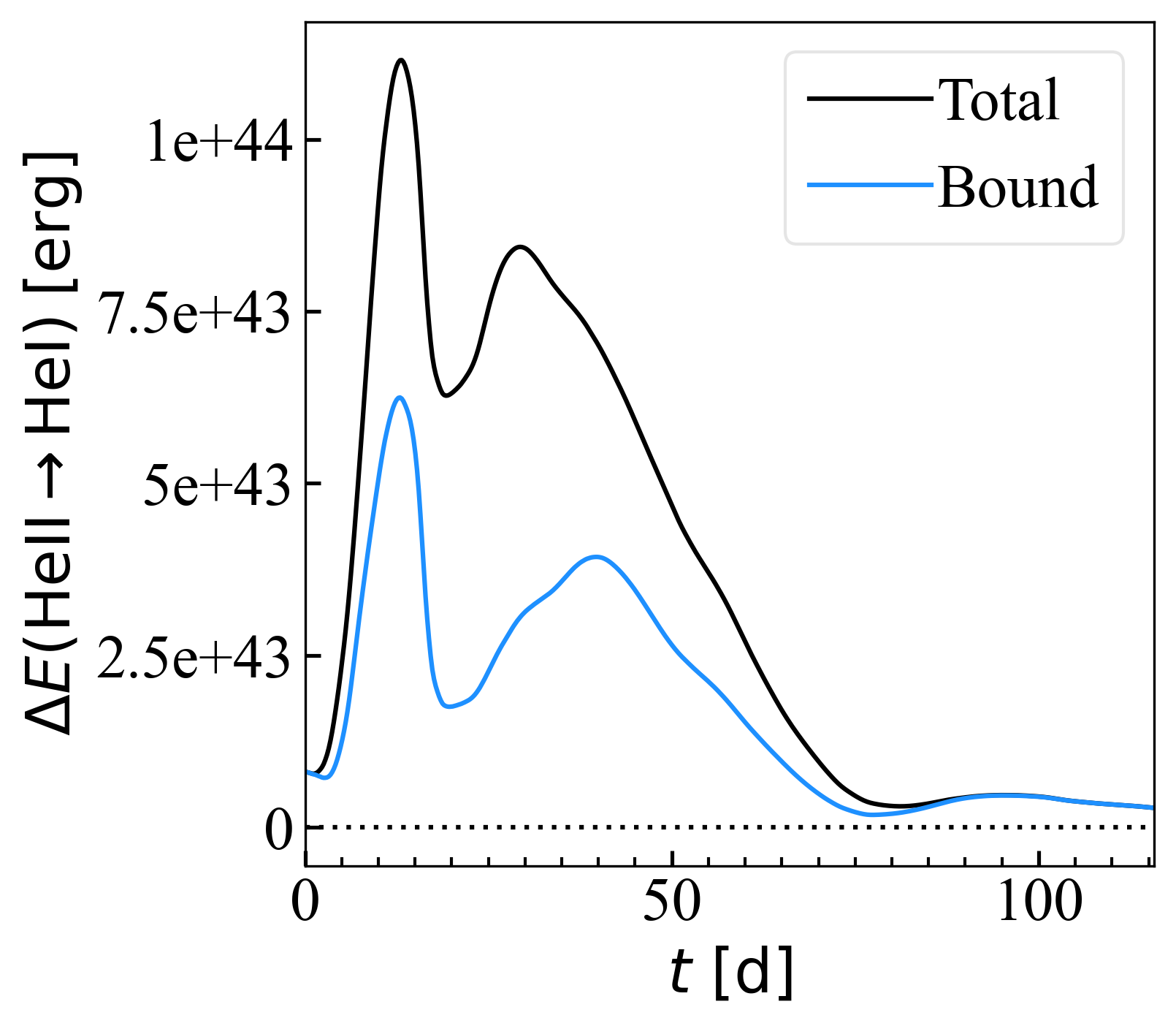}
\includegraphics[height=0.25\textwidth,clip=true,trim= 0 0 0 0]{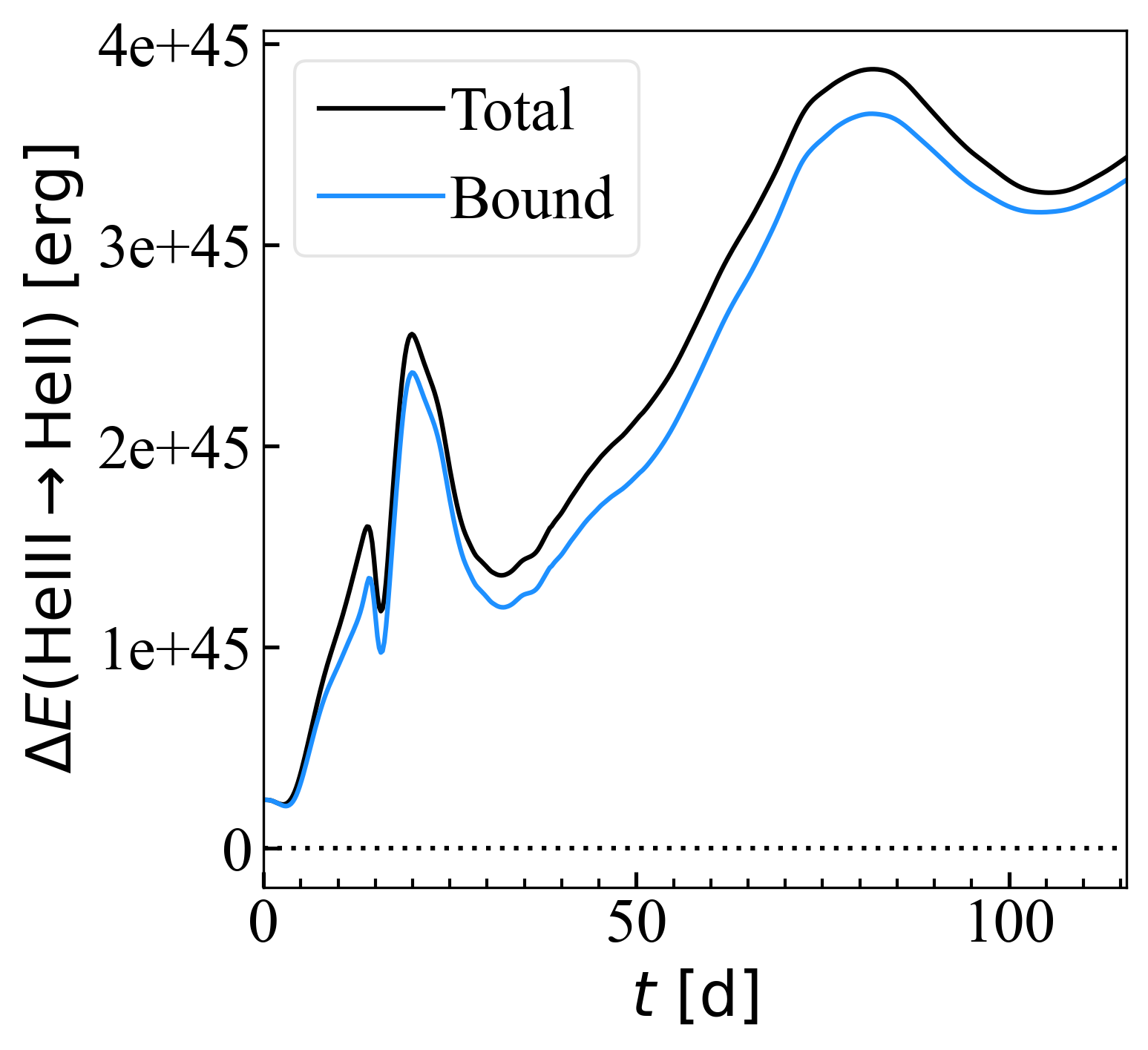}
\includegraphics[height=0.25\textwidth,clip=true,trim= 0 0 0 0]{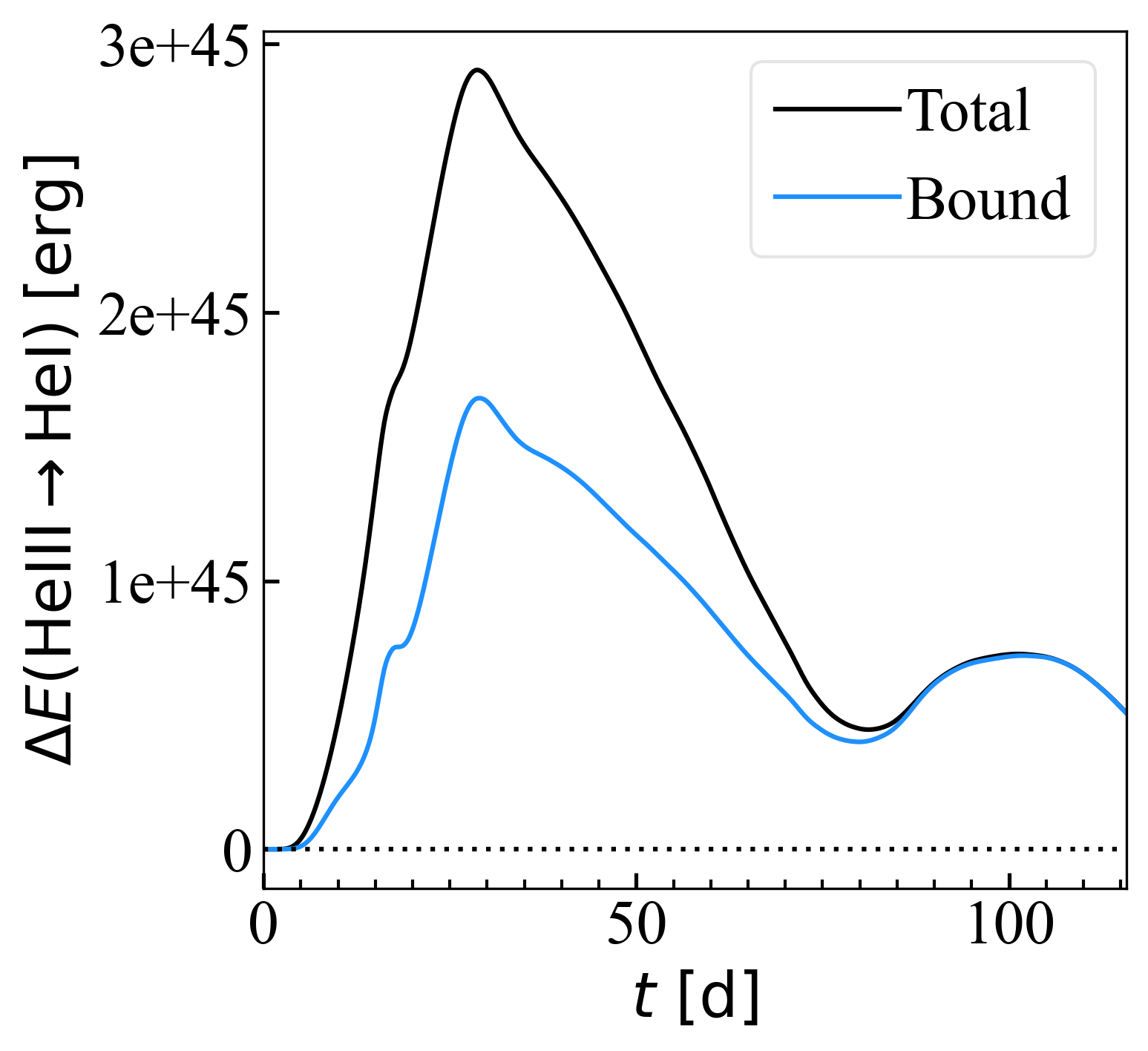}\\
\includegraphics[height=0.25\textwidth,clip=true,trim= 0 0 0 0]{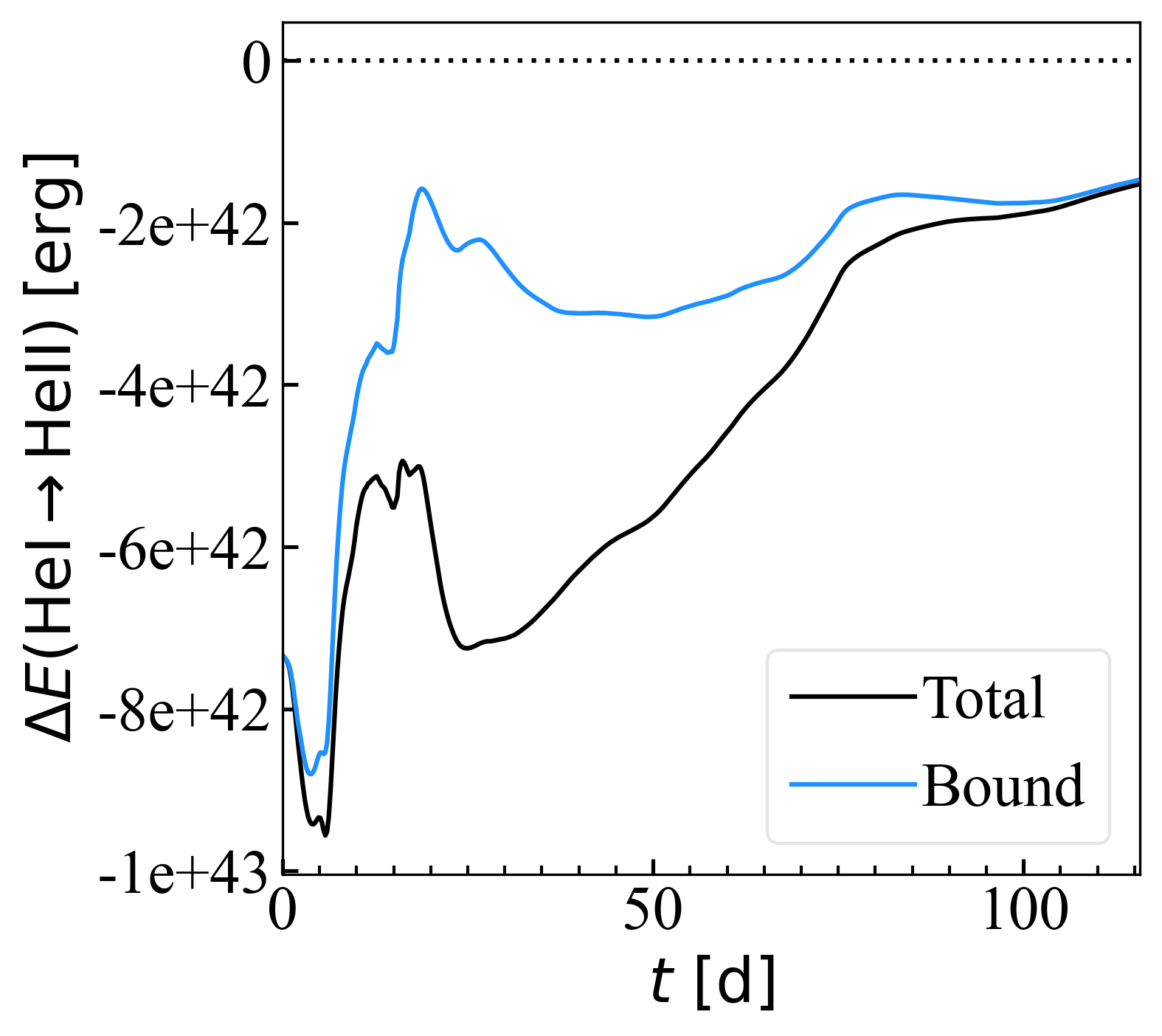}
\includegraphics[height=0.25\textwidth,clip=true,trim= 0 0 0 0]{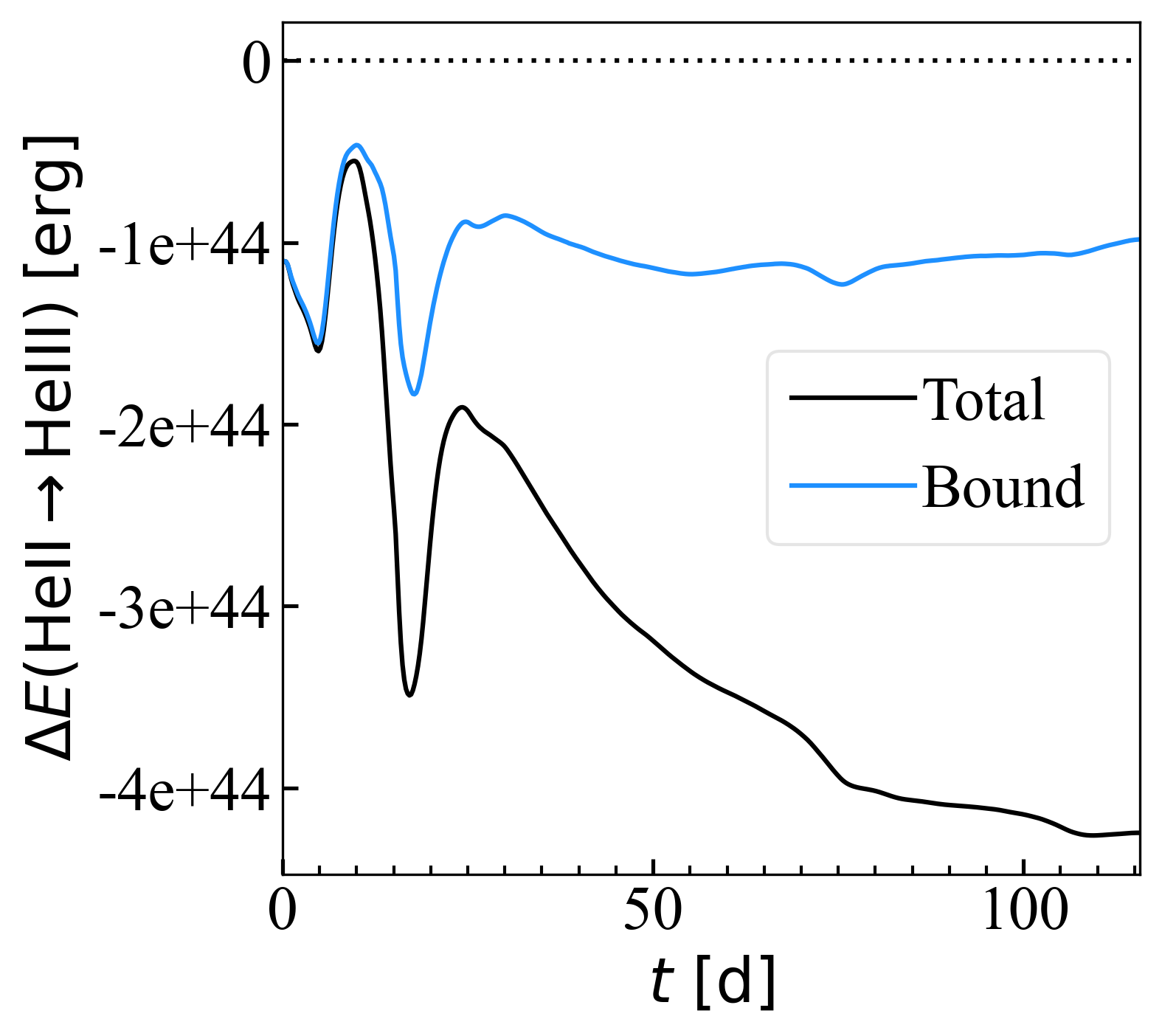}
\includegraphics[height=0.25\textwidth,clip=true,trim= 0 0 0 0]{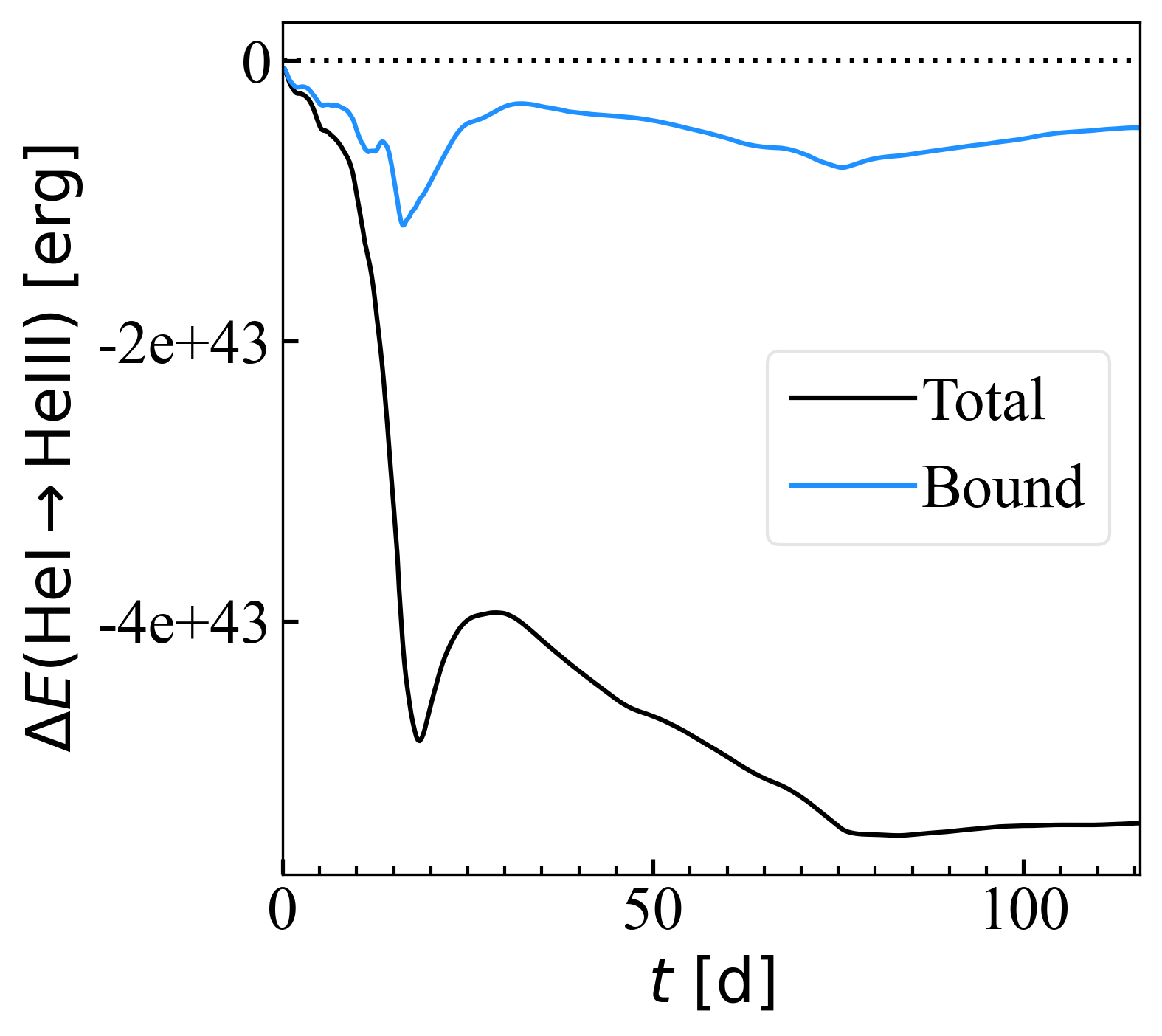}\\
\caption{Energy released by each ionic transition of helium. 
Recombination results in positive values (top row) whereas ionization results in negative values (bottom row).
The recombination of $\HeIII$ tracer gas into $\HeII$ releases the most energy into bound envelope gas, 
and the recombination of $\HeIII$ tracer gas into $\HeI$ also contributes significantly.
From the top row, we also see that reionization of $\HeI$ into $\HeII$ absorbs substantial energy during the simulation. 
The non-zero values at $t=0$ in some panels are explained in the main text.
        }            
\label{fig:DeltaE_He}
\end{figure*}

\begin{figure*}
\includegraphics[height=0.25\textwidth,clip=true,trim= 0 0 0 0]{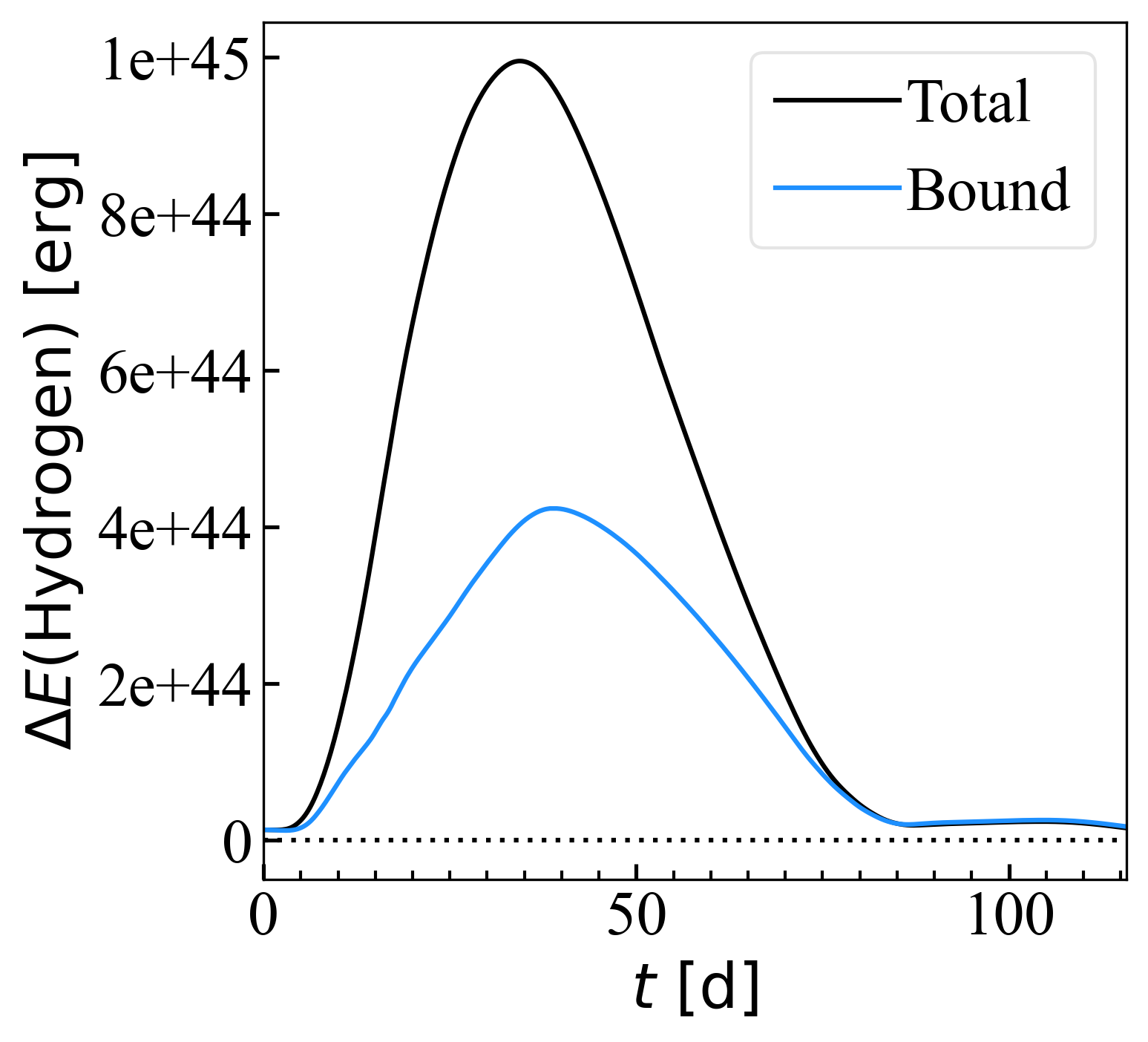}
\includegraphics[height=0.25\textwidth,clip=true,trim= 0 0 0 0]{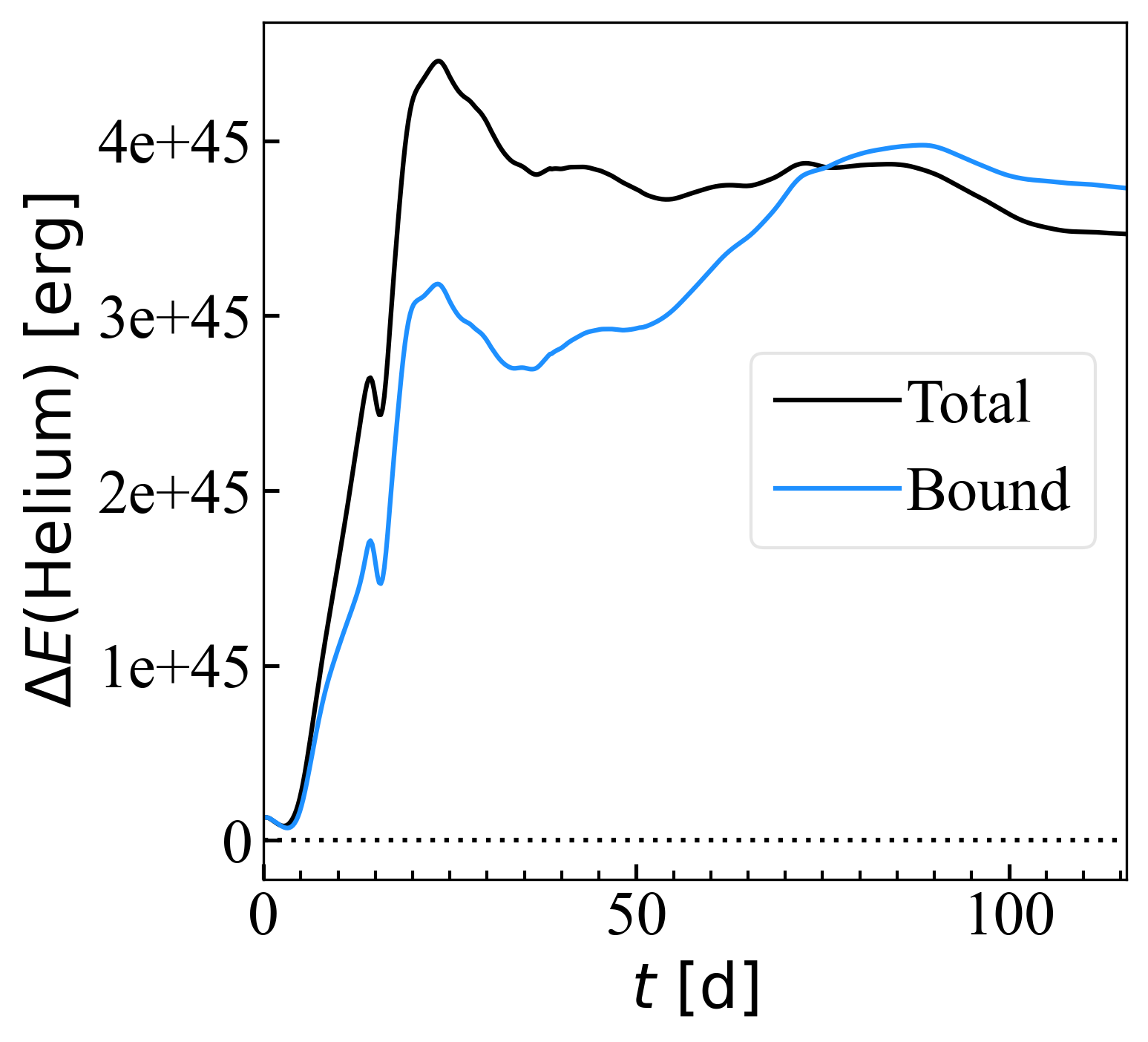}
\includegraphics[height=0.25\textwidth,clip=true,trim= 0 0 0 0]{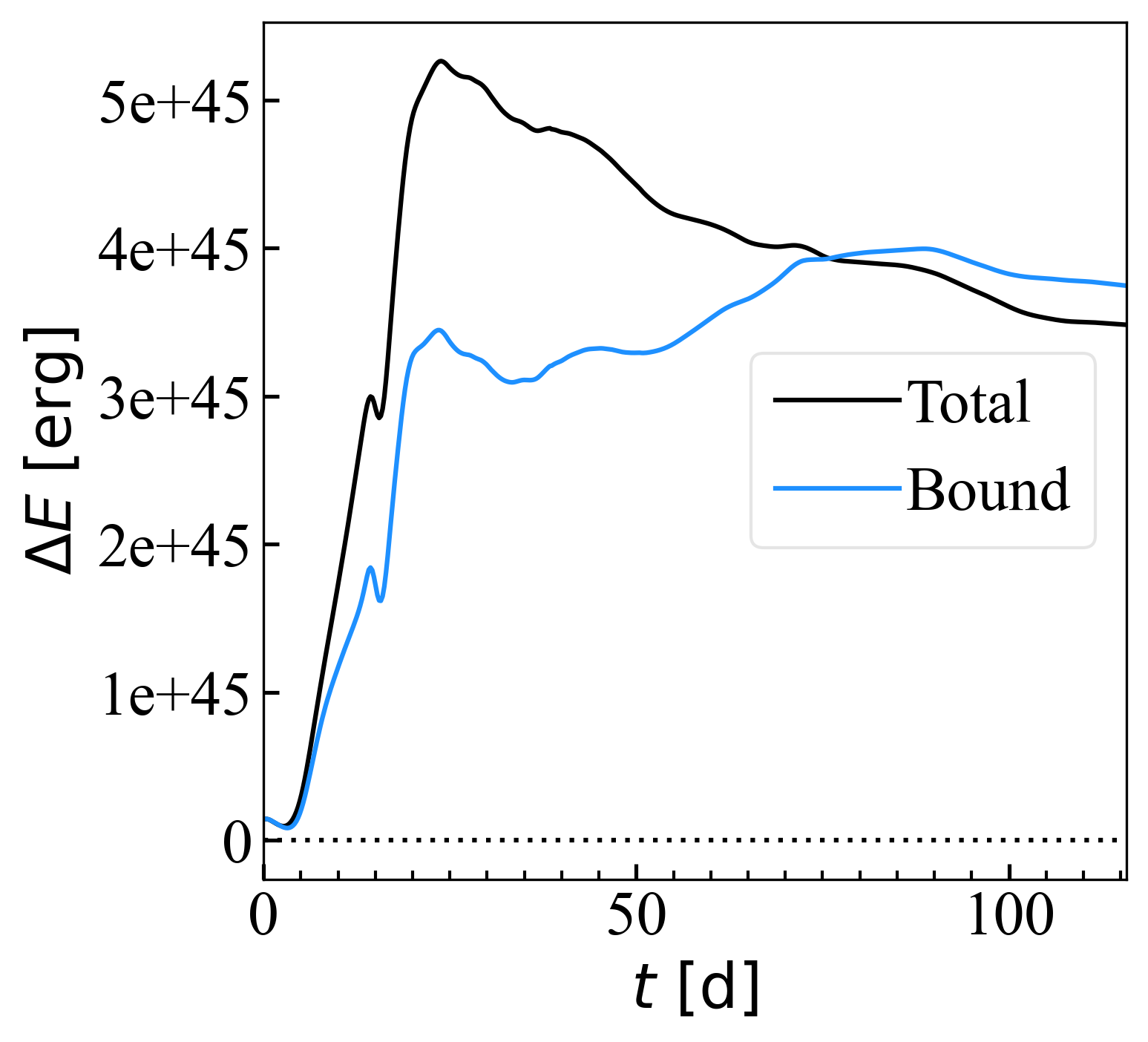}\\
\caption{Release of energy from recombination (left column) and ionization (middle column) of H (top row), He (middle row),
and their combination (bottom row). 
The net energy release is plotted in the right column.
Note that the ambient gas is excluded (as it should be).
        }            
\label{fig:DeltaE}
\end{figure*}

\begin{figure}
\includegraphics[width=1.0\columnwidth,clip=true,trim= 0 0 0 0]{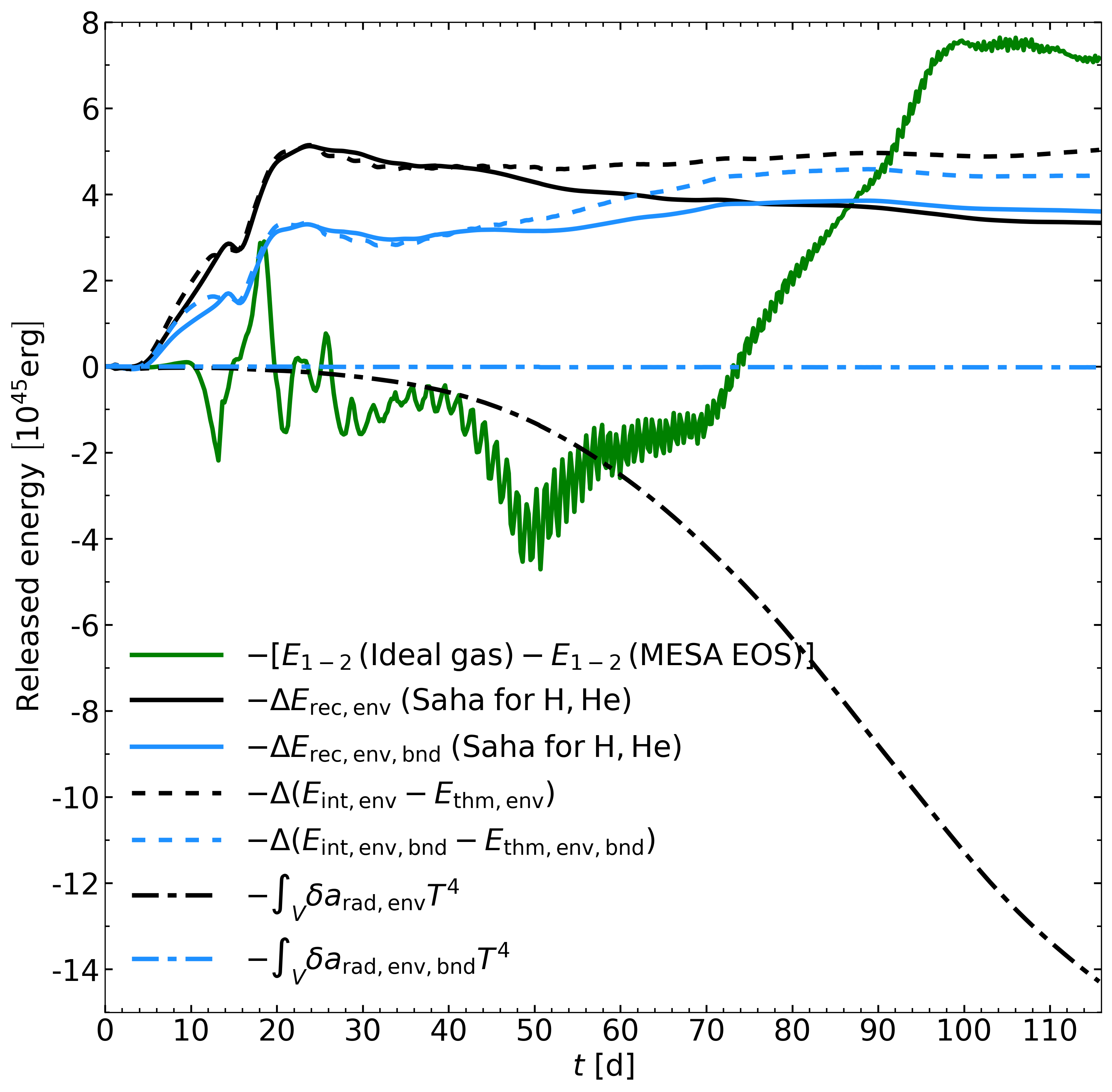}
\caption{Comparison of recombination energy release calculated using two different methods.
The first  uses the Saha equation and is shown as a solid black line, 
or a solid blue line for just the component which is bound at time $t$.
These lines are identical to those in the right panel of Fig.~\ref{fig:DeltaE}.       
The second method uses equation~\eqref{E_rec}, and is shown by dashed lines.
The green line shows the difference in the energy injected by the particles in the two runs,
with the ideal gas run hosting a deeper inspiral and thus greater energy release.
By the end of the simulation, 
the absence of recombination energy release in the ideal gas run is fully or partially compensated by the deeper inspiral.
}            
\label{fig:DeltaE_bnd_compare}
\end{figure}

\subsection{Envelope energization, unbinding and expansion}\label{sec:energization}
We now turn to Fig.~\ref{fig:Env_E_terms_bnd},
which shows the evolution of the volume-integrated energy components of the \textit{bound} envelope gas,
where the fiducial definition of unbound [equation~\eqref{unbound} and solid lines in 
Fig.~\ref{fig:EnvMass}] has been used.%
\footnote{As the bound envelope mass is not  the same between  runs (Sec.~\ref{sec:unbinding}),
we also considered the specific energy, 
but that plot is qualitatively  similar to Fig.~\ref{fig:Env_E_terms_bnd} and leads  to the same conclusions.}
The subscript `bnd' refers to that fraction of the energy contained in gas that is bound at time $t$.
The opaque curves represent Model~B and the semi-transparent curves represent Model~A in Fig.~\ref{fig:Env_E_terms_bnd}.
The particle energy $E_{1-2}$, in green, 
is the sum of the kinetic energies of the particles and their mutual gravitational potential energy.%
\footnote{In addition to the kinetic energy associated with the quasi-Keplerian motion, 
$E_{1-2}$ also includes the kinetic energy associated with the centre of mass motion.
This contribution lessens as the system evolves and by $t=40\da$ it is  negligible \citep{Chamandy+19a}.}

The blue curves show the gas thermal energy, the red curves show the bulk kinetic energy,
and the purple curves show the potential energy, including the gas self-gravity as well as the potential energy associated 
with the mutual attraction between the core particles and the bound gas.
The orange curves show the sum of blue, red and purple curves, 
and the black curves are the sum of green and orange curves.%
\footnote{The black curves do not include the energy of unbound envelope gas or ambient gas 
and so need not remain constant.
The discontinuities at $t\approx25\da$ and $t\approx50\da$ 
are due to the instantaneous reduction in the softening radius
(Sec.~\ref{sec:methods}) and the associated extra negative potential energy. Although unphysical, 
it does not affect our results.}

Let us compare the two runs at $t\approx50\da$, 
when the unbound mass is still significantly higher in Model~B than in Model~A (Fig.~\ref{fig:EnvMass}).
The purple lines in Fig.~\ref{fig:Env_E_terms_bnd} show that $|E_\mathrm{pot,env,bnd}|$ is smaller in Model~B at this time,
so the envelope has expanded more than in Model~A. 
This extra expansion was discussed in Sec.~\ref{sec:unbinding_spatial} 
and is likely caused by the extra energy supplied by recombination.
The total energy $E_\mathrm{env,bnd}$ is correspondingly higher (i.e.~less negative) in Model~B. 
And although Model~B incurs a slightly deeper inspiral up to this time as implied by the gap between the green lines, 
this deeper inspiral can only partially account for the wider energy gap between the orange lines.

However, the fact remains that the difference in the potential energy between the runs (purple) 
is larger than that of the total gas energy (orange) -- why?
The bulk kinetic energy, shown in red, is almost identical between the runs, 
but the thermal energy content is higher in the ideal gas run (blue),
so this explains why the gap between the orange curves is smaller than that between the purple curves.
Although recombination converts latent recombination energy into thermal energy,  
the extra heat from recombination is evidently used to expand the envelope, 
causing it to cool, in turn resulting in less overall thermal energy content compared to Model~A.

Next we compare the energy terms at the end of the simulations at $t\approx115\da$.
The total energy of the bound system of envelope and core particles is clearly higher in Model~B 
(see black lines in Fig.~\ref{fig:Env_E_terms_bnd}).
However, the energy of the gas is almost equal for the two runs (orange lines).
This is because the inspiral is deeper in Model~A, resulting in smaller (i.e.~more negative) 
orbital energy of the system of core particles (green lines).
The extra release of orbital energy in Model~A has evidently led to more expansion and cooling of the envelope than in Model~B.

The overall picture is that  energy released by recombination leads to extra expansion and cooling of the envelope. 
This reduces the drag force on the core particles, their inspiral, 
and their transfer of orbital energy from the cores to the envelope. 
\textit{This picture implies the existence of a self-regulating mechanism 
that limits the effectiveness of the recombination energy
for energizing and unbinding the envelope.}

\subsection{Spatially Integrated Recombination and Ionization}\label{sec:recombination}
In Sec.~\ref{sec:recombination_spatial} 
we saw that much recombination and ionization occur in Model~B as the envelope expands, 
particularly at late times.
To better understand these processes and their effect on unbinding,
we now explore the volume-averaged evolution of the recombination energy associated with the various ionic transitions,
where transitions are defined to include both single and double recombinations or ionizations.

\subsubsection{Helium}\label{sec:helium}
Fig.~\ref{fig:DeltaE_He} shows the cumulative energy release by the various ionic transitions of helium up to time $t$,
for all envelope gas (black) and bound envelope gas (blue).
Recombination is shown in the top row and ionization in the bottom row.
Positive values indicate net release of recombination energy 
and negative values indicate net gain of recombination energy due to ionization.
Fig.~\ref{fig:DeltaE_He} shows
that the $\mathrm{\HeIII\rightarrow \HeII}$ and $\mathrm{\HeIII\rightarrow \HeI}$ transitions are the most important,
with the former releasing the most recombination energy by the end of the simulation, 
and mostly from bound gas.
Net ionization of tracer gas during the simulation occurs over time, but mostly in unbound gas, 
and much less commonly than recombination.
\textit{Re}ionization of \textit{recombined} tracer gas \textit{is} however important, as discussed below.
For reference, 
Fig.~\ref{fig:DeltaM_He} in App.~\ref{sec:ionic} 
shows the mass of each ionic species for each of the three helium ion tracers,
and App.~\ref{sec:ionic} also provides details of how the energy release is calculated.

The top right panel of Fig.~\ref{fig:DeltaE_He} 
shows a peak in the energy released by $\mathrm{\HeIII\rightarrow \HeI}$ at $t\approx30\da$, and then a decline until $t\approx80\da$,
followed by a local peak at $t\approx100\da$.
At the first peak, about $40\%$ of the released energy comes from unbound  envelope gas, 
so much of the released energy is ``wasted'' in energizing already unbound gas.
The decline after the first peak is caused by reionization of $\HeI$ into $\HeII$ and, to a lesser extent, $\HeIII$.
This is evident from the right column of Fig.~\ref{fig:DeltaM_He}. 
In the middle panel of this column, 
we see that between $(30$--$80)\da$ the mass of bound $\HeII$ increases. 
Comparing the panels of this column, 
this increase is caused by both ionization of $\HeI$ and recombination of $\HeIII$.
Unbound $\HeI$ gas at $t\approx30\da$ is mostly  reionized to  $\HeIII$.
As discussed below, this helps to explain why Model~B does not unbind more mass than Model~A at late times.
Comparison with Fig.~\ref{fig:He_face-on_100-400} 
suggests that the reionized envelope gas is mainly in the outer envelope and energized by  the ambient medium,
whereas the recombining  bound gas is located deeper in the expanding envelope.
The middle panel of the bottom row of Fig.~\ref{fig:DeltaE_He} also shows
that much energy is used to ionize $\HeII$ tracer gas into $\HeIII$.
Ionization absorbs bulk kinetic, thermal and potential energy from the surroundings, hindering unbinding.
However, comparing the black and blue lines reveals that most of the ionization happens in already unbound gas.

The net recombination energy release for helium is shown in the middle panel of Fig.~\ref{fig:DeltaE}.
The peak occurs at $t\approx25\da$, when $\sim 2/3$ of released energy has been released by bound gas, 
which  can  be used to unbind the envelope.

\subsubsection{Hydrogen}\label{sec:hydrogen}
The net energy release from the recombination and ionization of hydrogen is shown in the left panel of Fig.~\ref{fig:DeltaE}.
Transitions between three ionic states ($\Hmol$, $\HI$ and $\HII$) are included in the calculation,
but $\Hmol$ does not play a significant role.
The energy released by hydrogen is small compared to that from helium,
despite the greater hydrogen mass fraction.
This is mainly because only a small fraction of the primarily $\HII$ envelope recombines into $\HI$.
Moreover, less than half of the hydrogen recombination energy release, 
at its peak at $t\approx35\da$, is from bound gas.
Thus, hydrogen less efficiently releases recombination energy than helium 
up to this time because the former releases most of this energy in already unbound material.
These results are qualitatively consistent with those of \citet{Lau+22a} and \citet{Lau+22b}, 
who studied  different binary systems from  ours.
Fig.~\ref{fig:DeltaE}, 
shows that most all of the $\HI$ formed from recombination of $\HII$  reionizes into $\HI$, 
likely due to the interaction between envelope and ambient gas. 

\subsubsection{Net effect from hydrogen and helium}\label{sec:hydrogen_helium}
Fig.~\ref{fig:DeltaE} shows the net recombination energy released by  hydrogen and helium.
The spatially integrated release continues up to $t\approx25\da$ in both total and bound gas.
Subsequently, $\HeII$, $\HeI$ and $\HI$ reonizations decrease the net energy released.
The net released recombination energy in the bound envelope gas plateaus higher than that of the total envelope gas 
at the end of the simulation,
as some of the unbound helium was ionized (Fig.~\ref{fig:DeltaE_He}, bottom row).

\begin{figure*}
\includegraphics[width=0.95\textwidth,clip=true,trim= 0 0 0 0]{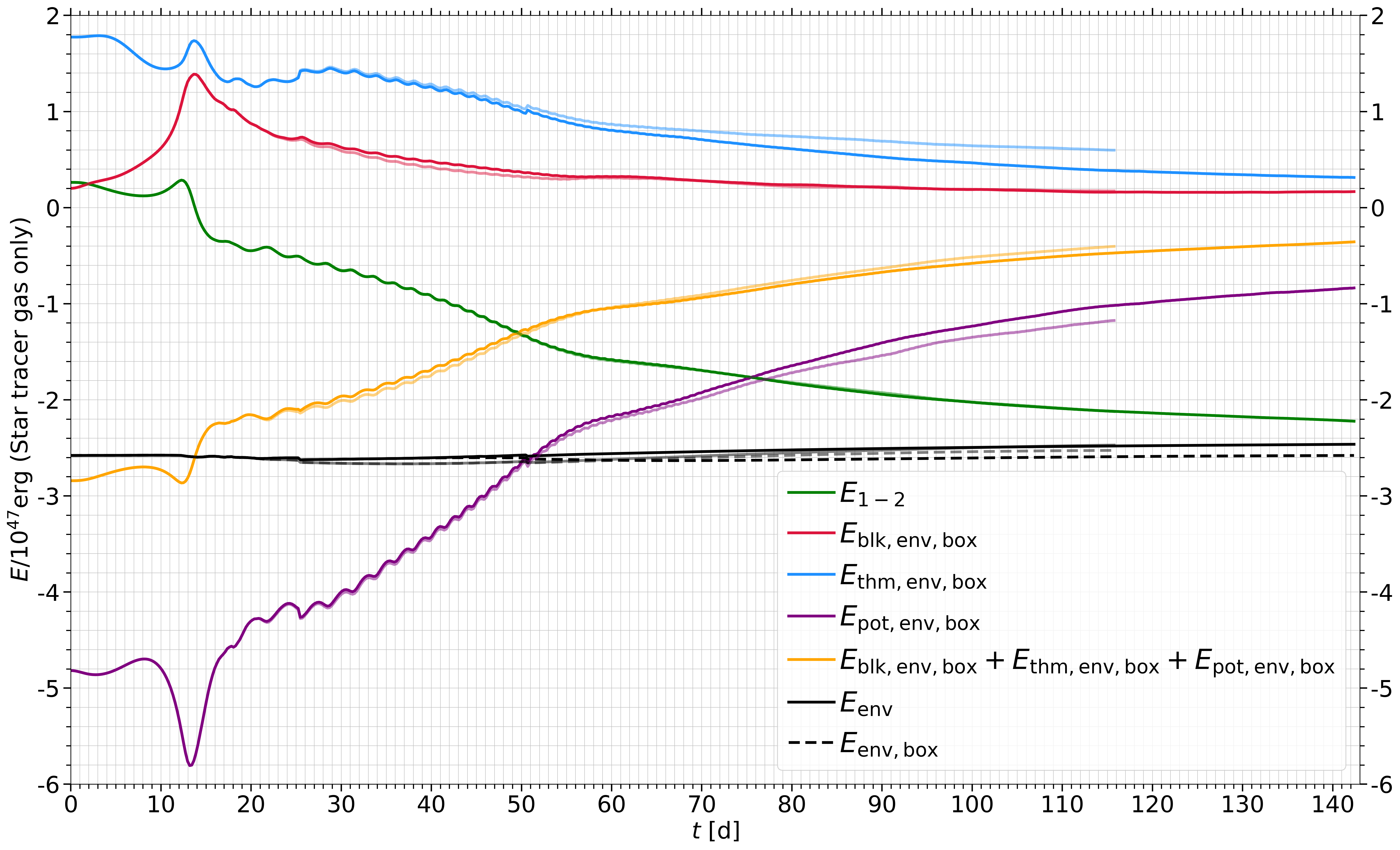}
\caption{Similar to Fig.~\ref{fig:Env_E_terms_bnd} except now comparing Model~A (ideal gas EOS, semi-transparent lines)
with Model~D (ideal gas EOS with reduced ambient, opaque lines), and including both bound and unbound envelope gas.
}            
\label{fig:Env_E_terms_ambient}
\end{figure*}

\subsubsection{Net release of recombination energy from all species}\label{sec:net_release}
Fig.~\ref{fig:DeltaE_bnd_compare} compares the recombination energy released by the envelope gas 
according to the above analyses (solid black and blue lines, as in Fig.~\ref{fig:DeltaE})
with the same quantities obtained in a different way (dashed black and blue).
To obtain the dashed black line,
the total recombination energy was computed by differencing the total internal and thermal energy of the envelope gas,
taking into account leftover radiation energy (Sec.~\ref{sec:eos}).
Thus, 
\begin{equation}
  \label{E_rec}
  E_\mathrm{rec,env} = E_\mathrm{int,env} -E_\mathrm{thm,env} -\delta E_\mathrm{rad,env},
\end{equation}
where $\delta E_\mathrm{rad,env}= \delta a_\mathrm{rad,env}T^4$ and $\delta a_\mathrm{rad,env}$ 
is the small difference in $a\rad$ of about $1.5\times10^{-4}a\rad$ discussed in Sec.~\ref{sec:eos}.
The thermal energy density is given by 
\begin{equation}
  \label{E_thm}
  \mathcal{E}\thm=\frac{3}{2}\frac{\rho \kB T}{\mu m_H}, 
\end{equation}
where the mean molecular mass $\mu$ is taken from the simulation, and calculated using the EOS tables.
The dashed black line in Fig.~\ref{fig:DeltaE_bnd_compare} shows the \textit{released} recombination energy, 
equal to negative the net change in $E_\mathrm{rec,env}$ at time $t$, denoted as $-\Delta E_\mathrm{rec,env}$.
The difference in the released orbital energy $\Delta E_{1-2}$ between Models~A and B 
is also plotted for comparison (green).
Details of the calculation are provided in App.~\ref{sec:ionic}.

The solid and dashed black curves in Fig.~\ref{fig:DeltaE_bnd_compare} nearly agree,
but with more recombination energy release obtained by the more direct method described just above.
Metals likely account for some of the difference, 
as they have ionization energy of about $2\times10^{45}\erg$.  
Most of this would remain locked up in ions, so this cannot account for the full excess.
``Leftover'' radiation energy (Sec.~\ref{sec:eos}) rises as gas is heated, 
and this energy must be taken from other forms of energy such as recombination energy.
This energy absorption, 
$-\int_V\delta a_\mathrm{rad,env}T^4$ with $\int_V$ denoting volume integration over the simulation domain,
is plotted as a dashed-dotted black line in Fig.~\ref{fig:DeltaE_bnd_compare}.
This extra energy component may also contributee to the discrepancy between solid and dashed black lines.
The shape of the black dashed line is sensitive to the value of $a\rad$ most appropriate for the tabulated EOS,
which is not precisely known.
In any case, the small discrepancy between solid and dashed lines does not impact  our main conclusions.

The quantity $-\Delta E_\mathrm{rec,env,bnd}$ is plotted as a blue dashed line in Fig.~\ref{fig:DeltaE_bnd_compare}.
Likewise $-\int_V\delta a_\mathrm{rad,env,bnd}T^4$ is plotted as a blue dashed-dotted line, 
but its contribution is negligible, which means that all of the leftover radiation energy is in unbound gas.
Since it does not affect bound gas, the leftover radiation energy is likely inconsequential for envelope unbinding.

\subsubsection{Comparison with orbital energy release}
Comparing the green line with the blue solid and dashed lines in Fig.~\ref{fig:DeltaE_bnd_compare} 
shows that the release of recombination energy into bound envelope gas is \textit{compensated} 
by extra orbital energy released in the ideal gas run, owing to deeper particle inspiral.
As argued in Secs.~\ref{sec:unbinding} and \ref{sec:energization}, 
this likely explains why recombination energy release in Model~B does not result in more unbound mass than in Model~A.

\begin{figure}
\includegraphics[scale=0.15,clip=true,trim= 290  100 238 200]{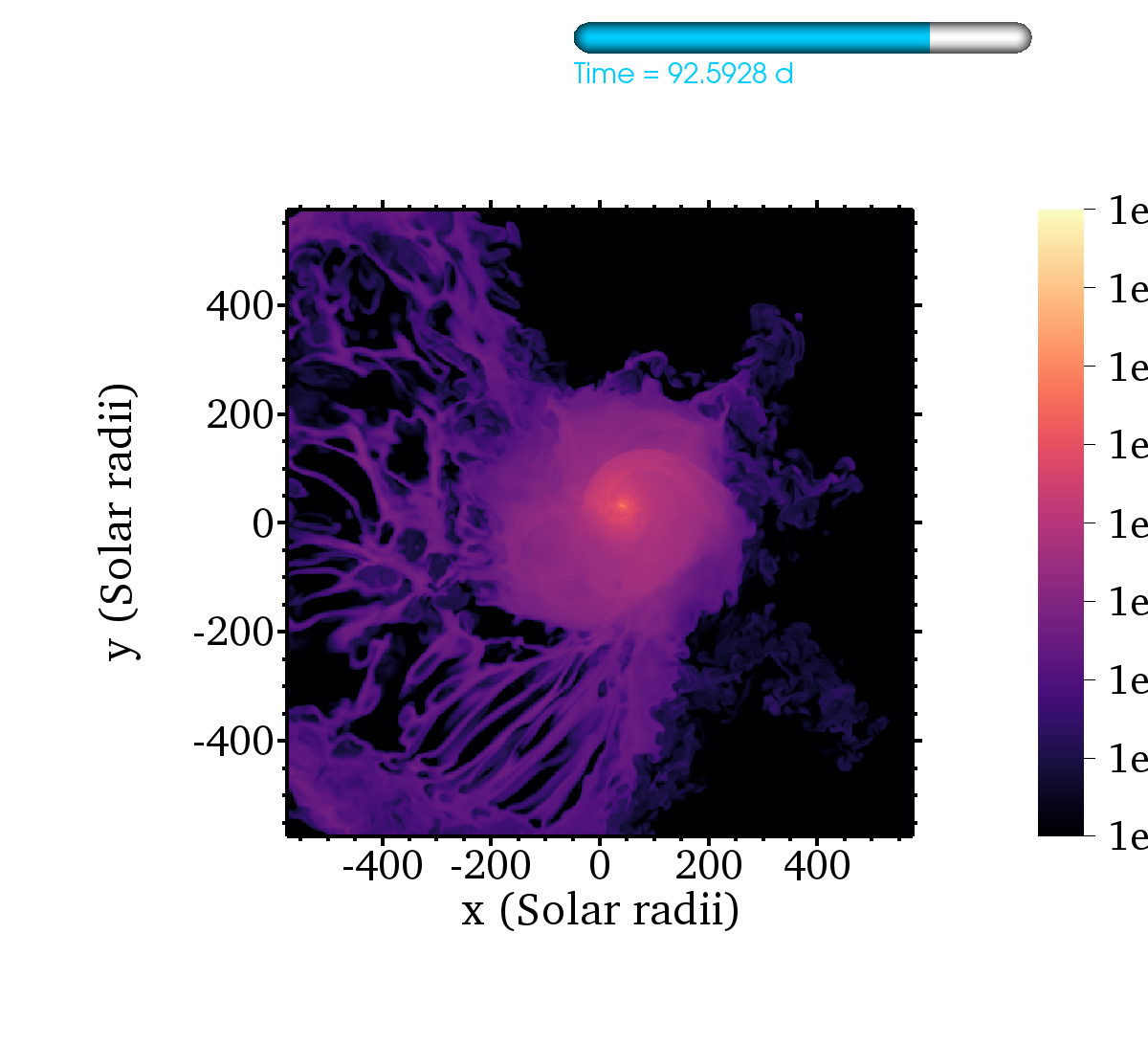}
\includegraphics[scale=0.15,clip=true,trim= 290  100   0 200]{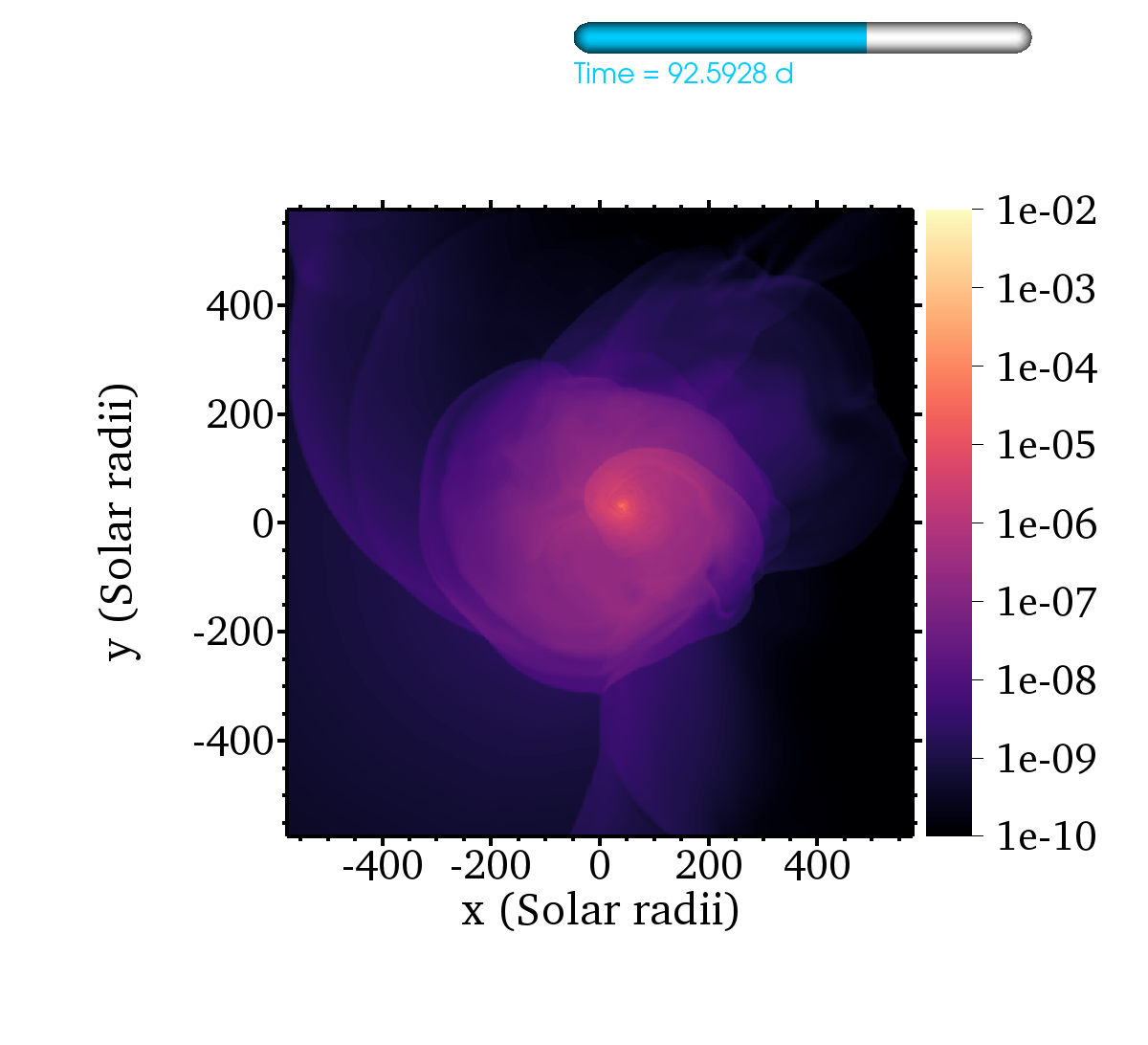}\\
\includegraphics[scale=0.15,clip=true,trim= 290  100 238 200]{E_bind_0400_282.png}
\includegraphics[scale=0.15,clip=true,trim= 290  100   0 200]{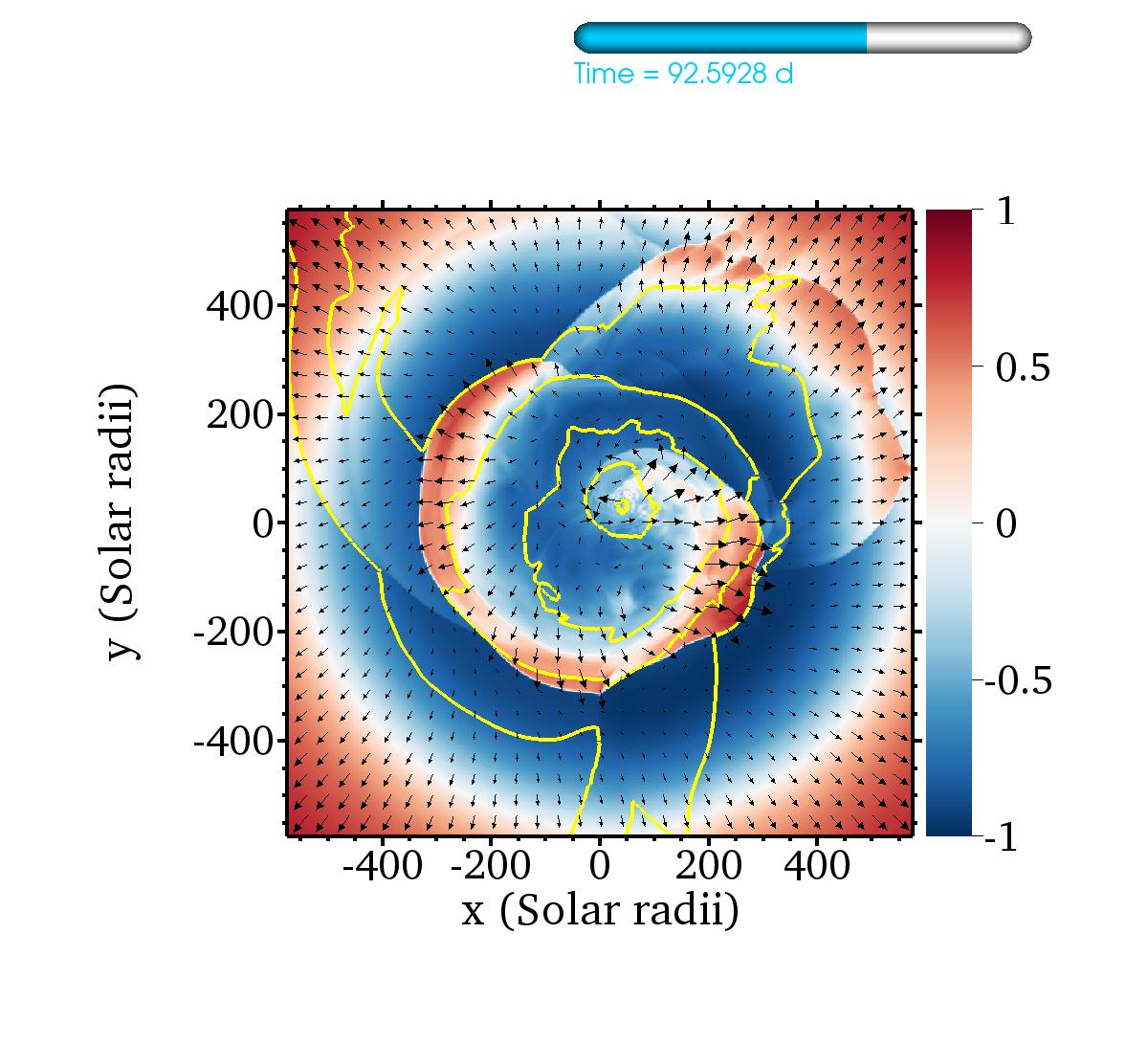}\\
\caption{Comparison of Model~A (left) and Model~D (right), 
which restarts from Model~A at $t\approx20\da$ but drastically reduces the ambient gas after a few days.
Panels show slices through the orbital plane.
Top row: Mass density in $\!\gcmcmcm$.
In Model~A, the gas is confined by the ambient, whereas in Model~D it expands freely.
Bottom row: $\widetilde{\mathcal{E}}\env$, defined in equation~\eqref{widetildeE}, 
 is a measure of binding energy density: red gas is unbound and blue gas is bound (see Sec.~\ref{sec:unbinding_spatial}).
In Model~A, there is mixing with the ambient medium (the ambient gas is uncoloured), 
whereas in Model~D the ambient medium is not present in the slice because it has been completely displaced by envelope gas.
}
\label{fig:ambient}
\end{figure}

\subsection{Role of the ambient medium}\label{sec:ambient_role}
To assess whether the ambient medium influences envelope unbinding, 
we compare Models~A and D, which are identical except for ambient gas removal in the latter. 
Fig.~\ref{fig:Env_E_terms_ambient} shows the evolution of various bound and unbound envelope gas energy components.
The subscript ``box'' means that only gas remaining in  the simulation domain is  counted.
The total energy of envelope gas including that lost through the simulation boundary is shown in black,
while that which remains is shown dashed.
The difference becomes significant after $t\approx50\da$ 
as only unbound gas (equation~\ref{unbound}) is lost.%
\footnote{We verified that similar plots for just  bound envelope gas 
or specific energy rather than total energy do not alter our conclusions.}

Between $t\approx25\da$, soon after the ambient reduction, and $t\approx40\da$, 
the bulk kinetic energy in Model~A reduces relative to that of Model~D.
This is caused by a reduction in the unbound component.
There is a corresponding deficit in the gas energy, shown in orange, and in the total energy, shown in black.
In Model~A, this is explained by a transfer of energy from the outgoing ejecta to the ambient medium \citep{Chamandy+19a}.

Between $t\approx40\da$ and $t\approx60\da$, 
the deficit in $E_\mathrm{blk,env,box}$ between the runs reduces to zero,
and the thermal energy becomes larger in Model~A as compared to Model~D.
From the middle and right columns of Fig.~\ref{fig:He_face-on_100-400}, 
it can be seen that mixing occurs between ambient and envelope gas during this time, 
likely from the Rayleigh-Taylor instability \citep{Chamandy+19a}.
This mixing accelerates and compresses envelope gas, 
raising thermal and bulk kinetic energy components  relative to Model~D.

After $t\approx60\da$, $E_\mathrm{blk,env,box}$ is about the same in the two runs,
but the envelope continues to gain thermal energy at the expense of potential energy in Model~A relative to Model~D.
In Model~D, ejected envelope gas can freely expand,
but in Model~A, the inertia and weight of the ambient stalls the expansion.
The expanding envelope collides with previously ejected overlying envelope, 
resulting in more shock heating, and limiting outward flow.
Density and $\widetilde{\mathcal{E}}\env$ snapshots at $t=92.6\da$ are presented in Fig.~\ref{fig:ambient},
with Model~A  and Model~D in the  left and right columns respectively.
The structure of the bound, inner envelope is  similar between the two runs,
but that of the outer envelope is very different.

\subsubsection{Effect of ambient on envelope unbinding and ionic state}\label{sec:ambient_unbinding}
The unbound mass evolution of Models~A and D can be compared using the red and yellow curves of Fig.~\ref{fig:EnvMass}. 
Beginning at $t\approx50\da$, the unbound mass curve of Model~A (solid red)  exceeds that of Model~D (solid yellow).
There is no corresponding surge in red relative to yellow in the dashed lines, 
which omit thermal energy in the definition of unbound. 
Therefore,  Model~A  unbinds more mass than Model~D due to extra heating of envelope gas.
This extra heating, discussed earlier,  is visible in Fig.~\ref{fig:Env_E_terms_ambient}.

In Fig.~\ref{fig:EnvMass}, the yellow solid and dashed curves of the reduced ambient run, Model~D, tell an interesting story.
Until $t\approx70\da$, these curves nearly overlap and are flat after the first periastron passage. 
Thus Model~D unbinding up to the first periastron passage at $t\approx13\da$ 
is driven by the bulk kinetic energy and thermal energy is unimportant.
Only when a dense ambient medium is present, as in Model~A, 
does some of this bulk kinetic energy convert to thermal energy, 
helping to unbind the envelope (compare red solid, yellow dashed and red dashed lines).
The solid yellow line  increases again at $t\approx70\da$, 
while the dashed yellow line stays flat and then  increases at $t\approx85\da$, 
though not  as fast as the solid line.
This indicates a second phase of unbinding begins at $t\approx70\da$ kickstarted 
by  thermalisation  of core particle orbital energy.
That the solid red and yellow lines have almost the same slope after $t\approx70\da$ 
shows that \textit{the rate of mass unbinding is roughly independent of the ambient density}, 
as long as the thermal energy is accounted for in the definition of unbound.

The additional thermal energy helps to unbind envelope gas in Model~A between 
$t\approx50\da$ and $t\approx70\da$ (red solid line in 
Fig.~\ref{fig:EnvMass}), 
but not in  Model~B (blue solid line), 
even though the ambient is equally dense and high-pressured with even more energy 
owing to the leftover radiation energy (see Sec.~\ref{sec:eos}).
Moreover, the envelope expands slightly faster at early times in Model~B owing 
to the release of recombination energy (Sec.~\ref{sec:spatial}),  
suggesting a stronger interaction between envelope and ambient gas.
Why does $M_\mathrm{env,unb}$ remain  flat during this time for Model~B (tabulated EOS), but increase for Model~A?
In Model~B, the extra energy transferred from ambient gas may be redirected into recombination energy
(i.e.~latent energy of reionized gas).
This is supported by the analysis of Sec.~\ref{sec:recombination}.
The left panel of Fig.~\ref{fig:DeltaE}, 
shows that all of the recombined hydrogen is reionized between $t\approx35\da$ and $t\approx85\da$.
Similarly, from the top-right panels of Fig.~\ref{fig:DeltaM_He} and Fig.~\ref{fig:DeltaE_He},
between $t\approx30\da$ and $t\approx80\da$, 
most of the $\HeI$ formed by  recombination of $\HeIII$ tracer gas reionizes into $\HeII$ or $\HeIII$.
But this reionization cannot account for all of the $\HeII$ formed, 
which means there is still recombination of $\HeIII$ into $\HeII$ taking place 
(compare Fig.~\ref{fig:DeltaM_He} middle-right panel).
But the release of this recombination energy happens almost exclusively in bound gas,  
likely  deep inside the envelope. 
In short, the  extra unbinding that happens between $t\approx50\da$ and $t\approx70\da$ in the ideal gas run (Model~A)
but not the tabulated EOS run (Model~B) likely occurs because the latter 
\textit{transfers ambient energy to reionize the gas rather than heat it}.
In Model~B, ambient energy is also likely being used to increase leftover radiation energy of unbound envelope gas,
Ambient energy is also likely being used to increase leftover radiation energy of unbound envelope gas,
as discussed in Sec.~\ref{sec:recombination}.

\subsection{Lack of Stalling and Extrapolation to Termination}\label{sec:extrapolation}
Despite Models~A, B and D having completed over $100$ orbits, the orbit-averaged separation steadily decreases,
as seen in the inset of Fig.~\ref{fig:separation} and the green lines in Figs.~\ref{fig:Env_E_terms_bnd} and \ref{fig:Env_E_terms_ambient}.
There is no evidence for ``stalling'' of the inspiral, as is often reported in the CE simulation literature.
From $t\approx70\da$ onward, 
the envelope unbinds at almost the same   rate of $\approx0.91\Msun\,\yr^{-1}$ 
in all three simulations
(slope of solid lines in Fig.~\ref{fig:EnvMass}).
Linearly extrapolating this constant rate, 
we estimate that the envelope would be ejected (i.e.~fully unbound) at $t\approx1.7\yr$ (where $t=0$ corresponds to our initial condition). 
This is about four times shorter than the estimate obtained using the same method for a similar 
simulation involving a more evolved (asymptotic giant branch; AGB) primary \citep{Chamandy+20}.
Interestingly, 
this ratio of $1/4$ happens to be equal to the ratio of the initial orbital periods of the RGB and AGB simulations.

Likewise, we can extrapolate the orbital inspiral down to the final separation $a\final$,
with the latter estimated using the CE energy formalism
\begin{equation}
  \Delta E\bind = \alpha\CE \frac{GM_2}{2}\left(\frac{M_\mathrm{1,c}}{a\final}-\frac{M_1}{a\init}\right),
\end{equation}
where the change in binding energy $\Delta E\bind$ equals
$GM_1(M_1-M_\mathrm{1,c})/\lambda R_1$, with $\lambda\approx1.3$ \citep{Chamandy+19a}. 
Let us focus on the longest run, Model~D.
From $t=115\da$ onward,  the nearly constant rate of core particle orbital energy loss 
is $\dot{E}_{1-2}\approx -3.9\times10^{44}\erg\,\da^{-1}$,
where $E_{1-2}\approx-GM_\mathrm{1,c}M_2/2a$.
Assuming that $\dot{E}_{1-2}$ remains constant, 
we can calculate the duration of the CE phase for different values of $\alpha\CE$.
The energy formalism yields for the final separation $a\final=0.3$, $0.8$, $1.5$ 
or $2.6\Rsun$ for $\alpha\CE=0.1$, $0.25$, $0.5$ or $1$, respectively \citep{Chamandy+19a}. 
We then estimate $12.5$, $4.6$, $2.0$ or $0.7\yr$ for the CE duration. 
Conversely, for a duration $t\approx1.7\yr$ as estimated above, we obtain $\alpha\CE\approx0.55$.
We can take the same approach using the orbital angular momentum of the particles 
in the particle centre-of-mass frame.
Approximating the orbit as circular,
$J_{z,1-2}\approx\sqrt{GM_\mathrm{1,c}M_2\mu_\mathrm{red} a}$, 
where $\mu_\mathrm{red}=M_\mathrm{1,c}M_2/(M_\mathrm{1,c}+M_2)$ is the reduced mass.
From $t=115\da$ onward, the torque $\dot{J}_{z,1-2}$ is approximately $-3.5\times10^{43}\g\,\cm^2\,\s^{-2}$.
Assuming that this rate of angular momentum transfer remains constant 
and that the CE terminates at $t\approx1.7\yr$, we obtain $\alpha\CE\approx0.27$.
Thus, extrapolations of energy and angular momentum both yield comparable values of $\alpha\CE<1$
which are broadly consistent with observation-based estimates \citep[e.g.][]{Scherbak+Fuller23}.
The same procedure can be applied to Models~A and B.
$\dot{E}_{1-2}$ and $\dot{J}_{z,1-2}$ are approximately constant from $100\da$ onward in Model~A 
and $92\da$ onward in Model~B.
Extrapolating $E_{1-2}$ yields $\alpha\CE=0.4$ for both runs, 
but extrapolating $J_{z,1-2}$ yields much smaller values.
This mismatch suggests that the estimates are less meaningful for Models~A and B than they are for Model~D.
Finally, we note that $\alpha\CE$ ultimately depends on the details of physical processes 
not yet included in numerical simulations, 
such as convective and radiative transport \citep{Wilson+Nordhaus19,Wilson+Nordhaus20,Wilson+Nordhaus22}.  

During the revision of this work, 
\citet{Valsan+23} reported an ideal gas simulation run using the moving mesh code \textsc{MANGA}
with parameters very similar to our Model~D.
However, their simulation ran for much longer, about $13\yr$.
They found that the envelope was fully ejected after about $4\yr$, 
and that $80\%$ of the envelope mass was ejected after $1.1\yr$.
The estimate of $\sim2\yr$ obtained from our naive extrapolation of the unbound mass
thus aligns quite well with their results.
However, \citet{Valsan+23} also reported that the separation 
of the central binary plateaus at about $5\Rsun$ after about $200\da$,
whereas in our Model~D the mean separation is $3.1\Rsun$ at the end of the simulation ($t=142\da$)
and is still decreasing.
This difference is unlikely to be caused by the slightly different initial conditions,
and suggests possible limitations in the various methods, which needs to be explored.
In Appendix~\ref{sec:comparison}, 
we provide comparisons between various models in the literature that use parameters similar to the ones used in the present work. 
In any case, extending Model~D up to envelope ejection would be interesting future work.

\section{Conclusions}\label{sec:conclusions}
We performed 3D hydrodynamic simulations involving a $2\Msun$ RGB primary 
and $1\Msun$ point particle companion
to study the role of ionization and recombination in envelope unbinding and orbital evolution.
We also studied the role of the ambient gas.
The runs were performed using the code AstroBEAR, 
and employed adaptive mesh refinement and spline softening for the point particles.
We compared four simulations, as summarized in Tab.~\ref{tab:runs},
to isolate the effects of ionization/recombination and ambient-envelope interaction.
Our main conclusions are as follows:
\begin{enumerate} 
  \item Released recombination energy somewhat helps to unbind the envelope, 
with a  $10$--$20$ per cent increase in unbound mass compared to the ideal gas run 
during the first half of the simulation. 
  \item However, release of recombination energy also expands the envelope, reducing the drag force, 
  and reducing the orbital energy release relative to the ideal gas case. 
Thus, we find a \textit{stabilizing feedback which limits the gain in unbound mass caused by the release of recombination energy}.
  \item Interaction between bound envelope gas, 
unbound ejecta and the ambient medium leads to transfer of ambient energy to the envelope. 
Some of this energy is used to reionize gas, 
which prevents it from being used to directly unbind the envelope. 
This is a \textit{second stabilizing feedback that limits the effectiveness of recombination energy 
for envelope unbinding when a dense circumstellar environment is present}.\label{item:ambient}
  \item By employing tracers in the code, we separately tracked gas \textit{initially} 
  dominated by each ionic species 
  (i.e.~$\HI$, $\HII$ or $\Hmol$ for hydrogen and $\HeI$, $\HeII$ or $\HeIII$ for helium).
Computing the ionization state using the Saha equation then 
enabled us to follow the net ionic transitions that had occured up until a given time.
Given that tracers are computationally inexpensive,  future work could, in addition, 
include tracers for metal species and even introduce new ionic tracers at every snapshot 
to better resolve the time evolution.
  \item We also tracked the various ionization transitions for bound and unbound gas separately, 
and found that at the time of peak cumulative release of recombination energy ($t\approx25\da$),
$\sim2/3$ was released by envelope gas that is still bound, 
suggesting that this recombination energy is used efficiently for unbinding.
  \item Recombination energy release was dominated by helium. 
Recombination energy release by hydrogen contributed at first, 
but hydrogen was later almost fully reionized.
Moreover, at the time of maximum recombination energy release of hydrogen, 
more than half of this release happened in already unbound gas, 
so was not used as efficiently as helium recombination energy for envelope unbinding.
Radiative losses, not included in our simulation, likely reduce this efficiency further.
  \item At late times ($>100$ orbits) we find a steady decrease in the period-averaged orbital separation, 
not a ``stalled'' inspiral as commonly reported in the literature. 
The approximately constant rates of mass unbinding, 
core particle orbital energy release and core particle orbital angular momentum release
at late times motivate a simple linear extrapolation to the termination of the CE phase.
This predicts a CE phase duration of about $2\yr$, and a best-guess $0.2\le \alpha\CE\le 0.6$.
\end{enumerate}

The second stabilizing effect summarized in item~(3) depends on the ambient gas.
The uniform ambient medium of density $10^{-9}\gcmcmcm$ in our Models~A, B and C 
was imposed for numerical expediency but also helps to place an upper bound on its influence.
In nature, the environment likely has material concentrated in the orbital plane,
formed from pre-dynamical mass loss during a Roche lobe overflow phase 
and  by the early part of the CE plunge phase which was not included in our simulations.
These earlier phases of mass loss have been studied in previous simulations,
where they result in torus-like mass distributions with densities 
of order $10^{-9}\gcmcmcm$ for systems similar to the one simulated in this work \citep{Macleod+18a,Macleod+18b,Reichardt+19}.
\citet{Macleod+18b},
simulated a case with secondary-to-primary mass ratio $q=0.3$, 
finding that the primary loses $\sim 10\%$ of its mass up to 
when the orbital separation equals the original primary radius,
and about half of this remains bound.
\citet{Macleod+Loeb20b} find that about $25\%$ of the \textit{secondary} mass is unbound during this time, 
roughly independent of $q$.
Taken together, this suggests that for the system we simulated ($q=0.5$), 
$\sim5$-$10\%$ of the original primary mass, or $\sim0.1$-$0.2\Msun$, 
could have been present in the circumstellar environment in a bound torus-like structure.

Indeed, observations  suggest that interaction between CE ejecta and pre-CE ejecta may be common.
\citet{Matsumoto+Metzger22}, citing \citet{Metzger+Pejcha17}, model the light curves of luminous red novae (LRN)
by considering both thermal emission and emission owing to recombination. 
However, they find that several LRNe are too bright to be explained by their model unless shocks
between CE ejecta and predynamical ejecta are invoked.
They  point out that such predynamical mass loss likely explains certain observations, 
including the slow early rise in the light curves of some LRNe.
Furthermore, similar shock-heating models have  been proposed for other transient events involving high-mass progenitors.
In many cases, narrow emission lines are observed, which may stem from the ejecta-circumstellar material interaction
leading to ionization and subsequent recombination \citep[e.g][]{Smith13, Smith+15, Pearson+23}.
Our simulations (Model~B) suggest that something analogous may happen during CE events 
involving low- to intermediate-mass progenitors.

Overall, our models suggest that the potency with which recombination 
assists envelope unbinding is blunted by two effects.
First, released recombination energy appears to expand the envelope, 
creating a negative feedback loop which opposes the release of orbital energy.
Secondly, reionization of recombined gas can act as an energy sink, 
preventing heating by the ambient environment that would otherwise assist envelope unbinding.
Future work should explore these self-regulation effects in more detail.

\section*{Acknowledgements}
We thank the referee for insightful comments on the manuscript that helped us to make significant improvements.
We thank Paul Ricker for early discussions about the role of recombination energy in common envelope evolution.
LC thanks the organizers of the Physics and Astrophysics of Common Envelopes meeting, 
held from May~30 to June~3, 2022, 
for setting the stage for fruitful discussion.
This work used the computational and visualization resources 
in the Center for Integrated Research Computing (CIRC) at the University of Rochester 
and the computational resources of the Texas Advanced Computing Center (TACC) at The University of Texas at Austin, 
provided through allocations TG-AST120060 and XRAC AST180039 
from the Extreme Science and Engineering Discovery Environment (XSEDE) \citep{xsede}, 
which are supported by National Science Foundation grant number ACI-1548562,
and the Advanced Cyberinfrastructure Coordination Ecosystem:
Services \& Support (ACCESS) program \citep{Access}, 
which is supported by National Science Foundation grants \#2138259, \#2138286, \#2138307, \#2137603, and \#2138296,
and through TACC Frontera Pathways allocation AST20034.
This work also used the computational resources of the Rosen Center for Advanced Computing (RCAC) at Purdue University
under XSEDE and ACCESS allocation TG-AST120060.
Financial support for this project was provided by the Department of Energy grants 
DE-SC0020432 and DE-SC0020434,
the National Science Foundation grants AST-1813298, AST-2009713 and AST-2319326, 
and the National Aeronautics and Space Administration grant 80NSSC20K0622.

\section*{Data Availability}
The data underlying this paper will be shared on reasonable request to the corresponding author.

\footnotesize{
\noindent
\bibliographystyle{mnras}
\bibliography{refs}
}

\appendix

\section{Comparison with the literature}\label{sec:comparison}
\begin{figure*}
\includegraphics[width=0.95\textwidth,clip=true,trim= 0 0 0 0]{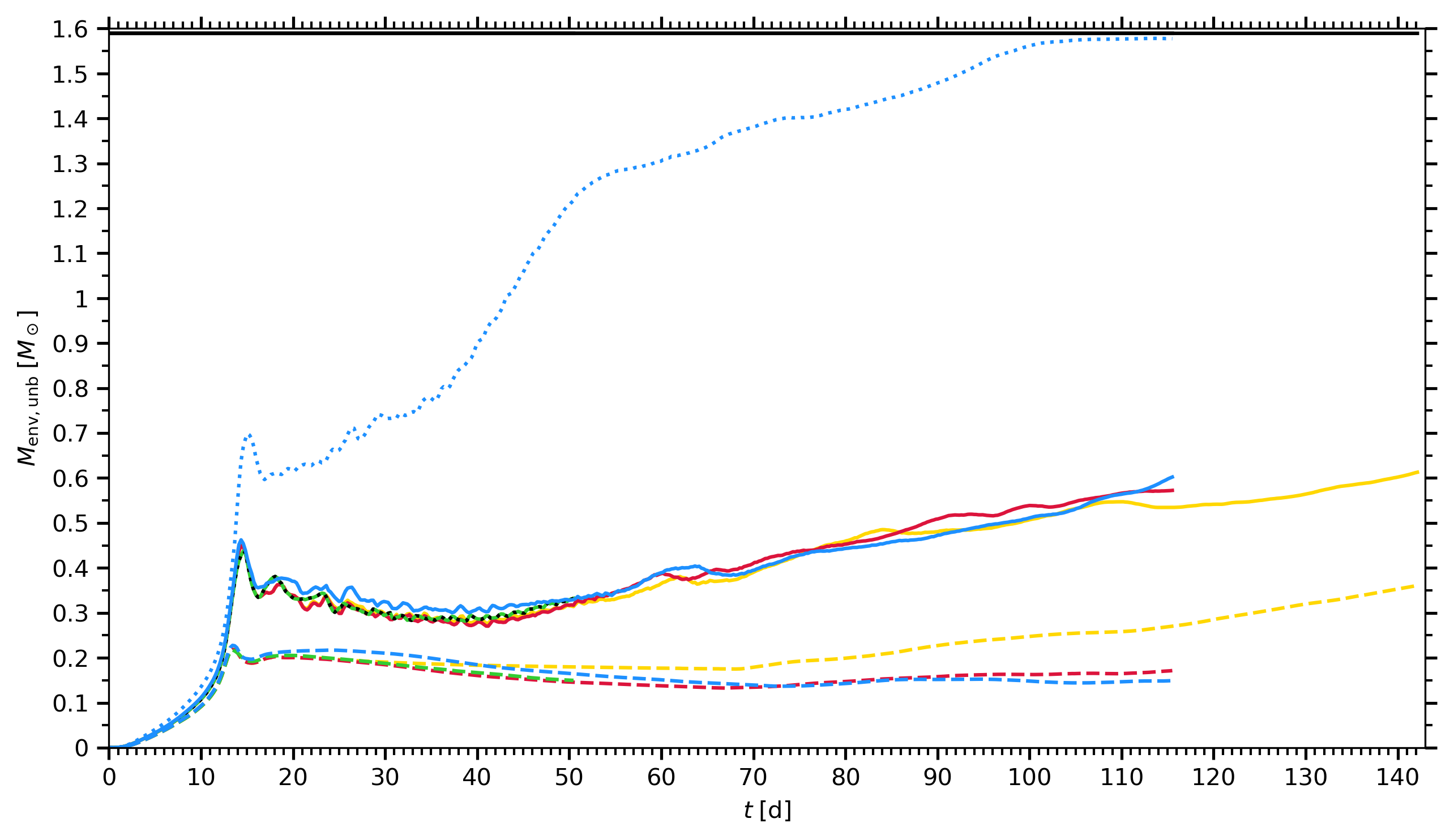}\\
\caption{Same as Fig.~\ref{fig:EnvMass} but omitting the factors of $2$ 
in the terms $\mathcal{E}_\mathrm{pot,env-1}$ and $\mathcal{E}_\mathrm{pot,env-2}$.
The horizontal black shows the total mass of envelope gas, 
accounting for the flux through the boundary.
}            
\label{fig:EnvMass_half}
\end{figure*}

Here we present a brief comparison with other works that presented simulations 
with parameter values very close (but not identical) to those used in the present work.
In Table~\ref{tab:comparison}, we show the core particle separation and unbound mass at $t=100\da$,
with or without the thermal energy term included in the definition of unbound.
To enable comparison, 
we show results for the definition of unbound that omits the factors of two in equation~\eqref{unbound}.
In Fig.~\ref{fig:EnvMass_half}, we plot the unbound mass as in Fig.~\ref{fig:EnvMass},
but now omitting factors of two in definition~\eqref{unbound}. 
Note that the definition of unbound mass still differs slightly in these different works
because of how the self-gravity potential energy contribution is calculated.

For further comparison between our model and that of \citet{Ohlmann+16a},
we refer the reader to appendix~A of \citet{Chamandy+19a}.

\begin{table*}
  \begin{center}
  \caption{Comparison with literature ($t=100\da$).
For this work, parameters are $M_1=1.96\Msun$, $M_2=0.978\Msun$, $M_\mathrm{1,c}=0.366\Msun$ and $R_1=48.1\Rsun$.
For \citet{Prust+Chang19} and \citet{Valsan+23}, 
parameters are $M_1=2\Msun$, $M_2=1\Msun$, $M_\mathrm{1,c}=0.379\Msun$ and $R_1=52\Rsun$.
For \citet{Ohlmann+16a} and \citet{Ohlmann16} parameters are $M_1=1.98\Msun$, $M_2=0.99\Msun$, 
$M_\mathrm{1,c}=0.38\Msun$ and $R_1=49\Rsun$.
For all models, the metallicity is $Z=0.02$ and MESA was used to prepare the initial RGB profile.
\citet{Ohlmann+16a} and \citet{Ohlmann16} used ambient density $\rho\amb=10^{-16}\gcmcmcm$, 
making their simulations more comparable to our Model~D,
whereas the ambient parameters of \citet{Prust+Chang19} were 
$\rho\amb=10^{-13}\gcmcmcm$ and $T\amb=10^5\K$ (L.~Prust, private communication).
Here $\Omega\init$ is the initial solid-body rotational angular velocity of the primary, 
relative to the orbital angular velocity,
and $a\init$ is the initial orbital separation. 
Other quantities, namely orbital separation $a$ and unbound mass fraction $m\unb/M\env$
with and without thermal energy included in the definition of unbound, 
are quoted at the time $t=100\da$.
Blank entries mean that the values were not given.
For the sake of comparison with other works, 
when determining the unbound mass in this table, 
we omit the factor of two in the definition~\eqref{unbound} for unbound gas (see Fig.~\ref{fig:EnvMass_half}).
          \label{tab:comparison}
          }
  \begin{tabular}{@{}cccccccc@{}}
    \hline
    Publication			&Model		&EOS		&$\Omega\init$	&$a\init\,[\!\Rsun]$&$a\,[\!\Rsun]$	&$m\unb/M\env$	&$m\unb/M\env$ (no thermal)			\\
    \hline                                                                                                              
    This work			&A		&$\gamma=5/3$	&$0$		&$49$		&$3.4$			&$0.34$		&$0.10$	\\
    				&B		&MESA, no rad.	&$0$		&$49$		&$3.5$			&$0.32$		&$0.09$	\\
    				&D		&$\gamma=5/3$	&$0$		&$49$		&$3.4$			&$0.32$		&$0.16$ \\
    \rule{0pt}{4ex}
    \citet{Prust+Chang19}	&Rotation	&MESA		&$0.95$		&$52$		&$4.3$			&$0.36$		&$0.03$	\\
    				&No rotation	&MESA		&$0$		&$52$		&$3.8$			&$0.30$		&$0.03$	\\
    \rule{0pt}{4ex}
    \citet{Valsan+23}           &Long-time-scale&$\gamma=5/3$ 	&$0.95$		&$52$		&$6.8$			&---		&$0.10$ \\
    \rule{0pt}{4ex}
    \citet{Ohlmann+16a}		&Ideal gas	&$\gamma=5/3$	&$0.95$		&$49$		&$4.6$			&---		&---	\\
    \rule{0pt}{4ex}
    \citet{Ohlmann16}		&Ideal gas	&$\gamma=5/3$	&$0.95$		&$49$		&$4.4$			&$0.06$		&$0.05$	\\
    				&OPAL		&OPAL		&$0.95$		&$49$		&$4.2$			&---		&$0.09$	\\
    				&No rotation	&$\gamma=5/3$	&$0$		&$49$		&$4.4$			&---		&---	\\
    \hline                  	 
  \end{tabular}
  \end{center}
\end{table*}

\section{Radial profiles of the 1D model used for the initial conditions}
\label{sec:MESA}
\begin{figure*}
  \includegraphics[width=0.6\textwidth,clip=true,trim= 30 0 0 0]{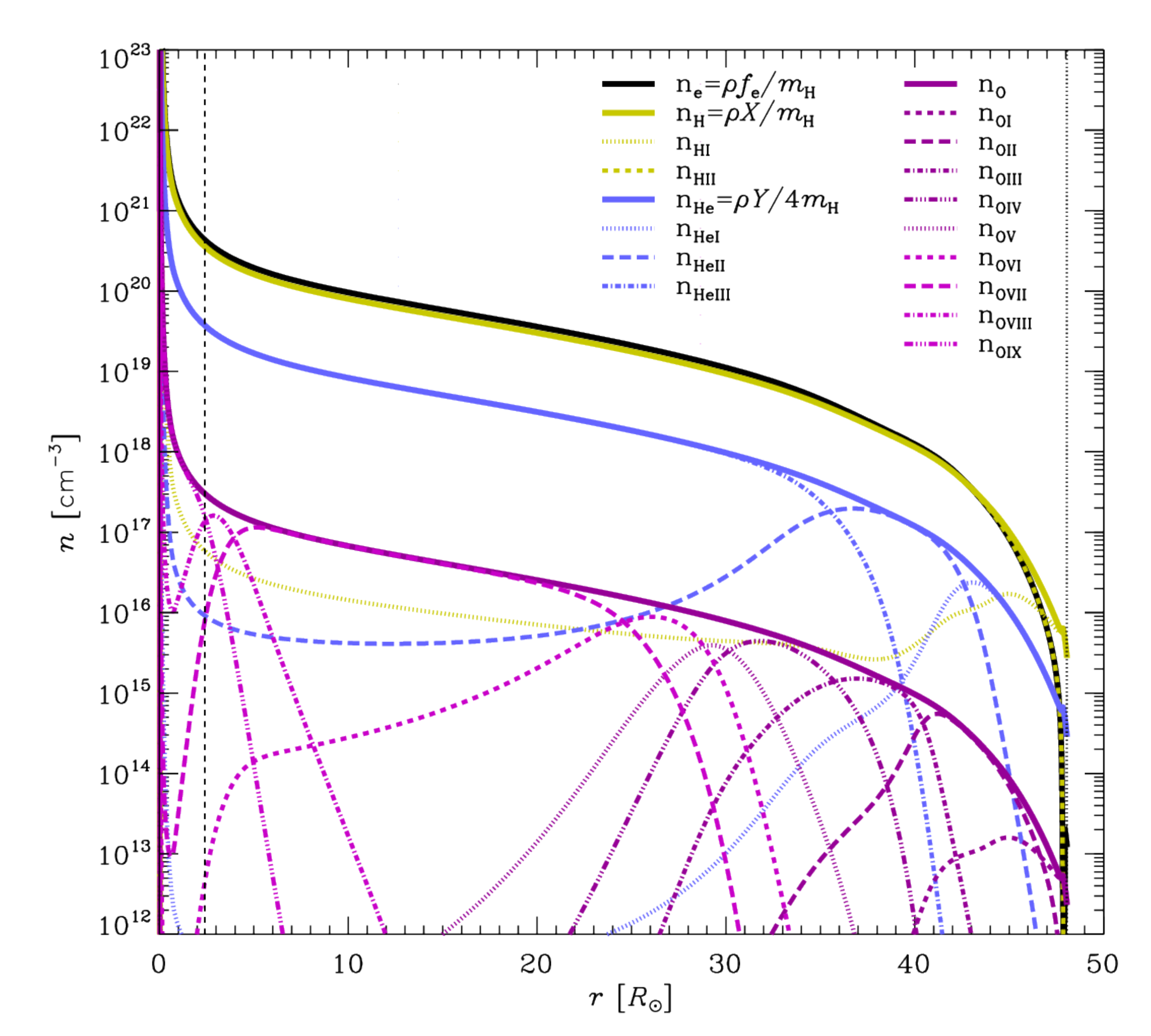}\\
  \caption{Number densities of various ion species in the original MESA model.
Mass fractions in the envelope are: 
$f\mass(^1\mathrm{H})=0.69$, 
$f\mass(^3\mathrm{He})=3\times10^{-4}$, 
$f\mass(^4\mathrm{He})=0.29$, 
$f\mass(^{12}\mathrm{C})=2\times10^{-3}$, 
$f\mass(^{14}\mathrm{N})=2\times10^{-3}$, 
$f\mass(^{16}\mathrm{O})=9\times10^{-3}$, 
$f\mass(^{20}\mathrm{Ne})=2\times10^{-3}$
and $f\mass(^{24}\mathrm{Mg})=4\times10^{-3}$. 
          }            
\label{fig:ne_r12778_O}
\end{figure*}

\begin{figure*}
  \includegraphics[width=0.9\textwidth,clip=true,trim= 30 0 0 0]{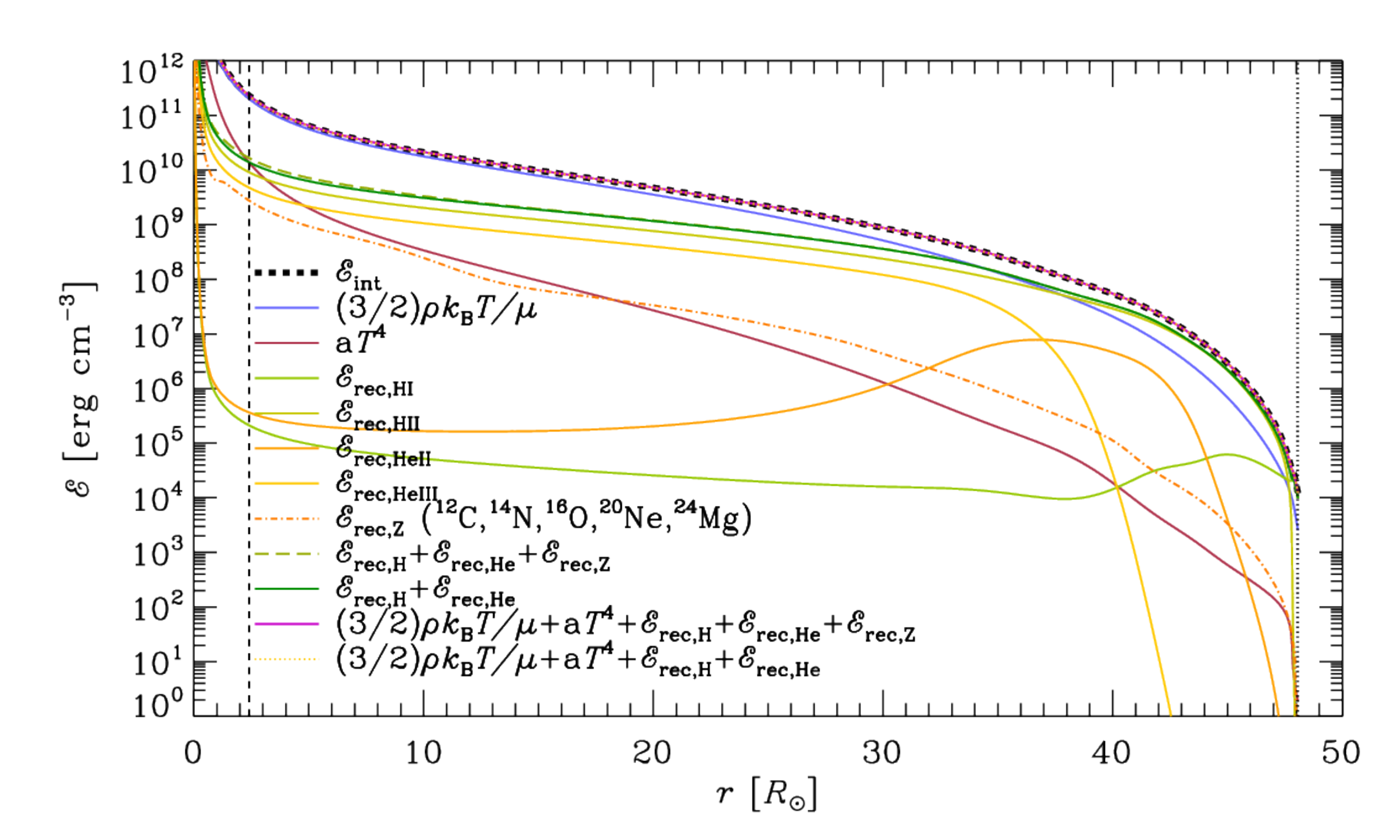}\\
  \caption{Internal energy density profiles for the initial primary star, computed using MESA. 
The internal energy profile (thick dashed black) is successfully reconstructed by summing 
the contributions from the thermal, radiation and recombination energy components. 
This is shown by the magenta line (second from bottom in the legend)
which precisely overlaps the black thick dashed line.
The dotted yellow line (bottom in legend) also precisely overlaps these lines,
showing that the contribution from the recombination energy due to metals is very small.
The vertical dashed and dotted lines show the softening radius $r\soft=2.41\Rsun$ and outer radius $R_1=48.1\Rsun$.
\label{fig:en_r12778}
          }            
\end{figure*}

In Fig.~\ref{fig:ne_r12778_O}, 
we show the number densities of the ionic species of the most abundant elements in the spherically symmetric model used for our initial condition.
In Fig.~\ref{fig:en_r12778} we plot the internal energy density profile for this MESA snapshot, 
as well as its various contributions. 
Recombination energy densities are calculated using the Saha equation.

The internal, thermal and radiation energy densities, 
respectively $\mathcal{E}_\mathrm{int}$, $\mathcal{E}_\mathrm{thm}$, $\mathcal{E}_\mathrm{rad}$,
are plotted directly from MESA output,
whereas the contributions from recombination energy alone are calculated by solving the Saha equation.
The sum of the thermal, radiation and recombination energies is plotted, 
The contributions from recombination energy alone are calculated by solving the Saha equation.
Finally, the sum of the thermal, radiation and recombination energies is plotted, 
both including and excluding the contribution from the recombination energy of metals, which is very small.
As expected, these curves overlap closely with the curve for $\mathcal{E}_\mathrm{int}$.
This shows that the internal energy density profile can be reconstructed using these energy terms, computed from first principles.%
\footnote{Numerical values of the partition function at $T=10^4\K$ from the Vizier database \citep{Ochsenbein+00},
were used whenever available. 
These differ from the low temperature limits for certain metal ions and molecular hydrogen,
but using the low temperature limits instead does not make any significant difference for this work.}

\section{Ionic transitions}
\label{sec:ionic}
\subsection{Tracers}
First, a few comments are in order about tracers.
As the tracer density is equal to the total gas density where the tracer is located,
we choose to place tracers of a given ionization state at locations where that ionization state
is higher than others of the same element.
Because the density of a given species depends exponentially on temperature,
transitions in the dominant species are rather sharp at $t=0$, 
so this neglect of partial ionization in the initial state is not an important limitation.
Because the mass fractions of all elements are constant within the envelope, as determined from MESA,
one can simply multiply by the hydrogen mass fraction ($0.69$) or helium mass fraction ($0.29$) 
to obtain the approximate density of the tracer (for example, gas that was originally $\mathrm{\HeII}$).
Also, because the pressure very close to the stellar surface is replaced by the (larger) ambient pressure,
but the density remains equal to that of the MESA profile,
the temperature at that location is higher than in the MESA snapshot.
This causes gas at the surface to be ionized from $t=0$, 
but tracers are set according to densities of the various ionic species in the MESA snapshot.
This is accounted for in the analysis.
Note that ionic tracers are defined such that only envelope gas is included, i.e.~ambient gas is excluded.

\subsection{Masses}
Fig.~\ref{fig:DeltaM_He} shows the amount of mass involved in the various ionic transitions of helium during the simulation.
The left, middle and right columns show the ionic transitions of $\mathrm{\HeI}$, 
$\mathrm{\HeII}$ and $\mathrm{\HeIII}$ tracer gas, respectively.
Recall that, for example, 
the $\mathrm{\HeI}$ tracer follows envelope gas for which $\mathrm{\HeI}$ was the most common ionization state 
in the 1D MESA snapshot of the RGB star.
From top to bottom, the rows indicate the mass of helium gas in a given tracer that is 
respectively $\mathrm{\HeI}$, $\mathrm{\HeII}$, or $\mathrm{\HeIII}$, at time $t$.
First, note that the $\mathrm{\HeIII}$ tracer (right column) 
contains more than $10$ times as much mass as the $\mathrm{\HeII}$ tracer,
which, in turn, contains more than $10$ times as much mass as the $\mathrm{\HeI}$ tracer.
Note also that the sum of the masses of all ionic states of a given tracer 
is equal to the tracer mass in the simulation domain at time $t$ (dashed green).
Finally, note that the same quantities are plotted for both the total mass (black) 
and only that part of it which is bound (blue),
where by ``bound'' we mean that it does not satisfy the local condition \eqref{unbound} 
to qualify as unbound gas at time $t$.
That part which is unbound is the difference between black and blue, which is not plotted to avoid clutter.
Comparing the blue and black curves in a given panel 
gives an \textit{indication} of what fraction of the recombination or ionization
has occurred in material that was bound at the time of said recombination or ionization.
We have no way of verifying whether gas that was bound (unbound) then is still bound (unbound) now,
so the correspondence is imperfect.
Nevertheless, given that only a small fraction of the envelope mass is unbound during the simulation
(Fig.~\ref{fig:EnvMass},
solid lines) the difference is likely to be small.

\subsection{Energetics}
\begin{figure*}
\includegraphics[height=0.25\textwidth,clip=true,trim= 0 0 0 0]{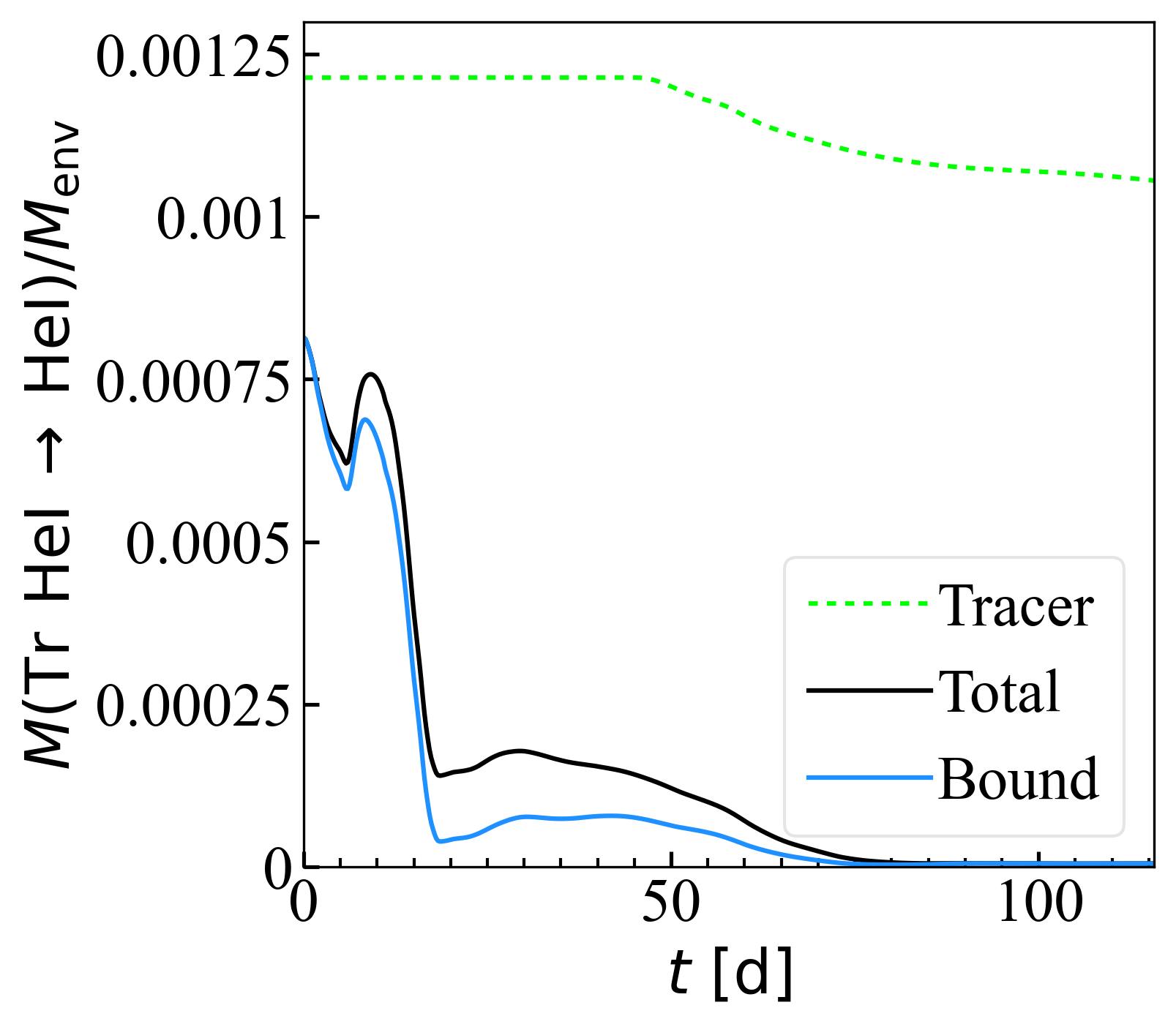}
\includegraphics[height=0.25\textwidth,clip=true,trim= 0 0 0 0]{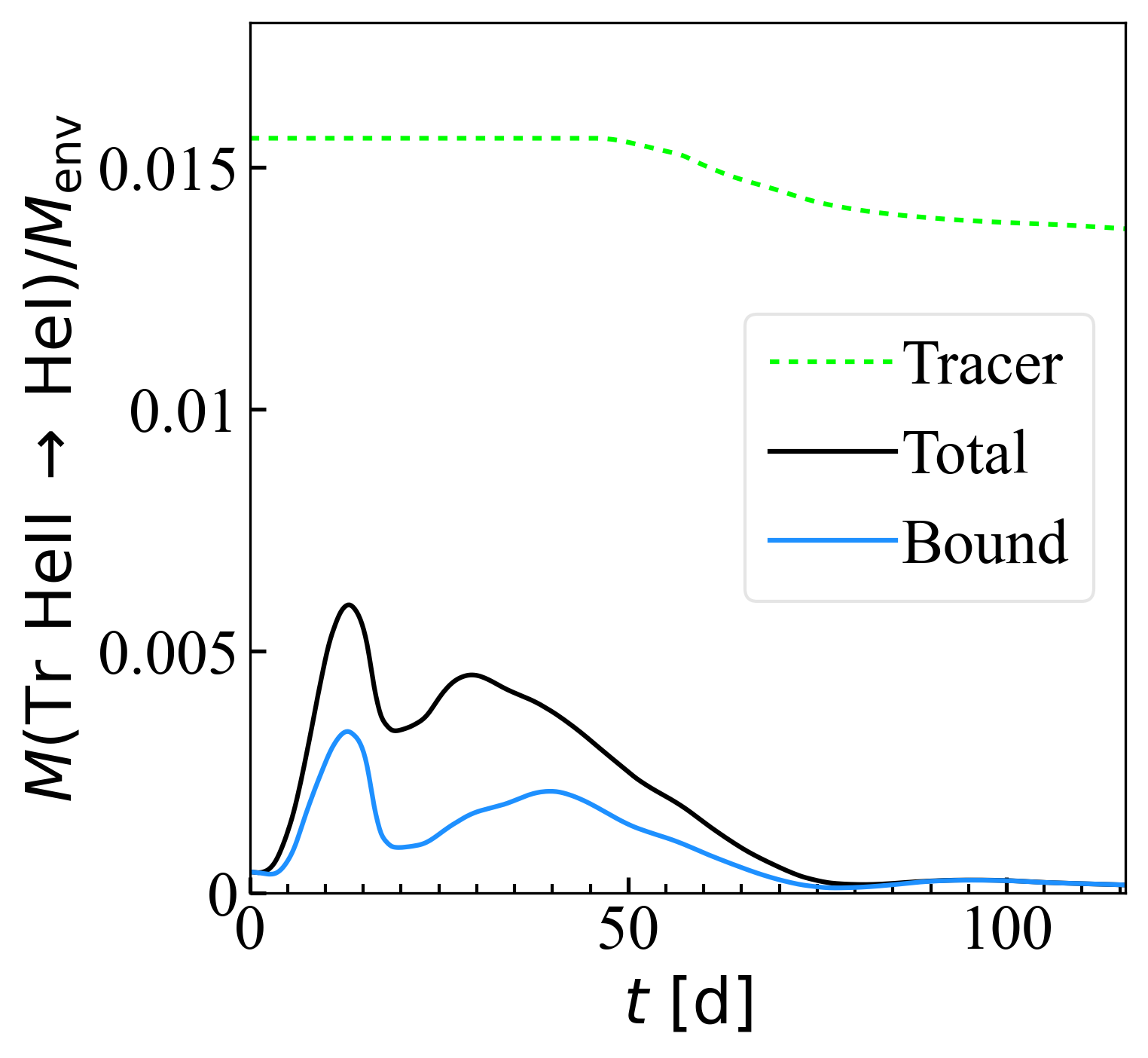}
\includegraphics[height=0.25\textwidth,clip=true,trim= 0 0 0 0]{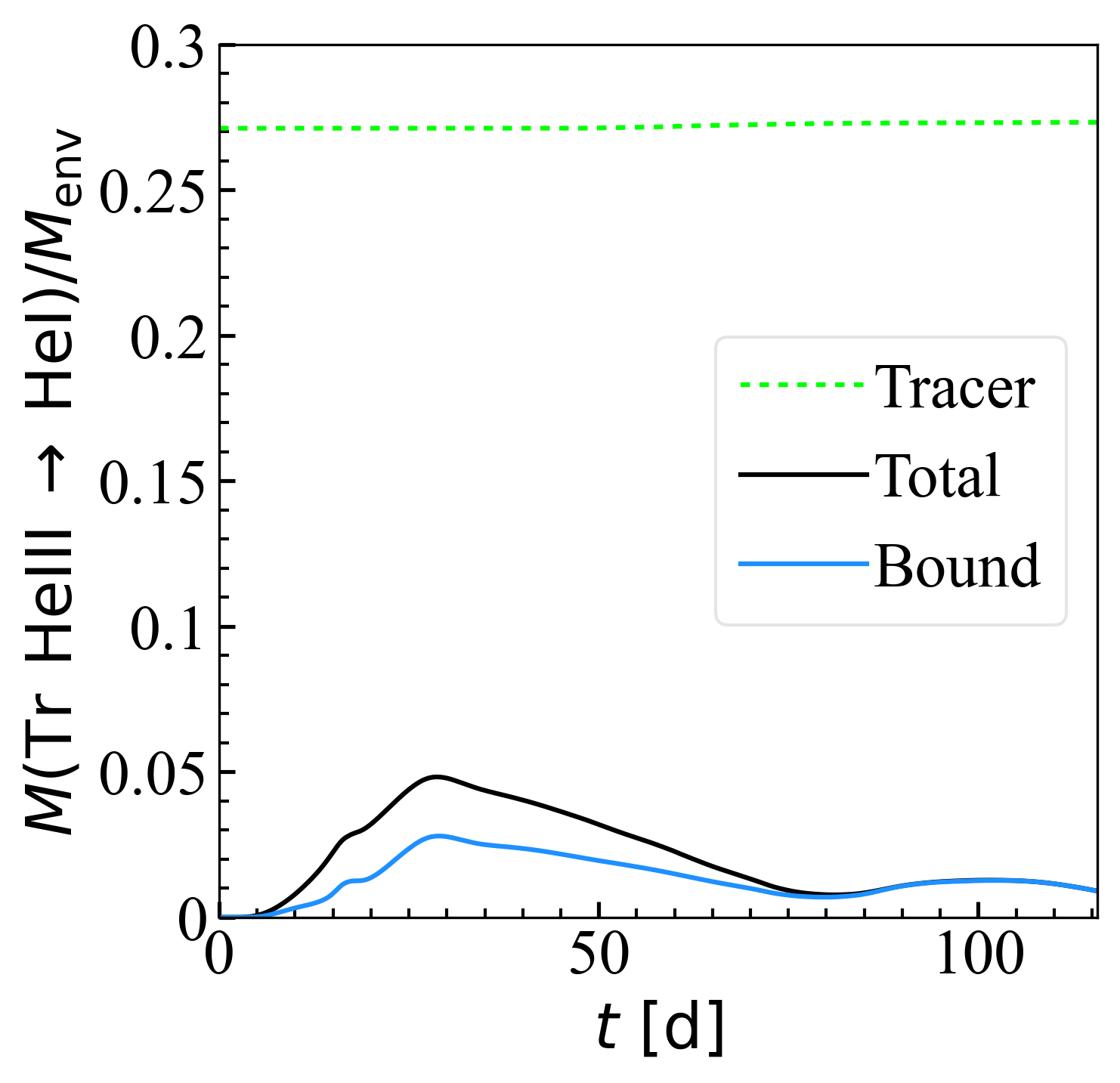}\\
\includegraphics[height=0.25\textwidth,clip=true,trim= 0 0 0 0]{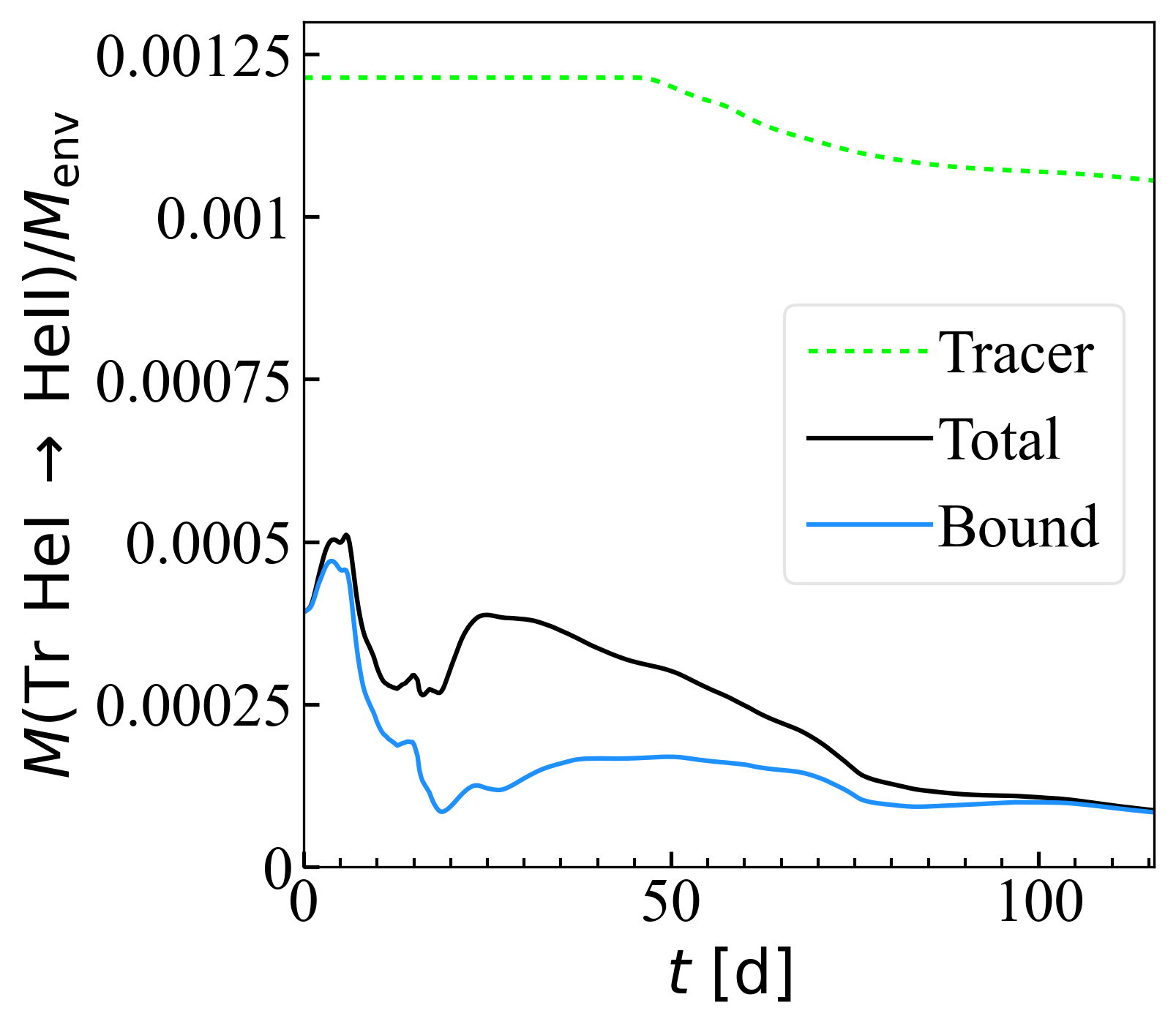}
\includegraphics[height=0.25\textwidth,clip=true,trim= 0 0 0 0]{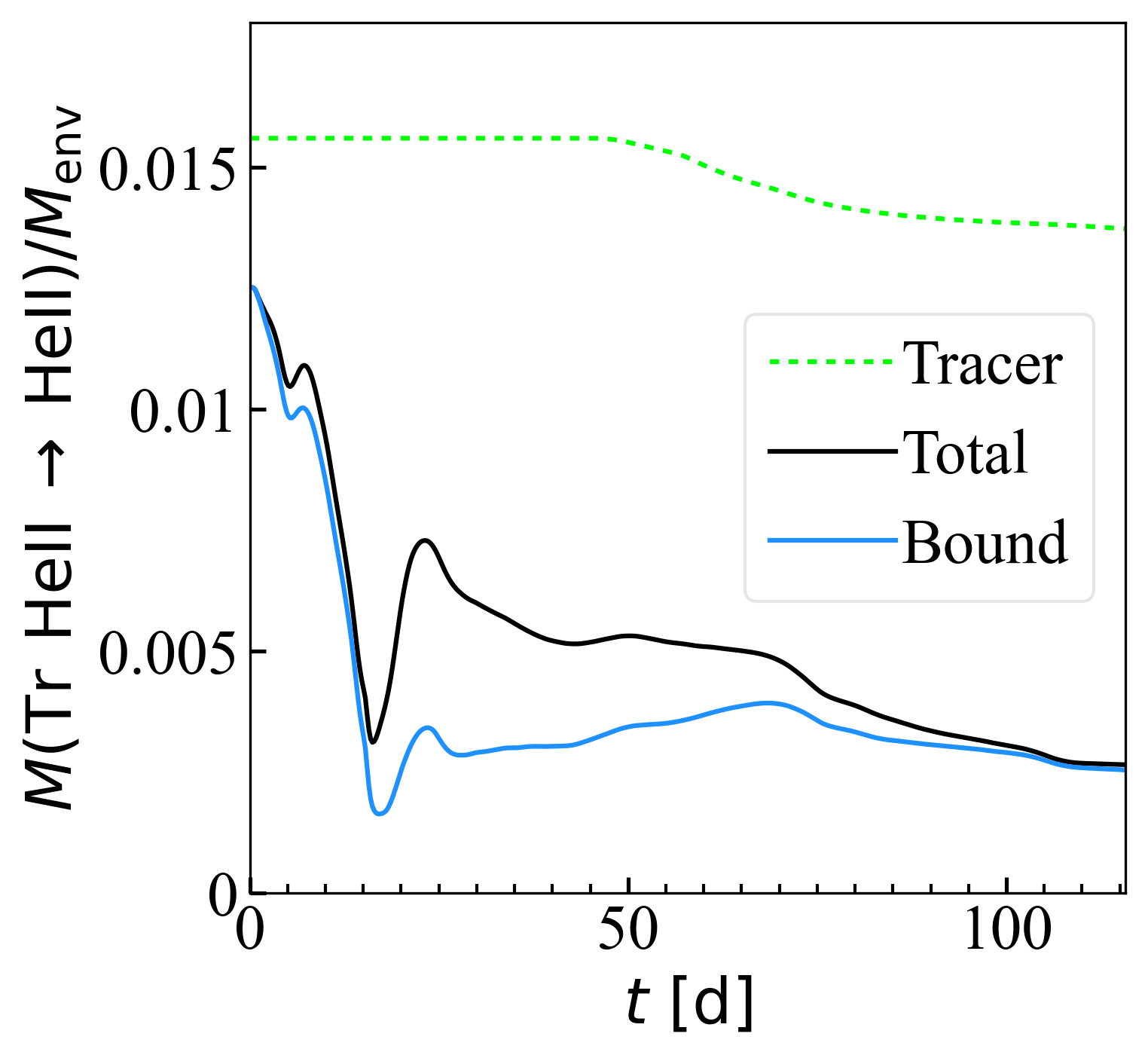}
\includegraphics[height=0.25\textwidth,clip=true,trim= 0 0 0 0]{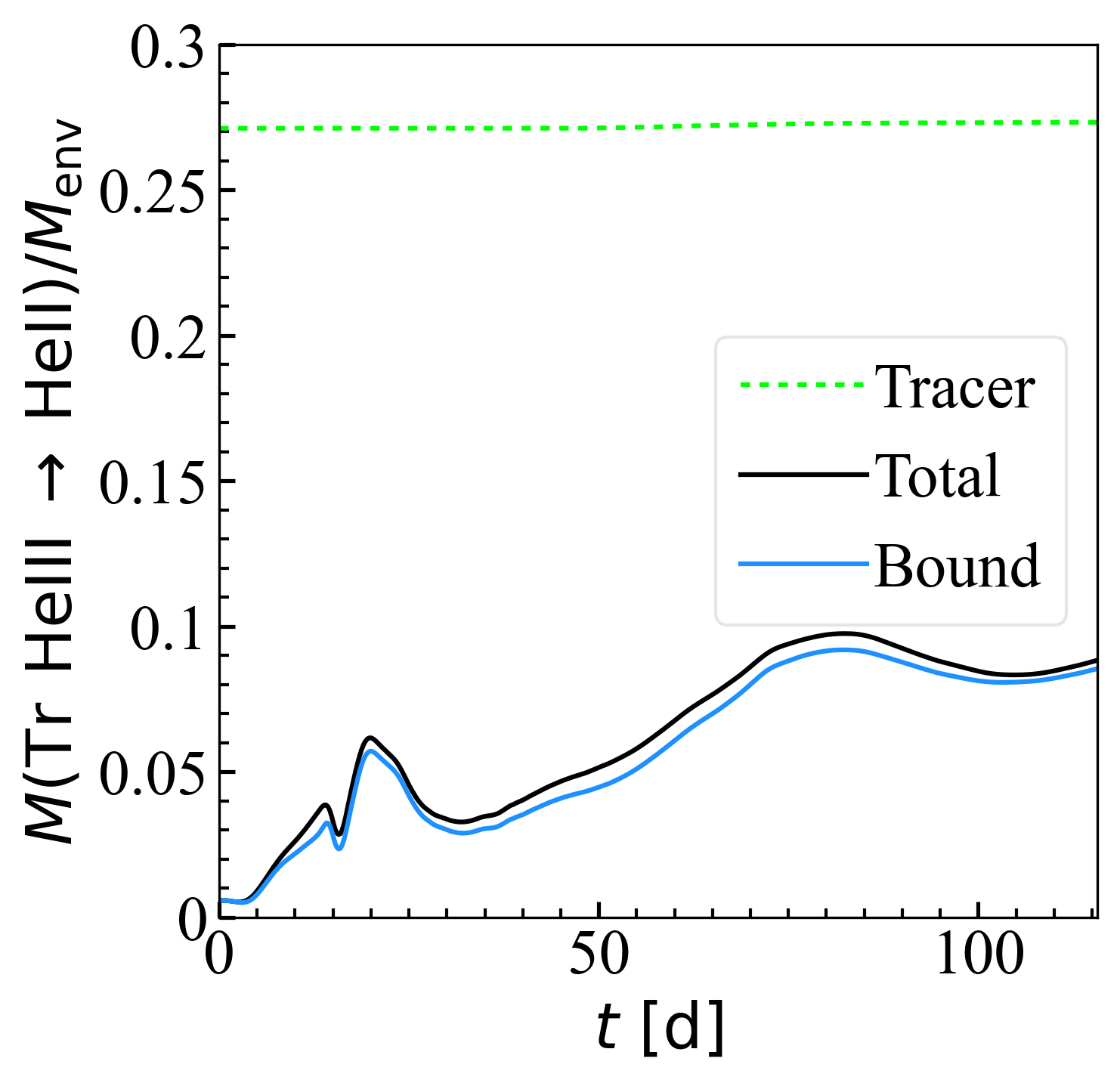}\\
\includegraphics[height=0.25\textwidth,clip=true,trim= 0 0 0 0]{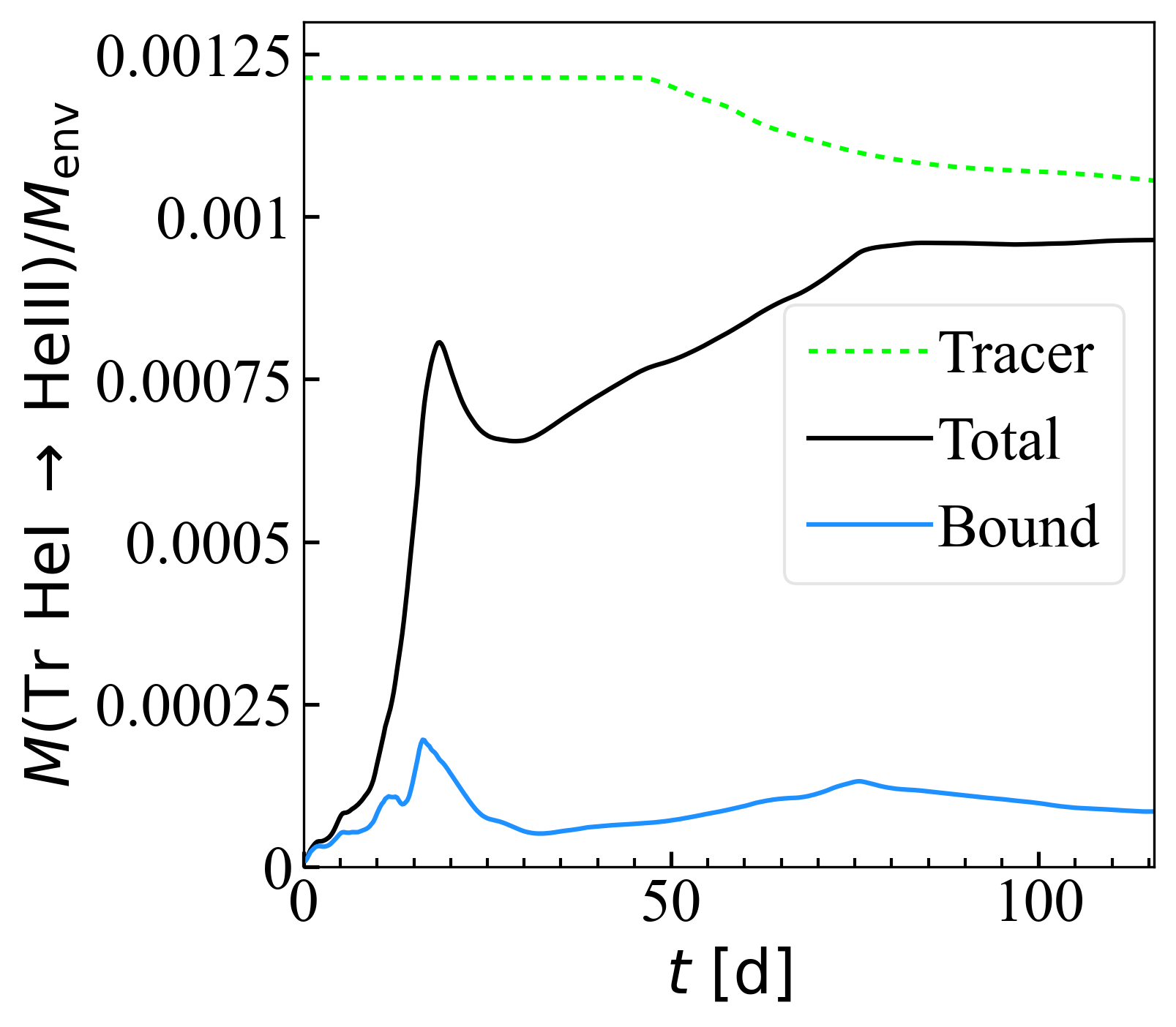}
\includegraphics[height=0.25\textwidth,clip=true,trim= 0 0 0 0]{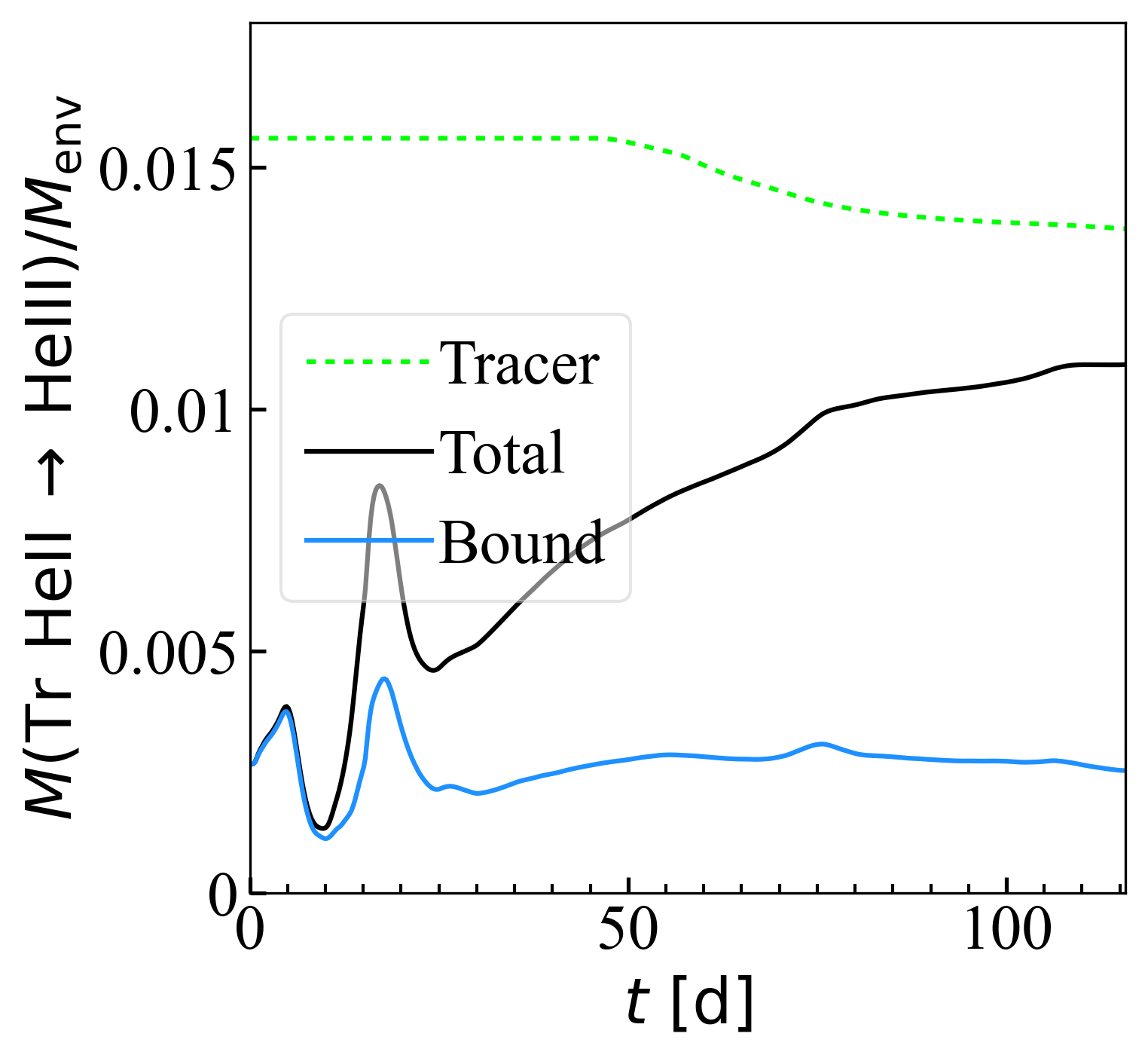}
\includegraphics[height=0.25\textwidth,clip=true,trim= 0 0 0 0]{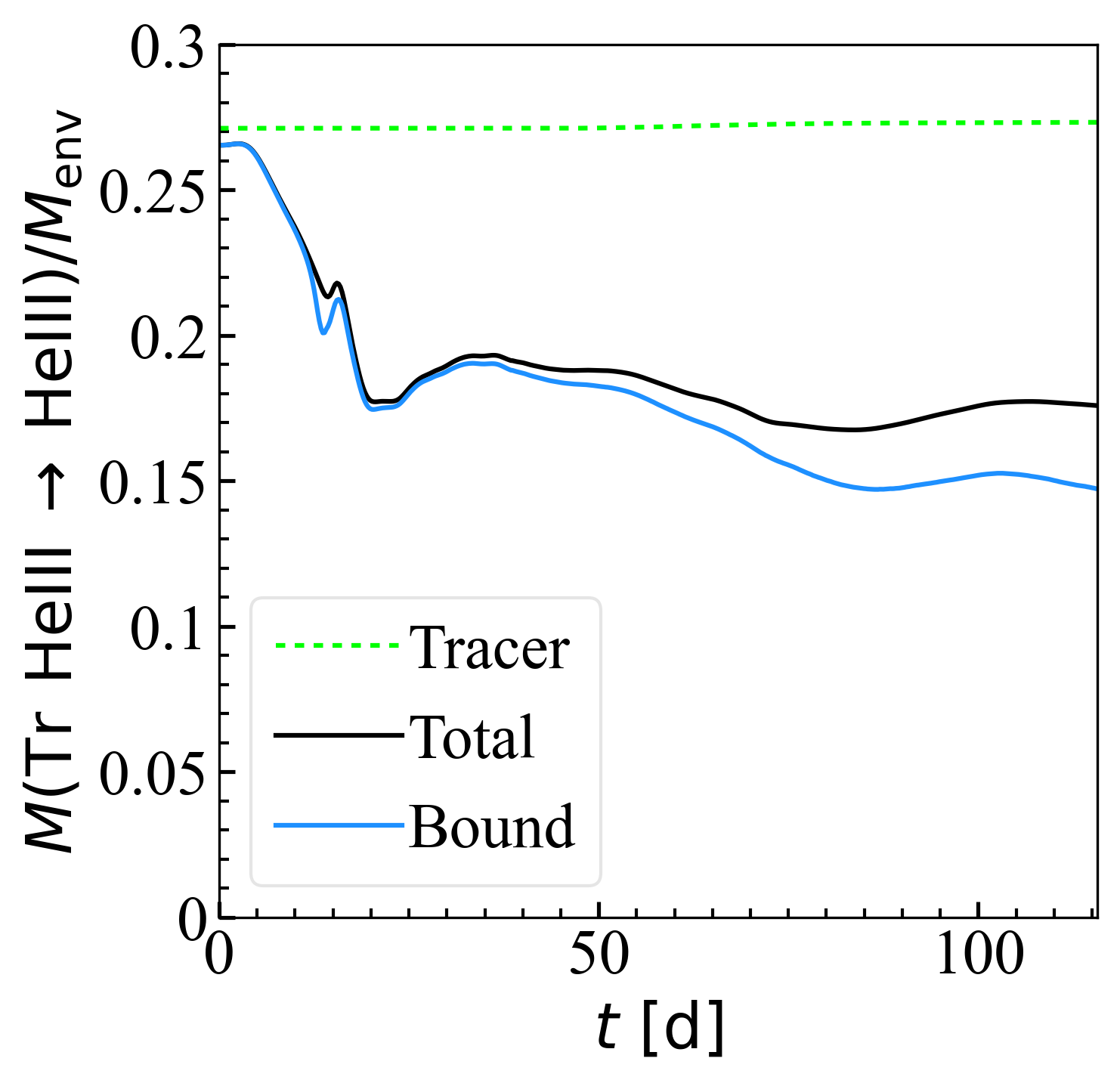}\\
\caption{Mass evolution for the various ionic transitions of helium in Model~B,
for all envelope gas (black) and envelope gas that is bound at time $t$ (blue).
Most of the helium gas is in the state $\HeIII$ at $t=0$,
and almost half of this $\HeIII$ recombines during the simulation to become $\HeII$ or $\HeI$.
The green line showing the total mass of the tracer reduces with time 
because envelope gas leaves the simulation domain through the boundary;
this outgoing envelope gas is almost completely comprised of unbound gas, 
by our fiducial definition of equation~\eqref{unbound}.
        }            
\label{fig:DeltaM_He}
\end{figure*}

Given that we are interested in explaining envelope unbinding, 
we are mainly interested in the energetics of the part of the envelope that is still bound at the time of recombination/ionization.
Because the release (acquisition) of recombination energy during recombination (ionization) is local in space,
we are interested in recombination/ionization that takes place in bound envelope gas.
Therefore, we often focus on the blue curves in the discussion,
but keep in mind that we technically mean gas that is bound at time $t$, 
not necessarily at the time of recombination/ionization.
From the right column of Fig.~\ref{fig:DeltaM_He}, 
we see that most of the recombination has happened in gas that is bound.
By contrast, while most of the $\mathrm{\HeI}$ and $\mathrm{\HeII}$ tracer gas ionizes into $\mathrm{\HeIII}$ 
(left and middle panels of bottom row),
there is almost no net ionization of the bound component.

The energy release shown in Fig.~\ref{fig:DeltaE_He} is calculated in the following way. 
For each cell in the simulation domain, 
we calculate the ionization state at time $t$, including partial ionization,
by making use of the Saha equation.
This allows us to compute the total recombination energy for all the cells of each ionization state.
Importantly, we do this for each ion tracer separately, 
adding up the recombination energy from all the cells of that tracer only.
This gives us the total recombination energy of a given ionic species $y$ that started out in ionic state $x$.
This is similar to what was done by eye by comparing the third and fourth rows 
of Figs.~\ref{fig:He_face-on_0-50} and \ref{fig:He_face-on_100-400} in Sec.~\ref{sec:recombination_spatial}.
However, now we are integrating over the full volume of envelope gas (rather than just a slice) 
and including partial ionization.
Recall that we were also able to estimate, by visual inspection of the fifth row of those figures, 
how much recombination takes place in envelope gas that remains bound.
Here we can choose to include only envelope gas that is bound at time $t$ in the integration, 
allowing us to plot the bound envelope gas separately.
Values are calculated with respect to the MESA model.
Initial values are nonzero for at least two reasons. 
First, the high temperature near the surface caused by imposing a higher pressure there (Sec.~\ref{sec:ambient})
causes extra ionization in the initial condition, as mentioned in Sec.~\ref{sec:recombination_spatial}.
Second, partial ionization regions are not accounted for by the tracers, 
which are assumed to represent gas of a single ionization state of H and He (Sec.~\ref{sec:tracers}).
This effect apparently causes a slight overestimate of the ionization of the MESA star,
and hence gives slightly positive values of the released recombination energy at $t=0$.

\subsection{Direct calculation of total recombination energy release}
Here we explain how the dashed curves in Fig.~\ref{fig:DeltaE_bnd_compare} are obtained.
To make a fair comparison with the solid curves which considered only tracer gas that is still present 
in the simulation domain at time $t$, 
we define $\Delta E_\mathrm{rec,env}$ to be the difference between the recombination energy in the simulation domain
and that of the \textit{same gas} at $t=0$.
This initial value (which depends on $t$) is estimated as follows.
We first calculate the mass of each ion tracer at time $t$.
Then we multiply by the hydrogen or helium mass fraction, 
divide by the hydrogen or helium atomic mass to obtain the number of atoms,
multiply by the appropriate ionization energy,
and sum the contributions from all hydrogen and helium ionic species.
For example, the initial recombination energy of $\mathrm{\HeIII}$ tracer gas at time $t$ is estimated as
\begin{equation}
  E_\mathrm{rec,\HeIII}(0) = \frac{M(\mathrm{\HeIII}) f_\mathrm{m}(\mathrm{He})}{\mHe} 
  (\chi_\mathrm{\HeI\rightarrow \HeII} + \chi_\mathrm{\HeII\rightarrow \HeIII}),
\end{equation}
where $M(\mathrm{\HeIII})$ is the mass of $\mathrm{\HeIII}$ tracer gas at time $t$, 
$f_\mathrm{m}(\mathrm{He})$ is the mass fraction of He ($=0.29$),
$\mHe$ is the mass of the helium atom,
and $\chi_\mathrm{\HeI\rightarrow \HeII}$ and $\chi_\mathrm{\HeII\rightarrow \HeIII}$
are the ionization energies associated with the relevant ionic transitions.
This gives us an estimate which neglects metals.
This estimate is then subtracted from $E_\mathrm{rec,env}(t)$ to obtain $\Delta E_\mathrm{rec,env}(t)$.
The initial value of $-\Delta E_\mathrm{rec,env}$ then turns out to be
negative rather than zero, implying the presence of extra recombination energy at $t=0$.
This initial value matches quite well with the recombination energy of metals in the star at $t=0$ 
that is calculated from the MESA model.
$\Delta E_\mathrm{rec,env}$ is then redefined to equal zero at $t=0$.

The part of $E_\mathrm{rec,env}$ associated with unbound gas is calculated using condition~\eqref{unbound}.
The same procedure is done for the unbound gas only to obtain $\Delta E_\mathrm{rec,env,unb}$, 
and then we calculate the bound component using $E_\mathrm{rec,env,bnd}=E_\mathrm{rec,env}-E_\mathrm{rec,env,unb}$.

\section{Role of radiation}\label{sec:radiation}
\begin{figure}
\includegraphics[scale=0.15,clip=true,trim= 290  100 238 200]{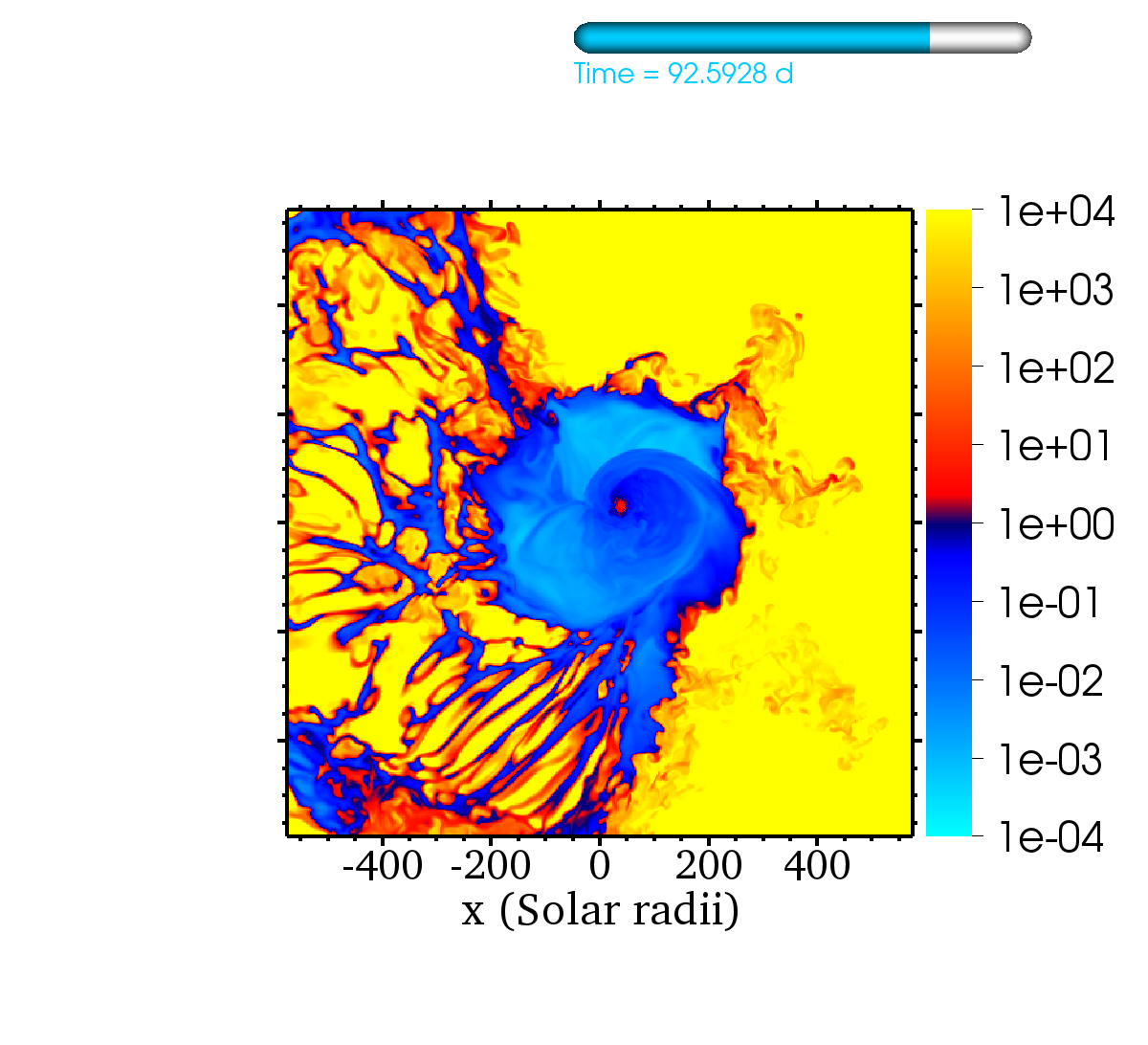}
\includegraphics[scale=0.15,clip=true,trim= 290  100 0 200]{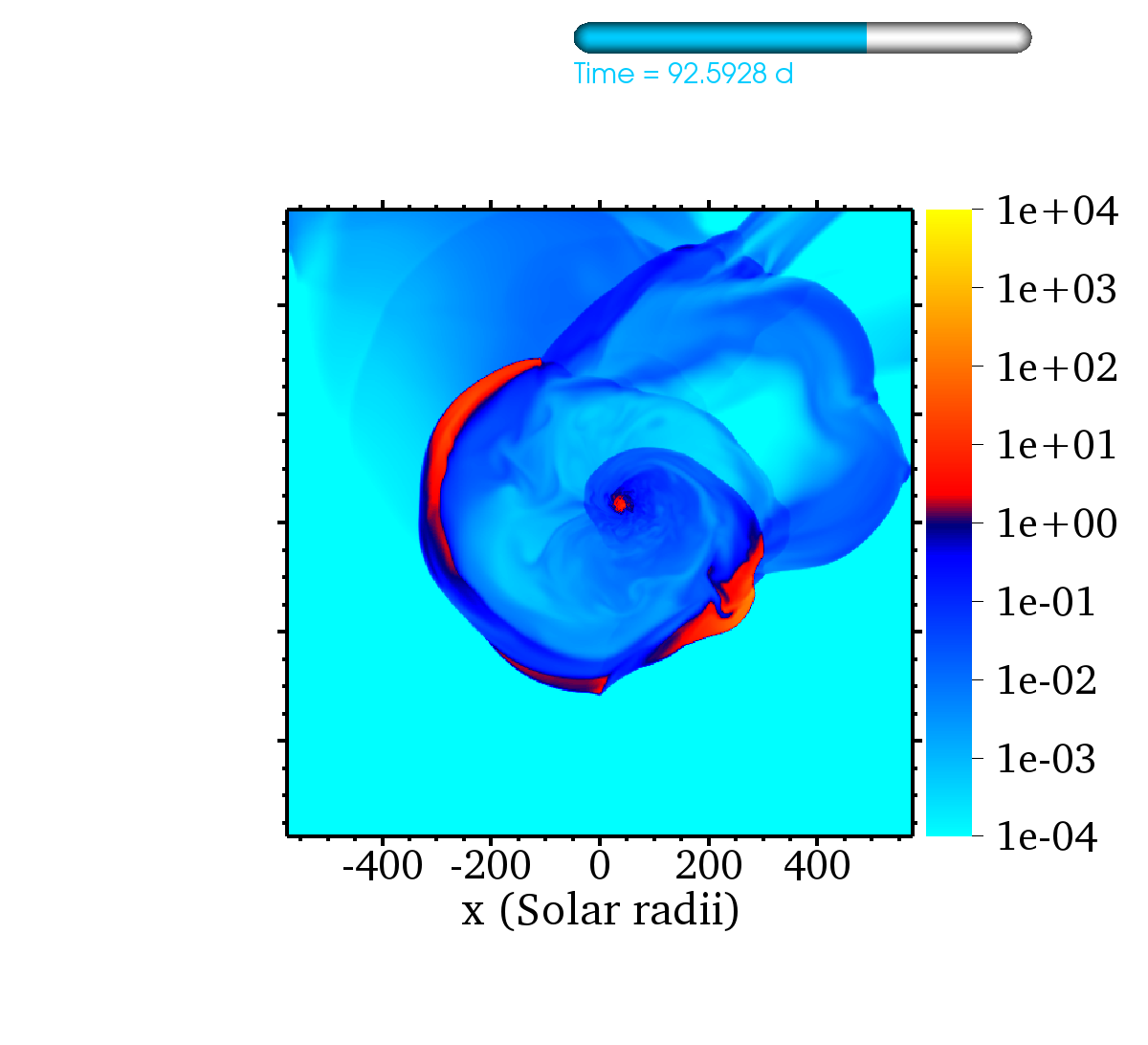}
\caption{Ratio of the quantity $a\rad T^4$ 
to the gas thermal energy density $3P\gas/2$ for the ideal gas run Model~A (left) 
and ideal gas reduced ambient run Model~D (right).
The radiation energy is not explicitly included in any of the simulations.
The temperature is estimated by assuming $\mu=0.62$, though the results are insensitive to this choice.
}
\label{fig:ambient_radiation}
\end{figure}

The leftover radiation energy that is present on account of the slightly different value of $a\rad$ mistakenly used when subtracting $a\rad T^4$
from the MESA EOS tables (see Sec.~\ref{sec:methods}) is accounted for in the analysis.
Considering that this leftover energy is just $0.015$ per cent of the actual radiation energy
that would be present assuming thermodynamic equilibrium,
Fig.~\ref{fig:DeltaE_bnd_compare} implies that $a\rad T^4$ can become exceedingly large in Models~A and B, where a dense ambient is present.
In fact, this term can greatly exceed the gas thermal energy density $3P\gas/2$ in regions where envelope gas and ambient gas are mixing.
In Fig.~\ref{fig:ambient_radiation} we plot the ratio 
\begin{equation}
  \frac{a\rad T^4}{\mathcal{E}\gas}=\frac{2}{3}a\rad\left(\frac{\mu\mH}{\rho\kB}\right)^4P\gas^3.
\end{equation}
at $t=92.6\da$ for Model~A (left) and Model~D (right).%
\footnote{Note that $a\rad T^4$ is calculated more directly in Model~B, since $\mu$ is obtained from the simulation. 
The plot of $a\rad T^4/\mathcal{E}\gas$ for Model~B (not shown) is very similar to the plot for Model~A, as expected.}
From the figure, we see that even in the reduced ambient run, Model~D, 
$a\rad T^4$ exceeds $3P\gas/2$ at the very centre where the core particles are located and in outward moving spiral density wakes.
Thus, even though radiation was negligible in the initial envelope, it may become important during CEE.
However, note that if radiation energy and pressure had been included in the simulation,
then regions of such large $a\rad T^4$ would be less likely because (i)~they would require extra energy to form,
and (ii)~radiation pressure would resist the concentration of radiation.
Moreover, for low enough optical depth radiation transport would also impede the formation of such regions.

\section{Deviation from perfect energy and angular momentum conservation}\label{app:conservation}
\begin{figure}
\includegraphics[width=1.0\columnwidth,clip=true,trim= 0 0 0 0]{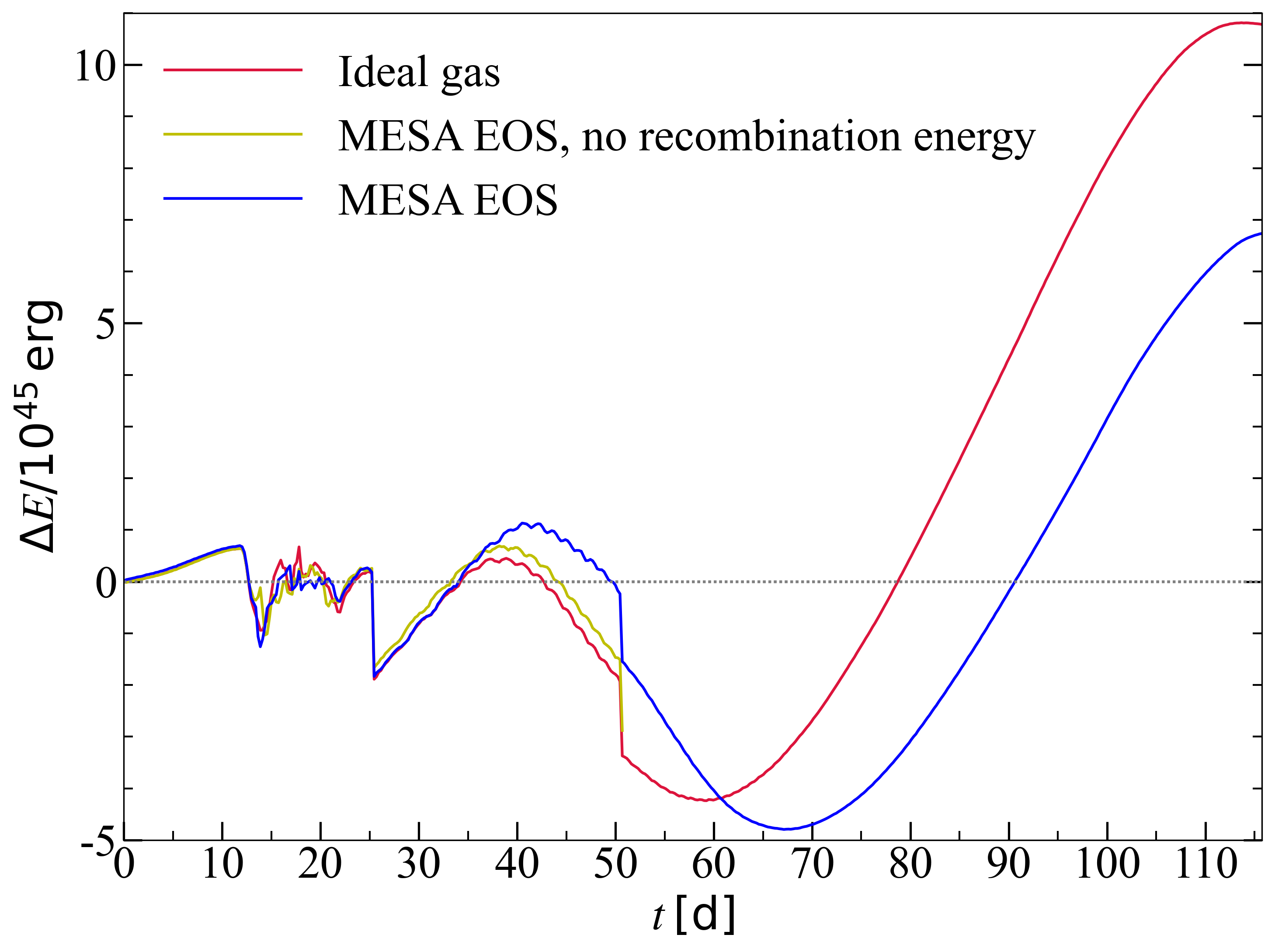}
\caption{Evolution of the total energy, accounting for energy contained in gas that passes through the domain boundary,
for Model~A (red), Model~B (blue) and Model~C (yellow).
Negative jumps in the energy at $t\approx25\da$ and $t\approx50\da$ are caused by 
the sudden halving of the softening radius of both core particles.
}            
\label{fig:energy_conservation}
\end{figure}

\begin{figure}
\includegraphics[width=1.0\columnwidth,clip=true,trim= 0 0 0 0]{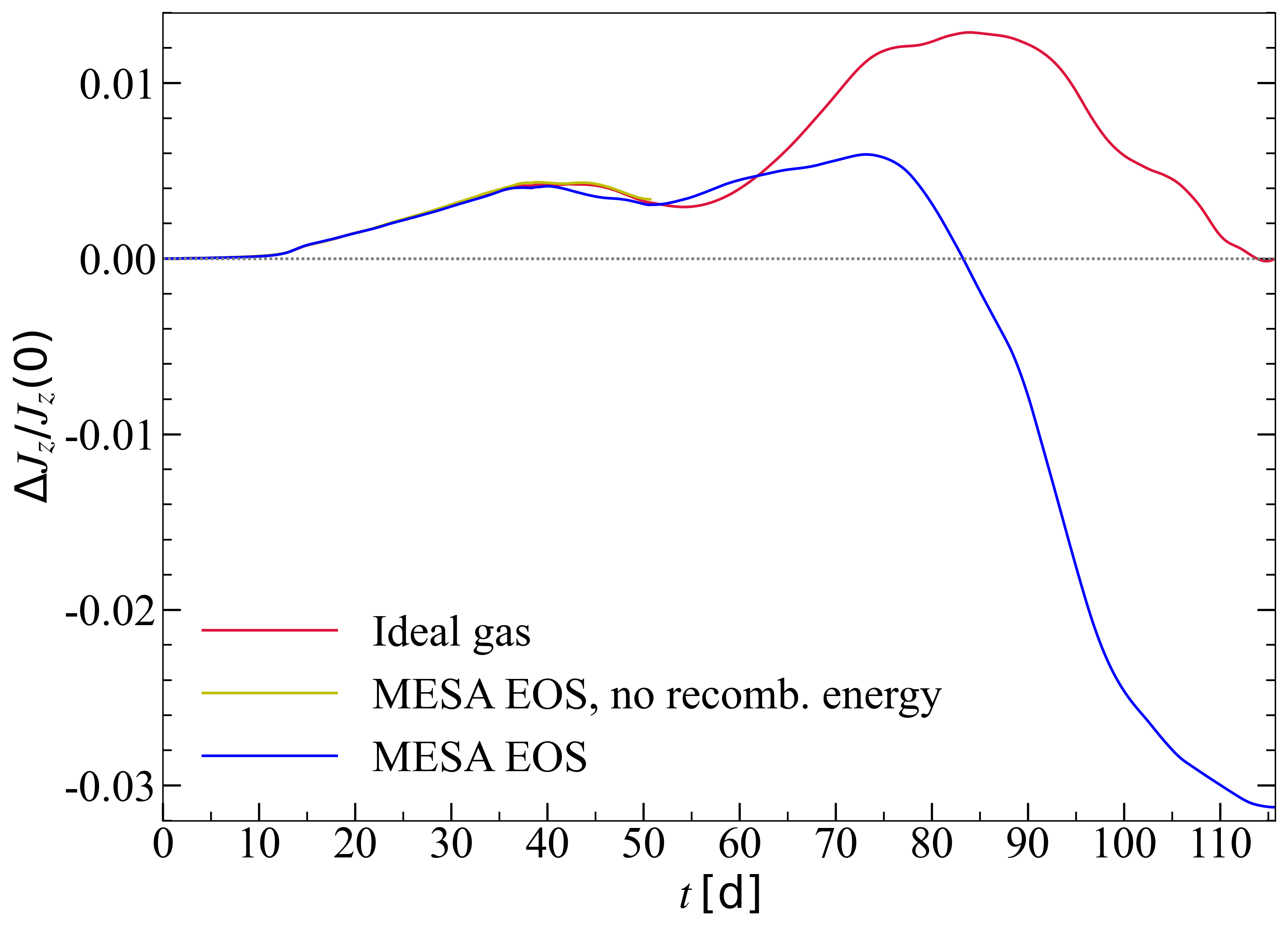}
\caption{Evolution of the total angular momentum, accounting for angular momentum contained in gas that passes through the domain boundary,
for Model~A (red), Model~B (blue) and Model~C (yellow).
}            
\label{fig:angmom_conservation}
\end{figure}

The evolution of the total energy and angular momentum in the simulation, 
including fluxes through the boundaries,
are shown in Fig.~\ref{fig:energy_conservation} and Fig.~\ref{fig:angmom_conservation}, respectively.
The total energy and angular momentum variation are small, 
and the differences between the runs smaller.
In some cases these differences (over the entire domain) 
can be comparable to differences discussed in this work 
(restricted to bound gas or the core particle orbit).
However, we note that the differences seen in 
Figs.~\ref{fig:energy_conservation} and \ref{fig:angmom_conservation}
would, in any case, have the wrong sign to explain the deeper inspiral in Model~A as compared to Model~B.
Model~D is not plotted because there is no way to account for the energy and angular momentum change 
caused by the removal of ambient gas.

Some non-conservation of energy and angular momentum 
is caused by the exclusion by the multipole Poisson solver 
of the potential associated with gas with $r>L_\mathrm{box}/2$.
Note that the sum of \textit{envelope} gas and particle orbital energy 
(black in Fig.~\ref{fig:Env_E_terms_ambient}) 
eventually increases with time in both simulations.
In Model~D the ambient energy is insignificant so transfer from the ambient cannot be causing the rise.
The rise may be caused by the potential energy gain (toward less negative values) 
due to gas moving out of the sphere $r=L_\mathrm{box}/2$.
Since more gas leaves the box and hence the sphere $r=L_\mathrm{box}/2$ in Model~D than Model~A, 
the extra energy gain from this effect in Model~D would then be comparable to the extra energy gained from the ambient in Model~A.

Imperfect energy and angular momentum conservation during the simulation
could significantly affect the estimate for $\alpha\CE$ of Section~\ref{sec:extrapolation} 
\textit{if} the \textit{particle} energy and angular momentum are significantly impacted,
which may or may not be the case.
To obtain a conservative estimate of how much this could change our estimate for $\alpha\CE$, 
we considered the unlikely scenario that all of the deviation from non-conservation is from the particle orbital energy.
We then recalculated the $\alpha\CE$ estimates for Model~D assuming energy and angular momentum non-conservation comparable to Models~A and B.
The gain in energy seen in Fig.~\ref{fig:energy_conservation} would then imply that $\dot{E}_{1-2}$ was in reality more negative, 
which leads to a smaller estimate for $\alpha\CE$ ($0.43$ instead of $0.55$).
The loss of angular momentum seen in Fig.~\ref{fig:angmom_conservation} would then imply that $\dot{J}_{z,1-2}$ was in reality less negative, 
which leads to a larger estimate for $\alpha\CE$ ($0.51$ instead of $0.22$).
In both cases, the estimates remain within the range $0.27<\alpha\CE<0.55$ mentioned in Section~\ref{sec:extrapolation}.

\section{Adiabatic index}
\begin{figure}
\centering
\includegraphics[scale=0.25,clip=true,trim= 200  100 0 200]{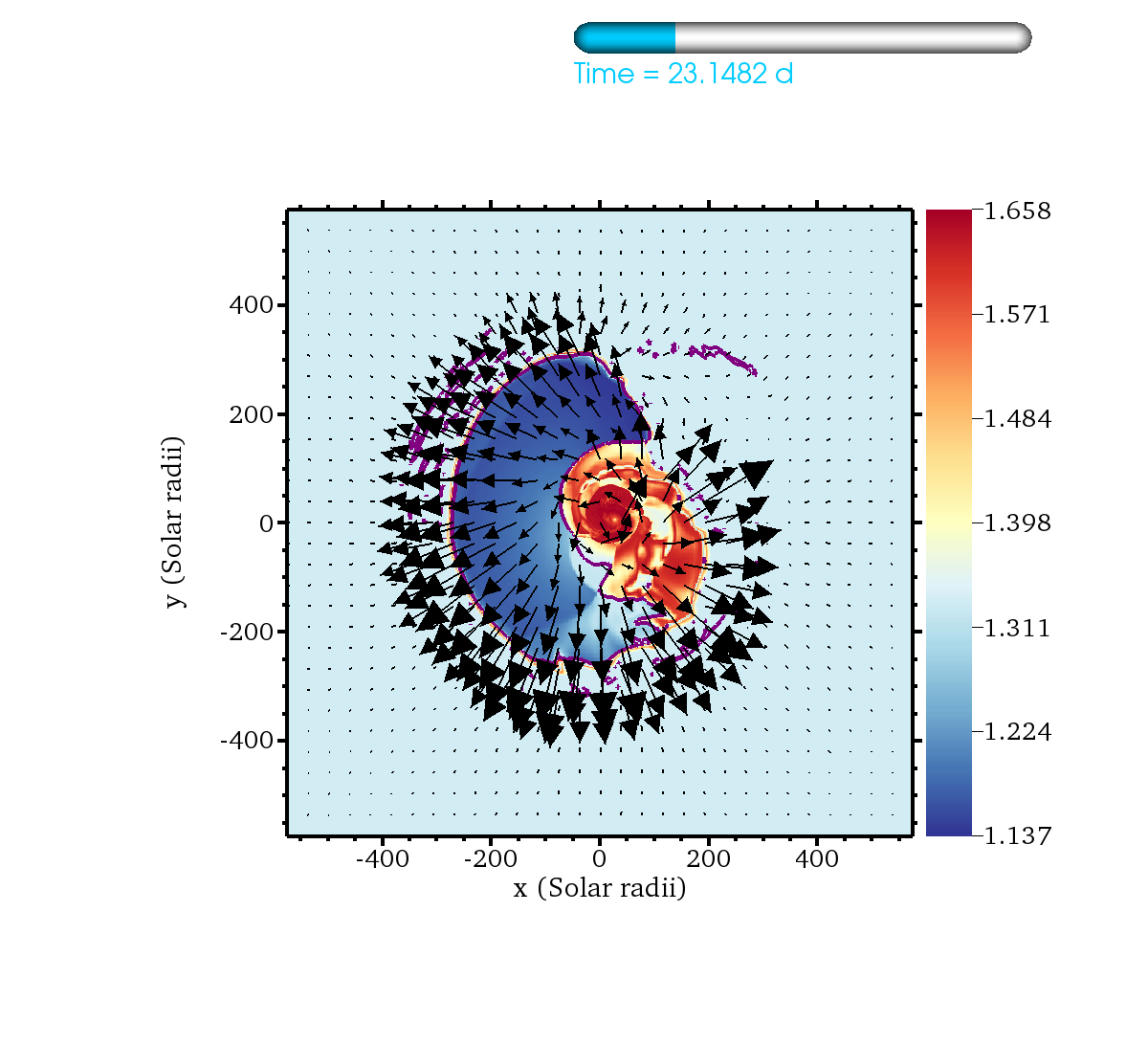}\\
\includegraphics[scale=0.25,clip=true,trim= 200  100 0 200]{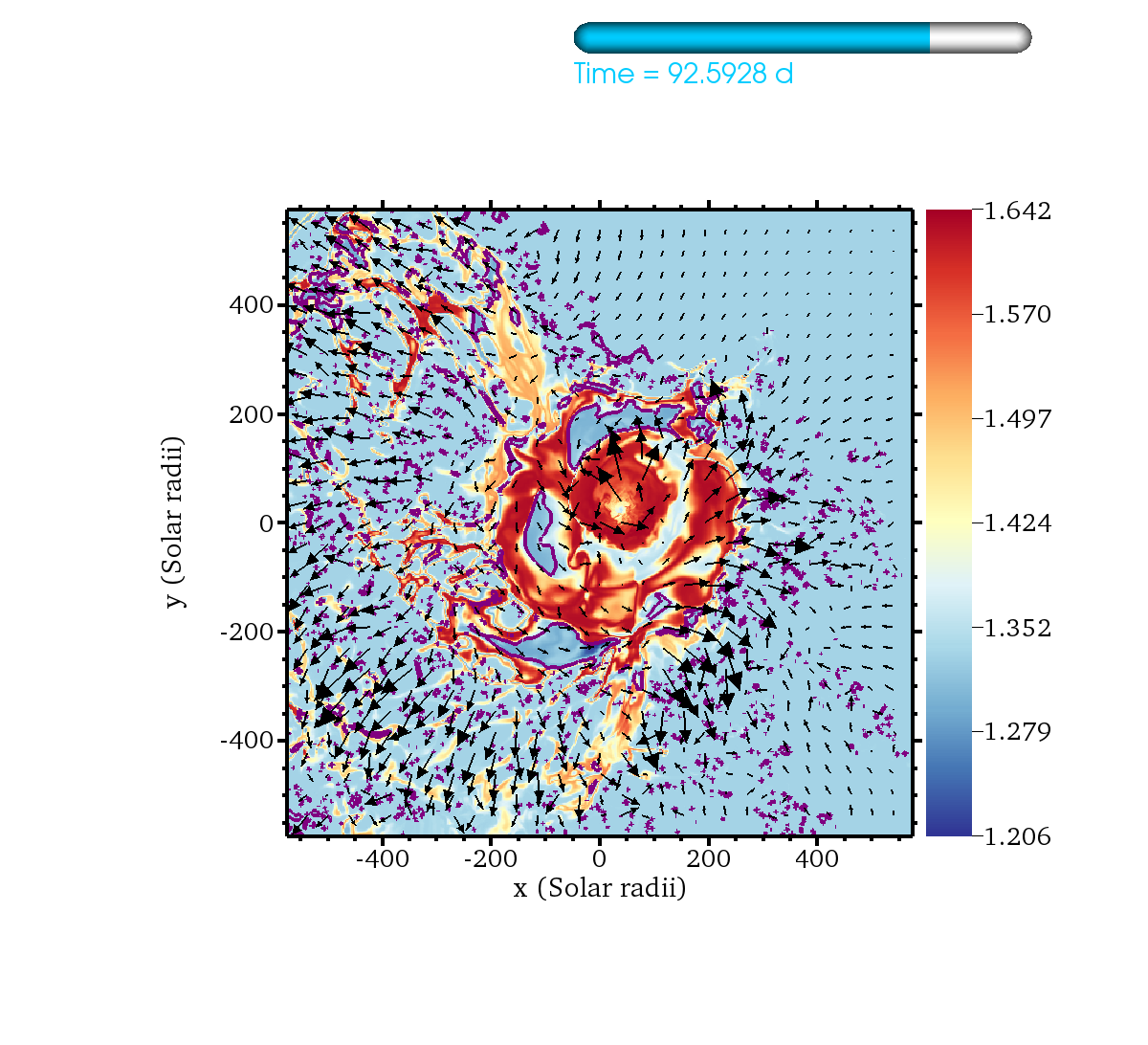}
\caption{The quantity $\Gamma_1=(d\log P/d\log\rho)_\mathrm{ad}$ plotted in the orbital plane at $t=23.1\da$ (top) and $t=92.6\da$ (bottom).
Purple contours show $\Gamma_1=4/3$ and vectors show the gas velocity in the lab frame, projected onto the orbital plane.}
\label{fig:Gamma1}
\end{figure}

In stars, low values of the adiabatic index $\Gamma_1=(d\log P/d\log\rho)_\mathrm{ad}$
can lead to dynamical instability.
For example, in the idealized case of homologous motion 
in a constant density, constant $\Gamma_1$ sphere,
instability sets in for $\Gamma_1<4/3$ \citep[e.g.][chapter~8]{Hansen+04}.
In Figure~\ref{fig:Gamma1}, we present $\Gamma_1$
in a slice through the orbital plane at $t=23.1\da$ (top) and $t=92.6\da$ (bottom), for Model~B.
Purple contours show $\Gamma_1=4/3$, and velocity vectors are also plotted.
While the high-density gas surrounding the particles has $\Gamma_1$ close to $5/3$, 
some gas does have $\Gamma_1<4/3$.
In both snapshots, the $\Gamma_1<4/3$ region traces rather precisely the region dominated by $\HeI$ 
(shown in the third row of Fig.~\ref{fig:He_face-on_100-400}).
The mechanism by which extra unbinding occurs in tabulated EOS runs as compared to ideal gas runs is difficult to precisely identify,
but the low values of $\Gamma_1$ may be playing a role.

\end{document}